\documentclass[a4paper,fleqn,usenatbib,useAMS]{mnras}

\usepackage{newtxtext,newtxmath}

\usepackage[T1]{fontenc}

\usepackage{graphicx}	% Including figure files
\usepackage{amsmath}	% Advanced maths commands
\usepackage{amssymb}	% Extra maths symbols

\newcommand{\snia}{SN~Ia}
\newcommand{\sneia}{SNe~Ia}

\let\ts=\thinspace
\newcommand{\one}{\ts {\sc i}}
\newcommand{\two}{\ts {\sc ii}}
\newcommand{\three}{\ts {\sc iii}}

\newcommand{\nifs}{\ensuremath{^{56}\rm{Ni}}}
\newcommand{\cofs}{\ensuremath{^{56}\rm{Co}}}
\newcommand{\fefs}{\ensuremath{^{56}\rm{Fe}}}
\newcommand{\msun}{\ensuremath{\rm{M}_{\odot}}}
\newcommand{\mtot}{\ensuremath{M_\mathrm{tot}}}
\newcommand{\mratio}{$M(\nifs)/$\mtot}

\newcommand{\kms}{\ensuremath{\rm{km\,s}^{-1}}}

\newcommand{\gcc}{\ensuremath{\rm{g\,cm}^{-3}}}
\newcommand{\ergs}{\ensuremath{\rm{erg\,s}^{-1}}}

\newcommand{\dmft}{\ensuremath{\Delta M_{15}(B)}}

\newcommand{\dmftuvoir}{\ensuremath{\Delta M_{15}(\rm{uvoir})}}
\newcommand{\mch}{\ensuremath{M_{\rm Ch}}}

\def\cmfgen{{\sc cmfgen}}

\title[The sub-Chandrasekhar-mass Type Ia SN~1999by]
{The detonation of a sub-Chandrasekhar-mass white dwarf at the origin of the low-luminosity Type Ia supernova 1999by}
\author[St\'ephane Blondin et al.]
{
St\'ephane Blondin,$^{1}$\thanks{E-mail: stephane.blondin@lam.fr}
Luc Dessart,$^{2}$
and D.~John Hillier$^{3}$
\\
$^{1}$Aix Marseille Univ, CNRS, LAM, Laboratoire d'Astrophysique de Marseille, Marseille, France\\
$^{2}$Unidad Mixta Internacional Franco-Chilena de Astronom\'ia (CNRS UMI 3386),
    Departamento de Astronom\'ia, Universidad de Chile,\\
    Camino El Observatorio 1515, Las Condes, Santiago, Chile\\
$^{3}$Department of Physics and Astronomy \& Pittsburgh Particle
Physics, Astrophysics, and Cosmology Center (PITT PACC), University of
Pittsburgh,\\ Pittsburgh, PA 15260, USA
}

\date{Accepted 2017 November 15. Received 2017 November 7; in original form 2017 September 8}
\pubyear{2017}

\begin{document}
\label{firstpage}
\pagerange{\pageref{firstpage}--\pageref{lastpage}}
\maketitle

%%%%%%%%%%%%%%%%%%%%%%%%%%%%%%%%%%%%%%%%%%%%%%%%%%%%%%%%%%%%%%%%%%%%%%

\begin{abstract}
While Chandrasekhar-mass (\mch) models with a low \nifs\ yield can
match the peak luminosities of fast-declining, 91bg-like Type Ia
supernovae (\sneia), they systematically fail to reproduce their
faster light-curve evolution.  Here we illustrate the impact of a low
ejecta mass on the radiative display of low-luminosity \sneia, by
comparing a sub-\mch\ model resulting from the pure central detonation
of a C-O White Dwarf (WD) to a \mch\ delayed-detonation model with the
same \nifs\ yield of 0.12\,\msun.  Our sub-\mch\ model from a
0.90\,\msun\ WD progenitor has a $\sim5$\,d shorter rise time in the
integrated UV-optical-IR (uvoir) luminosity, as well as in the
$B$-band, and a $\sim20$ per cent higher peak uvoir luminosity
($\sim1$\,mag brighter peak $M_B$). This sub-\mch\ model also displays
bluer maximum-light colours due to the larger specific heating rate,
and larger post-maximum uvoir and $B$-band decline rates. The
luminosity decline at nebular times is also more pronounced,
reflecting the enhanced escape of gamma rays resulting from the lower
density of the progenitor WD. The deficit of stable nickel in the
innermost ejecta leads to a notable absence of forbidden lines of
[Ni\two] in the nebular spectra. In contrast, the \mch\ model displays
a strong line due to [Ni\two] 1.939\,$\mu$m, which could in principle
serve to distinguish between different progenitor scenarios.  Our
sub-\mch\ model offers an unprecedented agreement with optical and
near-infrared observations of the 91bg-like SN~1999by, making a strong
case for a WD progenitor significantly below the Chandrasekhar-mass
limit for this event and other low-luminosity
\sneia.
\end{abstract}


%%%%%%%%%%%%%%%%%%%%%%%%%%%%%%%%%%%%%%%%%%%%%%%%%%%%%%%%%%%%%%%%%%%%%%

\section{Introduction}\label{sect:intro}

The Chandrasekhar mass for white dwarf (WD) stars ($\mch\approx
1.4$\,\msun) no longer appears to represent a fundamental quantity for
Type Ia supernova (\snia) progenitors.  The standard scenario involves
a C-O WD that undergoes runaway carbon fusion as it approaches
\mch\ through accretion from a non-degenerate binary companion star
\citep{Whelan/Iben:1973}. However, recent studies have demonstrated a
reasonable agreement with observations through detonations of single
sub-\mch\ WDs \citep[e.g.][]{Sim/etal:2010}, or in double-WD mergers
whose combined mass can either exceed
\mch\ \citep[e.g.][]{Pakmor/etal:2013} or remain below this limit
\citep{vanKerkwijk/etal:2010}. Such models provide viable alternatives
to the standard scenario and can account for the observed \snia\ rate
\citep[e.g.][]{Ruiter/etal:2011,Ruiter/etal:2013}. Variations in the
mass of the exploding WD result in a range of \nifs\ yields that can
reproduce the observed diversity in peak luminosity.

In particular, low-luminosity \sneia\ similar to the prototypical
SN~1991bg (and hence termed 91bg-like \sneia; see
\citealt{Taubenberger/etal:2008} for a review) appear difficult to
reconcile with a Chandrasekhar-mass ejecta.  These 91bg-like
\sneia\ share similar properties, with peak integrated UV-optical-IR
(hereafter uvoir) luminosities $\lesssim 3\times
10^{42}$\,\ergs\ (cf. $> 10^{43}$\,\ergs\ for more typical events),
corresponding to a \nifs\ yield of only $\sim0.1$\,\msun.  Their rapid
light-curve evolution around maximum light, along with the earlier
appearance of an emission-line dominated spectrum characteristic of
the nebular phase, prompted their association with
sub-\mch\ progenitors
(\citealt{Filippenko/etal:1992b,Leibundgut/etal:1993,Ruiz-Lapuente/etal:1993}).
Subsequent spectroscopic modeling of nebular-phase spectra constrained
the innermost ejecta layers to have a substantially lower density than
\mch\ models \citep{Mazzali/etal:2011,Mazzali/Hachinger:2012}, a
property that naturally follows from the explosion of a lower-mass WD
progenitor.

The merger of two equal-mass $\sim0.9$\,\msun WDs also results in a
lower density ejecta \citep{Pakmor/etal:2010}. The low \nifs\ mass
synthesized in the detonation of the merged system is compatible with
the low peak luminosity of 91bg-like \sneia. However, the predicted
light curves are too broad owing to the large ejecta mass ($\sim
1.8$\,\msun). \cite{Pakmor/etal:2013} speculate that the merger of a
less massive system consisting of a 0.9\,\msun\ C-O WD and a He WD
companion could reconcile their model with low-luminosity \sneia,
although this remains to be demonstrated with radiative-transfer
simulations.

The ejecta mass therefore stands out as a key discriminant between
these different progenitor scenarios, hence the importance of studying
its impact on the radiative display of low-luminosity \sneia.  In
\cite{Blondin/etal:2013} we were able to reproduce the maximum-light
properties of the 91bg-like Type Ia SN~1999by with a
\mch\ delayed-detonation model, but subsequently found this model to
decline too slowly past maximum light.  In contrast, a sub-\mch\ model
with the same \nifs\ yield not only matches the peak luminosity, but
also declines more rapidly past maximum, with a $\sim0.6$\,mag larger
$B$-band decline rate, \dmft. Such a model is thus in better agreement
with the faint end of the width-luminosity relation
\citep{Blondin/etal:2017a}.  Here we test the potential for this
sub-\mch\ model to reproduce the full photometric and spectroscopic
evolution of SN~1999by from a few days past explosion ($\sim 10$\,d
before maximum) until well into the nebular phase ($\sim 180$\,d past
maximum). We also search for unambiguous observational signatures of a
low progenitor mass for low-luminosity \sneia.

This paper is organized as follows: In Section~\ref{sect:model} we
present our input hydrodynamical models for the explosion.  We then
discuss the impact of a low ejecta mass on the radiative display of
low-luminosity \sneia\ in Section~\ref{sect:lowmass}, by comparing a
sub-\mch\ model to a \mch\ delayed-detonation model with the same
\nifs\ yield of 0.12\,\msun. The present study is similar in spirit to
the recent work of \cite{Wilk/etal:2017}, who studied the impact of
the WD mass for more luminous \snia\ models yielding
$\sim0.6$\,\msun\ of \nifs, and including ejecta masses below, at, and
above \mch. In Section~\ref{sect:99by}, both the sub-\mch\ and
\mch\ models are confronted to optical and near-infrared (NIR)
observations of the low-luminosity SN~1999by. We discuss possible
progenitor scenarios leading to the detonation of a sub-\mch\ WD and
present our conclusions in Section~\ref{sect:ccl}.

%%%%%%%%%%%%%%%%%%%%%%%%%%%%%%%%%%%%%%%%%%%%%%%%%%%%%%%%%%%%%%%%%%%%%%

\section{Input hydrodynamical models}\label{sect:model}

\begin{table}
\footnotesize
\caption{Basic properties of the \mch\ delayed-detonation model
DDC25 of {\protect\cite{Blondin/etal:2013}} and the sub-\mch\ model SCH2p0 of {\protect\cite{Blondin/etal:2017a}}. The
\nifs\ mass is given at $t_{\rm exp}\approx0$. All other yields
correspond to 0.75\,d past explosion.
}\label{tab:sch2p0_ddc25_comp}
\begin{tabular}{l@{\hspace{8.5mm}}c@{\hspace{8.5mm}}c@{\hspace{8.5mm}}c}
\hline
Property & Unit & DDC25 & SCH2p0 \\
\hline
\multicolumn{4}{c}{\it Global properties}\\
$M_{\mathrm{tot}}$ & M$_\odot$ & 1.41 & 0.90 \\
$E_{\mathrm{kin}}$ & erg & 1.18\,(51) & 8.14\,(50) \\
$E_{\mathrm{kin}}$$/$$M_{\mathrm{tot}}$ & erg g$^{-1}$ & 4.20\,(17) & 4.54\,(17) \\
$\langle \varv_m \rangle$ & km s$^{-1}$ & 8132 & 8815 \\
$M(\nifs)_{t=0}$ & M$_\odot$ & 0.117 & 0.116 \\
$M(^{56}\mathrm{Ni})_{t=0}$$/$$M_{\mathrm{tot}}$ & $\cdots$ & 0.083 & 0.129 \\
$\varv_{99}(\nifs)$ & km s$^{-1}$ & 8559 & 10520 \\
$M_{99}(^{56}\mathrm{Ni})$ & M$_\odot$ & 0.792 & 0.607 \\
$M(^{58}\mathrm{Ni})$ & M$_\odot$ & 2.30\,($-$2) & 1.36\,($-$3) \\
$M(^{54}\mathrm{Fe})$ & M$_\odot$ & 6.61\,($-$2) & 1.88\,($-$2) \\
\hline
\multicolumn{4}{c}{\it Elemental yields at 0.75 d past explosion}\\
$M(\mathrm{Ni})$ & M$_\odot$ & 0.137 & 0.109 \\
$M(\mathrm{Fe})$ & M$_\odot$ & 9.77\,($-$2) & 2.21\,($-$2) \\
$M(\mathrm{Ti})$ & M$_\odot$ & 1.13\,($-$4) & 1.58\,($-$5) \\
$M(\mathrm{Sc})$ & M$_\odot$ & 3.09\,($-$7) & 1.02\,($-$7) \\
$M(\mathrm{Ca})$ & M$_\odot$ & 2.40\,($-$2) & 3.01\,($-$2) \\
$M(\mathrm{S})$ & M$_\odot$ & 0.239 & 0.173 \\
$M(\mathrm{Si})$ & M$_\odot$ & 0.481 & 0.281 \\
$M(\mathrm{Mg})$ & M$_\odot$ & 3.64\,($-$2) & 2.01\,($-$2) \\
$M(\mathrm{Na})$ & M$_\odot$ & 1.53\,($-$4) & 3.93\,($-$5) \\
$M(\mathrm{O})$ & M$_\odot$ & 0.278 & 0.177 \\
$M(\mathrm{C})$ & M$_\odot$ & 2.17\,($-$2) & 7.42\,($-$3) \\
\hline
\multicolumn{4}{c}{\it Maximum-light properties$^{a}$}\\
$t_{\mathrm{rise}}(\mathrm{uvoir})$ & day & 21.0 & 15.9 \\
$t_{\mathrm{rise}}(B)$ & day & 19.8 & 14.6 \\
$L_{\mathrm{uvoir, peak}}$ & erg s$^{-1}$ & 2.62\,(42) & 3.17\,(42) \\
$\dot{E}_{\mathrm{decay}}$ & erg s$^{-1}$ & 2.10\,(42) & 2.80\,(42) \\
$\dot{E}_{\mathrm{dep}}$ & erg s$^{-1}$ & 2.06\,(42) & 2.72\,(42) \\
$\dot{e}_{\rm dep}=\dot{E}_{\rm dep}/M_{\rm tot}$ & erg s$^{-1}$ g$^{-1}$ & 7.36\,(8) & 1.52\,(9) \\
$Q_{\gamma}$ & $\cdots$ & 1.24 & 1.13 \\
$Q_{{\rm Katz,uvoir}}$ & $\cdots$ & 0.53 & 0.52 \\
$M_{B\mathrm{, peak}}$ & mag & $-$16.44 & $-$17.27 \\
$B-V$ & mag & 1.33 & 0.76 \\
$v_\mathrm{abs}$(Si{\ts {\,\sc ii}}~6355\,\AA) & km s$^{-1}$ & $-$8600 & $-$10224 \\
$v_\mathrm{1/2,opt}$ & km s$^{-1}$ & 9122 & 10848 \\
\hline
\multicolumn{4}{c}{\it Magnitude decline rates}\\
$\Delta M_{15}(\mathrm{uvoir})$ & mag & 0.62 & 0.85 \\
$\Delta M_{15}(B)$ & mag & 1.01 & 1.64 \\
d$M_{\mathrm{uvoir}}/$d$t|_{\mathrm{+50}}$ & mag day$^{-1}$ & 0.029 & 0.036 \\
d$M_{\mathrm{uvoir}}/$d$t|_{\mathrm{+100}}$ & mag day$^{-1}$ & 0.024 & 0.023 \\
\hline
\end{tabular}
\flushleft
$^{a}$Apart from $t_{\rm rise}(B)$ and $M_{B,{\rm peak}}$, all other maximum-light properties are given at uvoir maximum.\\
{\bf Notes:} Numbers in parenthesis correspond to powers of ten.\\
{\bf Meaning of various symbols:}
$\langle \varv_m \rangle$ is the mass-weighted mean velocity.
$\varv_{99}(\nifs)$ is the velocity of the ejecta shell that bounds 99\% of the total \nifs\ mass.
% Apart from $t_{\rm rise}(B)$ and $M_{B,{\rm peak}}$, all other maximum-light properties are given at uvoir maximum.
$\dot{E}_{\rm decay}$ is the instantaneous decay power, of which $\dot{E}_{\rm dep}$ is deposited in the ejecta.
$\dot{e}_{\rm dep}$ is the specific heating rate (noted $\dot{e}_{\rm decay}$ in \citealt{Blondin/etal:2017a}), corresponding to the instantaneous deposited decay power ($\dot{E}_{\rm dep}$) divided by the total mass (\mtot).
$Q_{\gamma}$ is the ratio of the uvoir luminosity to the instantaneous decay power. Arnett's rule states that $Q_{\gamma}=1$ {\protect\citep{Arnett:1979,Arnett:1982a}}.
$Q_{{\rm Katz,uvoir}}$ is the ratio of the integral of the time-weighted uvoir luminosity to the integral of the time-weighted decay luminosity up until the time of uvoir maximum {\protect\citep[see][]{Katz:2013}}.
$v_{{\rm abs}}$(Si{\ts {\,\sc ii}}~6355\,\AA) is the velocity at maximum absorption of the Si{\ts {\,\sc ii}}~6355\,\AA\ line; 
$v_{{\rm 1/2,opt}}$ corresponds to the velocity location above which half of the optical flux emerges (see Section~\ref{sect:col}).
$\mathrm{d}M_\mathrm{uvoir}/\mathrm{d}t|_{+X}$ is the instantaneous uvoir magnitude decline rate at $X$\,d past uvoir maximum.

\end{table}

%%% FIGURE: Elem distrib
\begin{figure}
\centering
\includegraphics{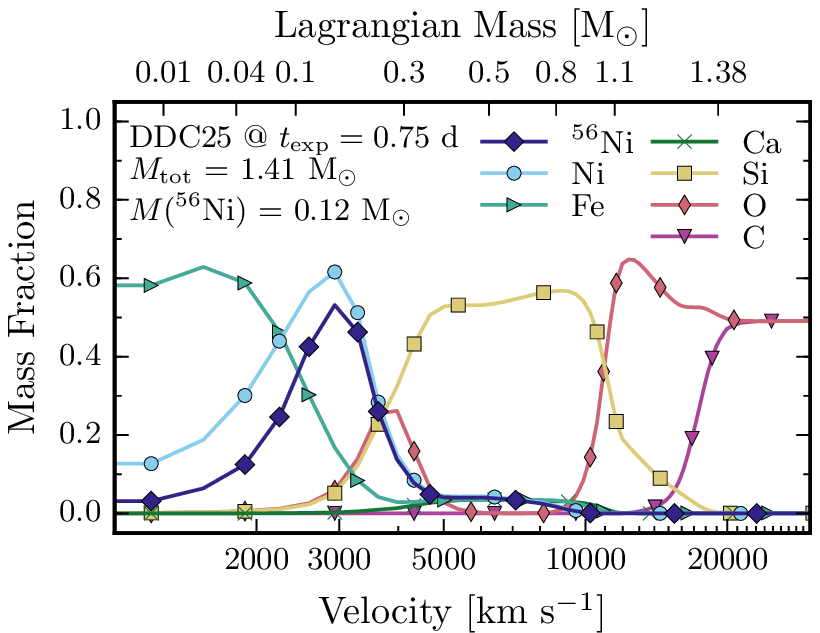}\vspace{.35cm}
\includegraphics{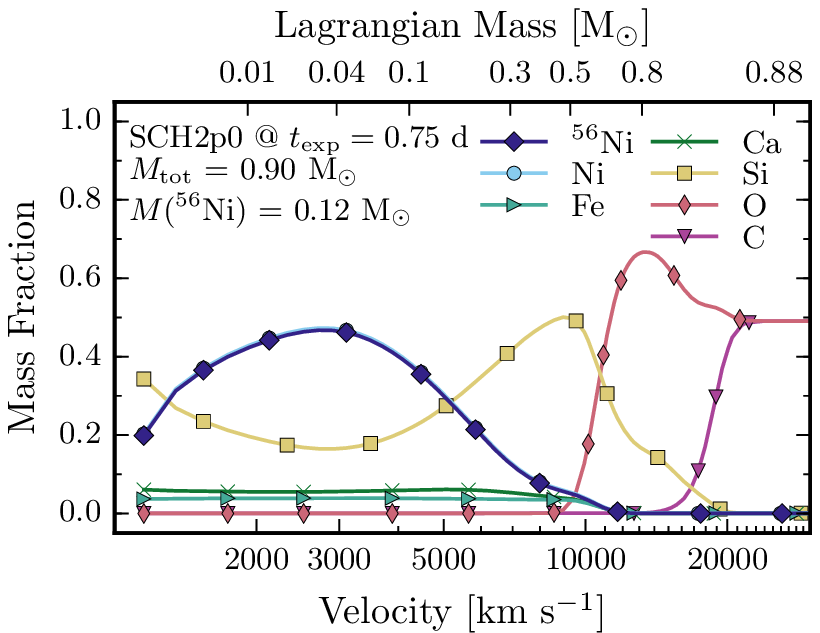}\vspace{.35cm}
\includegraphics{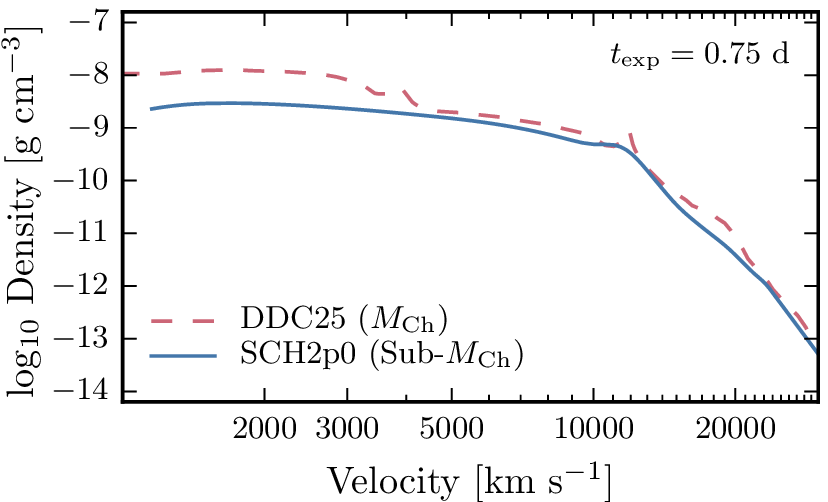}
\caption{\label{fig:elem_distrib}
Abundance profiles at 0.75\,d past explosion for the
\mch\ delayed-detonation model DDC25 (top) and the sub-\mch\ model
SCH2p0 (middle), between 1000 and 30000\,\kms\ (note the logarithmic
velocity scale; tickmarks are placed every 1000\,\kms). The upper
abscissa gives the Lagrangian mass coordinate.  Almost all the Ni in
the sub-\mch\ model is in the form of the radioactive isotope \nifs,
while the \mch\ model is dominated by stable IGEs below
$\sim0.1$\,\msun.  The bottom panel shows the density profile for both
models at this time.
}
\end{figure}

The sub-\mch\ model studied here results from the pure central
detonation of a sub-\mch\ WD in hydrostatic equilibrium, composed of
equal amounts of $^{12}$C and $^{16}$O by mass, with traces of
$^{22}$Ne and solar composition for all other isotopes.
Importantly, we do not consider the presence of an external He shell,
required in the double-detonation scenario to trigger a detonation in
the C-O core. This model has
already been presented in \cite{Blondin/etal:2017a} alongside a larger
grid of sub-\mch\ models spanning WD masses between 0.88\,\msun\ and
1.15\,\msun, corresponding to \nifs\ yields between 0.08\,\msun\ and
0.84\,\msun.  In what follows, we focus on model SCH2p0 resulting from
the detonation of a 0.90\,\msun\ WD with a \nifs\ yield of
0.12\,\msun\ (Table~\ref{tab:sch2p0_ddc25_comp}). Given the low
\nifs\ mass, the luminosity is expected to reach a peak value
comparable to low-luminosity, 91bg-like events. The \nifs\ yield is
within 0.001\,\msun\ of the \mch\ delayed-detonation model DDC25 of
\cite{Blondin/etal:2013}.\footnote{We have recomputed this model in
  the same vein as model DDC15 presented in \cite{Blondin/etal:2015},
  by applying a small radial mixing to the hydrodynamical input to
  smooth sharp variations in composition.  Apart from the two-step
  \nifs$\rightarrow$\cofs$\rightarrow$\fefs\ decay chain, we also
  treat eight additional two-step decay chains associated with
  $^{37}$K, $^{44}$Ti, $^{48}$Cr, $^{49}$Cr, $^{51}$Mn, $^{52}$Fe,
  $^{55}$Co, $^{57}$Ni, and a further six one-step decay chains
  associated with $^{41}$Ar, $^{42}$K, $^{43}$K, $^{43}$Sc, $^{47}$Sc,
  $^{61}$Co \citep[see][]{D14_Tech}. Therefore, the model differs
  slightly from the version published in \cite{Blondin/etal:2013}.} In
this study, we will repeatedly compare both models with one another
to isolate the impact of the WD mass on the radiative display for a
given \nifs\ mass.

The abundance profiles of selected species for both models are shown
in Fig.~\ref{fig:elem_distrib}, along with the density profile at
0.75\,d past explosion.\footnote{This time corresponds to the start of
  the radiative-transfer calculations for the sub-\mch\ model
  SCH2p0. Those for the \mch\ model were started at 0.5\,d past
  explosion.} The abundance profiles are nearly indistinguishable
beyond $\sim 10000$\,\kms.  This velocity corresponds to a mass
coordinate of $\sim 0.96$\,\msun\ for the \mch\ model ($\sim 69$ per
cent of the total mass), and to $\sim 0.56$\,\msun\ for the
sub-\mch\ model ($\sim 62$ per cent of the total mass). The mass
contained beyond $\sim 10000$\,\kms\ is thus $\sim 0.1$\,\msun\ larger
for the \mch\ model ($\sim 0.44$\,\msun) compared to the
sub-\mch\ model ($\sim 0.34$\,\msun).  In these outer layers, the
pre-expansion during the initial deflagration phase in the \mch\ model
results in similar combustion densities as in the sub-\mch\ WD
progenitor.

The inner $\sim0.1$\,\msun\ of the \mch\ model is dominated by stable
iron-group elements (IGEs). Stable isotopes dominate the Ni abundance
($^{58}$Ni in the mass range 0.04--0.1\,\msun, $^{60}$Ni in the mass
range 0.02--0.04\,\msun, and $^{62}$Ni below 0.02\,\msun), as well as
the Fe abundance ($^{54}$Fe above a mass coordinate of
$\sim0.05$\,\msun, primordial $^{56}$Fe --- i.e. not from \nifs\ decay
--- in the mass range 0.01--0.04\,\msun, and $^{58}$Fe below
$\sim0.01$\,\msun). The \mch\ model is thus characterized by an inner
region depleted in \nifs, commonly referred to as a \nifs\ ``hole''
(see Section~\ref{sect:nihole}).

In the inner $\sim0.1$\,\msun\ of the sub-\mch\ model, however, the
lower combustion density (due to the lower density of the progenitor
WD) results in an underproduction of stable IGEs. In these layers, the
radioactive \nifs\ isotope constitutes $\sim99$ per cent of the total
Ni abundance, the remainder consisting of roughly two-thirds of
radioactive $^{57}$Ni and one-third stable $^{58}$Ni. The iron mass
fraction is a factor of ten lower than in the \mch\ model in these
inner layers ($X_{\rm Fe}\lesssim 0.06$ cf. $\sim 0.6$) and consists
of near-equal parts of radioactive $^{52}$Fe and stable $^{54}$Fe.  A
non-negligible amount of intermediate-mass elements (IMEs; here
illustrated with Ca and Si) survives in these inner layers, contrary
to the \mch\ model where the combustion proceeds to the iron peak.

While the \nifs-rich layers extend to a larger mass coordinate in the
\mch\ model (0.79\,\msun\ cf. 0.61\,\msun\ in the
sub-\mch\ model for the layer containing 99 per cent of the
total \nifs\ mass; see Table~\ref{tab:sch2p0_ddc25_comp}), this mass
coordinate corresponds to a smaller fraction of the total WD mass
($\sim56$ per cent cf. $\sim67$ per cent). The result is a mass buffer
above the \nifs-rich layers that is a factor of two larger in the
\mch\ model ($\sim 0.6$\,\msun) compared to the sub-\mch\ model ($\sim
0.3$\,\msun).  Such variation in the \nifs\ distribution has important
consequences for the light-curve evolution and spectroscopic
properties.

Almost the entire C-O WD is burnt in both models, with the $\lesssim 1$ per
cent of unburnt material located at velocities beyond
$\sim20000$\,\kms. As a result, the ratio of asymptotic kinetic
energies is comparable to the ratio of WD masses, and the {\it
  specific} kinetic energy ($\equiv E_{\rm kin}/\mtot$) is $\sim10$
per cent larger in the sub-\mch\ model, as is the mass-weighted mean
velocity (due to the factor of $\sim5$ lower binding energy of the
progenitor WD; see Table~\ref{tab:sch2p0_ddc25_comp}).

%%%%%%%%%%%%%%%%%%%%%%%%%%%%%%%%%%%%%%%%%%%%%%%%%%%%%%%%%%%%%%%%%%%%%%

\section{The effects of a sub-\mch\ ejecta on the radiative display}\label{sect:lowmass}

\subsection{Radiative-transfer simulations}\label{sect:rt}

As in our previous \snia\ studies
\citep{Blondin/etal:2013,Blondin/etal:2015,Blondin/etal:2017a,D14_CoIII,D14_PDD,D14_Tech}
we use the 1D, time-dependent, non-LTE radiative-transfer code
\cmfgen\ of \cite{Hillier/Dessart:2012} to compute the light curves
and spectra based on our input hydrodynamical models. We use the same
outputs as published in \cite{Blondin/etal:2017a}, and refer the
reader to that paper for more details. The energy from radioactive
decays is assumed to be deposited locally during the first 10-15 days
past explosion depending on the model. At later times, we solve for
the transport of $\gamma$-rays produced in such decays using a Monte
Carlo procedure that takes into account (inelastic) Compton scattering
and photoelectric absorption but neglects pair production \citep[see
  Appendix~A in][]{Hillier/Dessart:2012}. Several decays result in the
emission of positrons, which are assumed to deposit their energy
locally.

We generate filtered light curves by integrating our synthetic spectra
over a given bandpass, weighted by the transmission function. For the
uvoir luminosities, we integrate the full synthetic spectrum between
the far UV ($\gtrsim 50$\,\AA) and the far IR ($\lesssim
50$\,$\mu$m). The term ``bolometric'' luminosity is often used to
refer to the uvoir luminosity, although this denomination then
neglects the escaping high-energy radiation in the form of
$\gamma$-rays resulting from the \nifs\ and other decay chains. While
the $\gamma$-ray luminosity is in general negligible up until past
maximum light, the recent observations of SN~2014J have revealed
\nifs\ $\gamma$-ray lines within the first 20 days past explosion
\citep{Diehl/etal:2014}. In what follows, we distinguish between the
uvoir, $\gamma$-ray, and bolometric luminosities, such that $L_{\rm
  bol} = L_{\rm uvoir} + L_{\gamma}$. The light curves for the two
models considered here are available in tabular form in
Appendix~\ref{sect:lctabs}.

\subsection{Impact on the UV-optical-IR evolution}\label{sect:luvoir}

%%% FIGURE: Lbol and gamma-ray escape fraction
\begin{figure}
\centering
\includegraphics{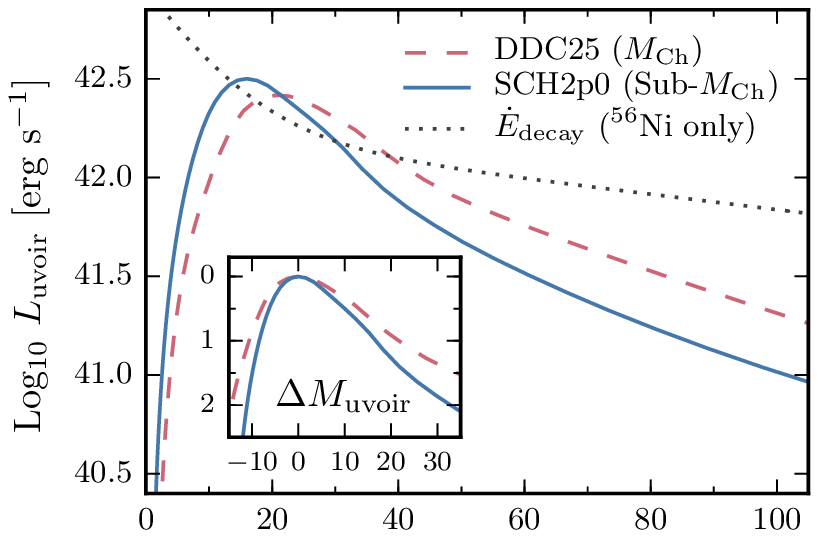}\vspace{-.2cm}
\includegraphics{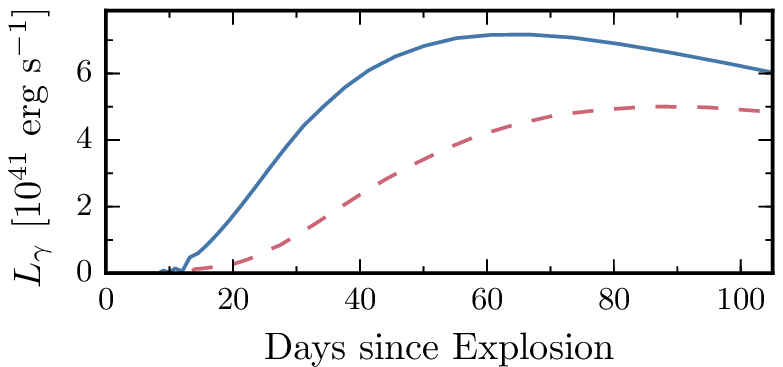}\vspace{-.45cm}
\includegraphics{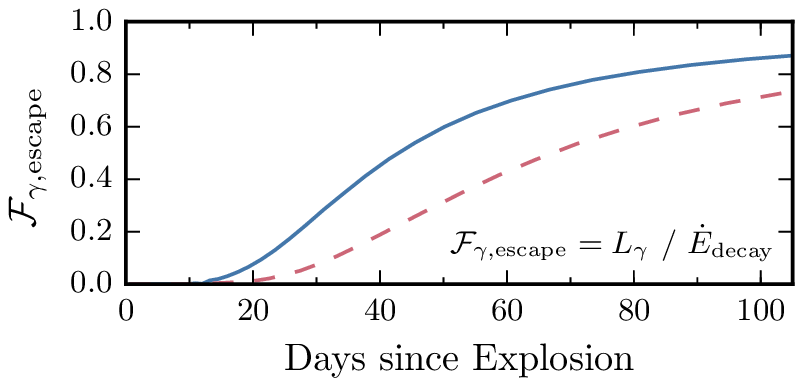}\vspace{-.45cm}
\includegraphics{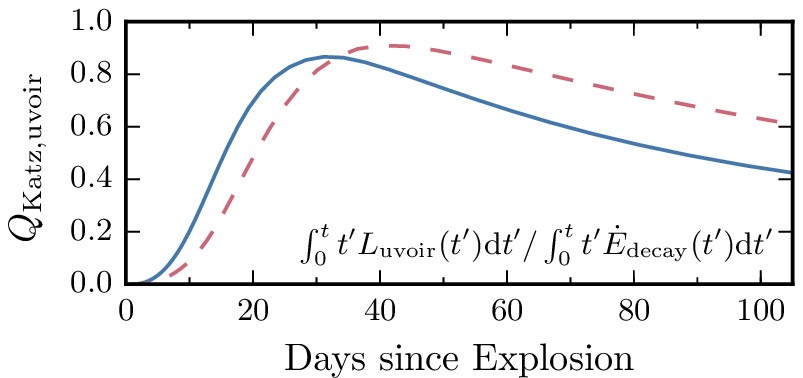}\vspace{-.45cm}
\includegraphics{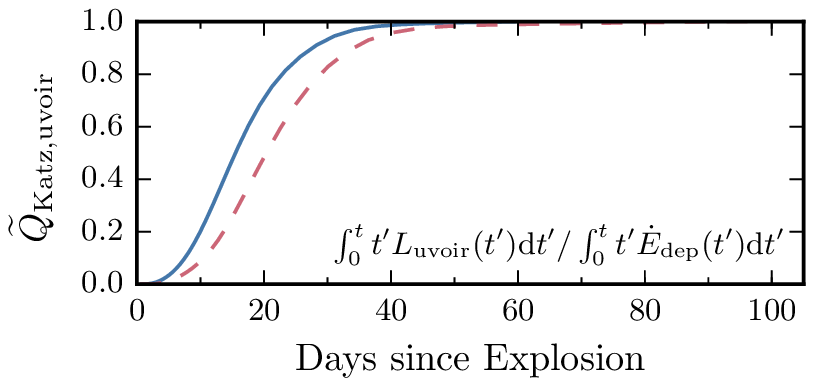}
\caption{\label{fig:luvoir}
Evolution of the uvoir (top) and $\gamma$-ray (second panel)
luminosities for the \mch\ delayed-detonation model DDC25 (dashed
line) and the sub-\mch\ model SCH2p0 (solid line). The dotted line in
the top panel corresponds to the decay power for 0.12\,\msun\ of
\nifs, and the inset shows the uvoir light curves normalized to the
peak luminosity, the time axis now corresponding to days from uvoir
maximum. The third panel shows the $\gamma$-ray escape fraction, which
also corresponds to the $\gamma$-ray contribution to the true
bolometric luminosity once the ejecta turn optically thin.  The fourth
panel shows the evolution of the ratio of the integral of the
time-weighted uvoir luminosity to the integral of the time-weighted
decay power ($\dot{E}_{\rm decay}$), noted $Q_{\rm Katz,uvoir}$
\citep{Katz:2013}. The bottom panel is the same as the fourth panel,
but using the decay power actually deposited in the ejecta
($\dot{E}_\mathrm{dep}$), yielding the modified $\widetilde{Q}_{\rm
  Katz,uvoir}$ ratio (equation~\ref{eqn:qtildekatzuvoir}).
}
\end{figure}

The detonation of a sub-\mch\ WD results in a larger outward extent of
the \nifs\ distribution compared to the delayed detonation of a
\mch\ progenitor (Fig.~\ref{fig:elem_distrib}, upper panels).  While
the kinetic energy is $\sim 30$ per cent lower compared to the
\mch\ model, its value per unit mass is within $\sim 10$ per cent,
resulting in a similar expansion rate.  As a result, the diffusion
timescale is typically shorter, and hence so are the rise times to
peak uvoir luminosity ($\sim 16$\,d for the sub-\mch\ model cf. $\sim
21$\,d for the \mch\ model; see Table~\ref{tab:sch2p0_ddc25_comp} and
Fig.~\ref{fig:luvoir}).  Despite having the same \nifs\ mass, the
sub-\mch\ model peaks at a $\sim20$ per cent larger luminosity, since
it radiates a similar amount of energy over a shorter
time.\footnote{Alternatively, one can say that both models radiate a
  similar fraction of the total decay energy by uvoir maximum, as seen
  from the similar value of the $Q_{{\rm Katz,uvoir}}$ ratio in
  Table~\ref{tab:sch2p0_ddc25_comp}.} The same effect is seen in the
higher-luminosity models of \cite{Wilk/etal:2017}.

Before peak, all the energy associated with radioactive decays is
deposited locally for both models.  However, the larger outward extent
of the \nifs\ distribution in the sub-\mch\ model results in a larger
$\gamma$-ray mean free path, which favours the earlier and enhanced
escape of $\gamma$-rays from the lower-mass ejecta
(Fig.~\ref{fig:luvoir}, second and third panels).

The post-maximum uvoir decline is thus also faster in the
sub-\mch\ model ($\dmftuvoir=0.85$\,mag, cf. 0.62\,mag for the
\mch\ model), due to the lower {\it rate} of energy deposition.  A low
ejecta mass thus offers a natural explanation for the narrower uvoir
light curves of low-luminosity \sneia\ (see Section~\ref{sect:99by}).

Owing to the non-negligible $\gamma$-ray escape fraction soon after
maximum light in both models, the uvoir luminosity deviates
significantly from the true bolometric luminosity. As a result, the
ratio of the integral of the time-weighted uvoir luminosity to the
integral of the time-weighted decay power (defined by
\citealt{Katz:2013} and noted $Q_{\rm Katz,uvoir}$ here), never quite
reaches unity (corresponding to full $\gamma$-ray trapping), and
starts to decline once the $\gamma$-ray escape fraction exceeds
$\sim$10--20 per cent (Fig.~\ref{fig:luvoir}, fourth panel). The use
of this ratio to infer the \nifs\ mass for observed \sneia\ thus
requires a $\gamma$-ray transport calculation to infer the decay power
actually deposited in the ejecta ($\dot{E}_\mathrm{dep}$). The
modified ratio

\begin{equation}
\label{eqn:qtildekatzuvoir}
\widetilde{Q}_{\mathrm{Katz,uvoir}}(t) = \int_0^t t' L_\mathrm{uvoir}(t') \mathrm{d}t'
\left/ \int_0^t t' \dot{E}_\mathrm{dep}(t') \mathrm{d}t'\right.
\end{equation}

\noindent
($\dot{E}_\mathrm{decay}$ replaced with $\dot{E}_\mathrm{dep}$ in the
denominator) indeed eventually reaches unity $\gtrsim 40$ days past
explosion in both models (Fig.~\ref{fig:luvoir}, bottom panel), and
illustrates the accuracy of energy conservation in our
radiative-transfer simulations. Alternatively, one could use the true
bolometric luminosity (i.e. including the contribution from
$\gamma$-rays) and the decay power to compute:

\begin{equation}
\label{eqn:qtildekatzbol}
Q_{\mathrm{Katz,bol}}(t) = \int_0^t t' L_\mathrm{bol}(t') \mathrm{d}t'
\left/ \int_0^t t' \dot{E}_\mathrm{decay}(t') \mathrm{d}t'\right.,
\end{equation}

\noindent
which indeed reaches unity in both models around the same time as
$\widetilde{Q}_{\mathrm{Katz,uvoir}}$. Unfortunately, only a single
\snia\ to date has measured $\gamma$-ray fluxes (SN~2014J;
\citealt{Diehl/etal:2014,Churazov/etal:2015}).

At later times ($\gtrsim 50$\,d past explosion), the uvoir luminosity
simply tracks the instantaneous rate of energy deposition. The rate at
which the luminosity declines at these times thus reflects the rate of
change of the $\gamma$-ray escape fraction.  While this rate is
initially larger for the sub-\mch\ model (Fig.~\ref{fig:luvoir},
second panel), it becomes smaller than that of the \mch\ model by
50\,d past explosion. As a result, the instantaneous uvoir magnitude
decline becomes comparable for both models at 100\,d past uvoir
maximum, while it was $\sim25$ per cent larger for the sub-\mch\ model
at 50\,d past maximum (cf. values of
$\mathrm{d}M_\mathrm{uvoir}/\mathrm{d}t|_{+50}$ and
$\mathrm{d}M_\mathrm{uvoir}/\mathrm{d}t|_{+100}$ in
Table~\ref{tab:sch2p0_ddc25_comp}). Throughout this time however, the
{\it absolute rate} of $\gamma$-ray escape remains larger for the
sub-\mch\ model, and hence its uvoir luminosity remains lower than the
\mch\ model.

%%%%%%%%%%%%%%%%%%%%%%%%%%%%%%%%%%%%%%%%%%%%%%%%%%%%%%%%%%%%%%%%%%%%%%
\subsection{Impact on the colour evolution}\label{sect:col}

%%% FIGURE: Temperature and ionfrac_Co @ vhalfopt vs. time
\begin{figure}
\centering
\includegraphics{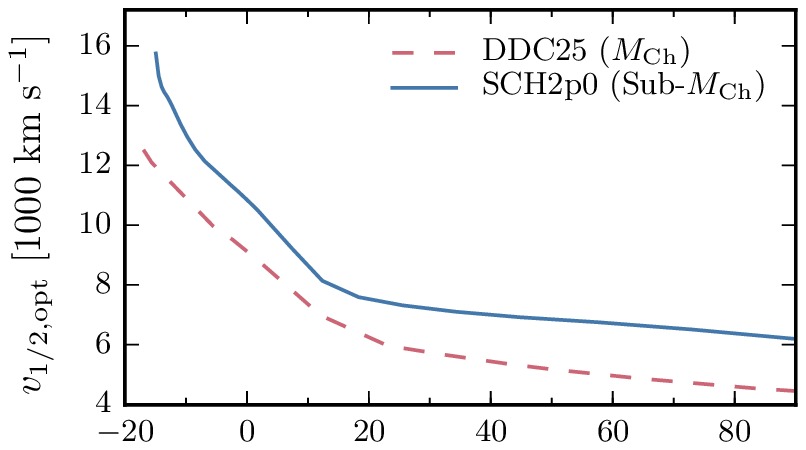}\vspace{-.5cm}
\includegraphics{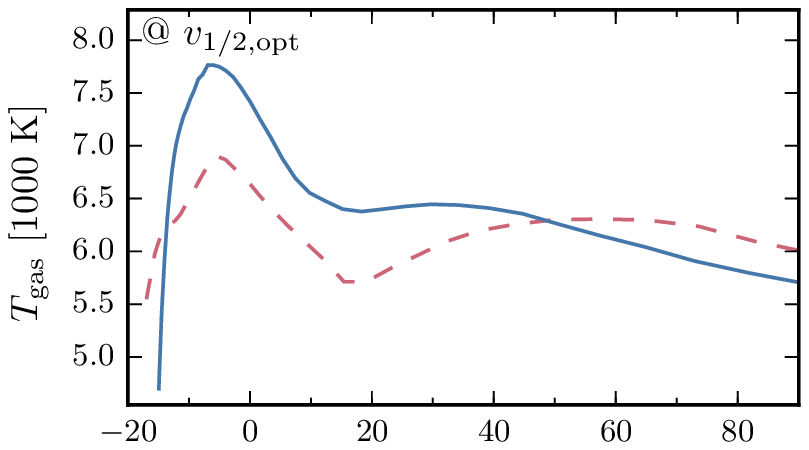}\vspace{-.5cm}
\includegraphics{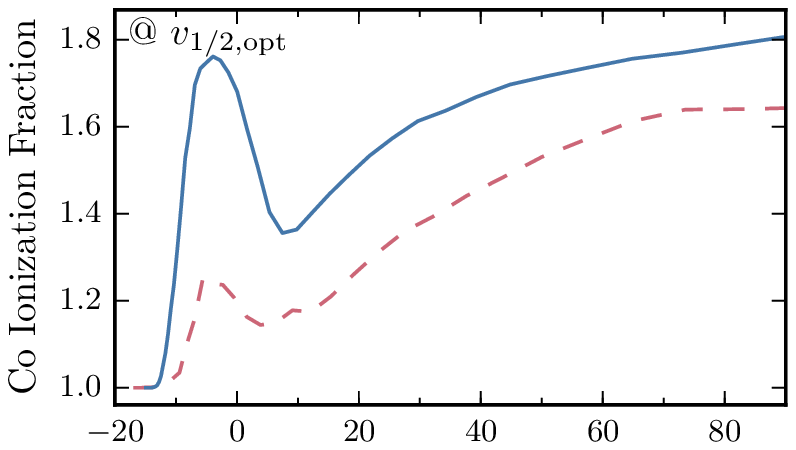}\vspace{-.5cm}
\includegraphics{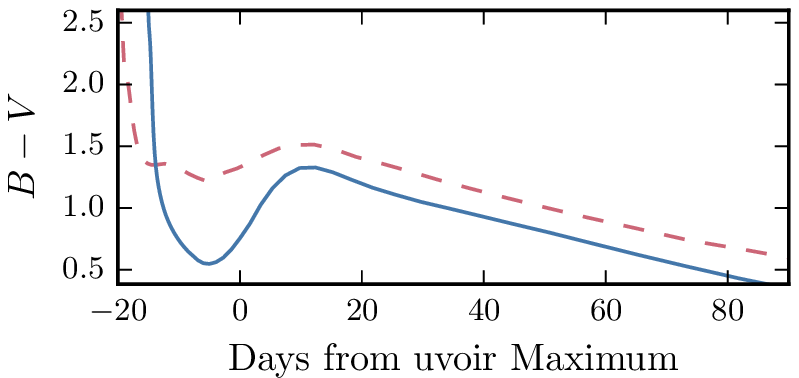}
\caption{\label{fig:temp_ionfrac_bmv}
Top panel: Evolution of the velocity location above which half of the
optical flux (defined here in the range 3000--10000\,\AA) emerges,
noted $\varv_{\rm 1/2, opt}$, for the \mch\ delayed-detonation model
DDC25 (dashed line) and the sub-\mch\ model SCH2p0 (solid line),
between $-20$\,d and $+90$\,d from uvoir maximum. Second and third
panels: evolution of the gas temperature and Co ionization fraction
(defined as $\sum_i{iX^{i+}}/\sum_i{X^{i+}}$, where $X^{i+}$ is the
mass fraction of ionization stage $i$ for Co) at $\varv_{\rm 1/2,
  opt}$. Bottom panel: evolution of the $B-V$ colour.
}
\end{figure}

As in \cite{Blondin/etal:2015}, we locate the spectrum-formation
region using the velocity, $\varv_{\rm 1/2, opt}$, above which half of
the optical flux (defined here to be in the range 3000--10000\,\AA)
emerges (Fig.~\ref{fig:temp_ionfrac_bmv}, top panel).  The more
extended \nifs\ distribution in the sub-\mch\ model results in a more
rapid and efficient heating of the corresponding ejecta layers before
maximum light (Fig.~\ref{fig:temp_ionfrac_bmv}, second panel). The gas
temperature at this location reaches a maximum around $-5$\,d from
maximum in both models, but it is higher for the sub-\mch\ model
($\sim 7800$\,K) compared to the \mch\ model ($\sim 6900$\,K), due to
the larger specific heating rate from the higher \nifs-to-total mass
ratio ($\dot{e}_{\rm dep} \approx 1.5 \times 10^9$ cf. $7.4 \times
10^8$\,erg\,s$^{-1}$\,g$^{-1}$ for the \mch\ model; see
Table~\ref{tab:sch2p0_ddc25_comp} and discussion in
\citealt{Blondin/etal:2017a}).

The higher ionization state of the gas in the sub-\mch\ model, here
illustrated with cobalt, reflects the higher temperature in the
spectrum-formation region (Fig.~\ref{fig:temp_ionfrac_bmv}, third
panel; see also \citealt{Wilk/etal:2017}). The temperature increase at
early times induces a shift in the ionization state of IGEs (from
singly to doubly ionized) around $-10$\,d from uvoir maximum,
resulting in a Co\three-dominated gas around maximum light (i.e. Co
ionization fraction greater than 1.5) where the \mch\ model remains
largely dominated by Co\two.

This difference in ionization state has a strong impact on the
$B$-band flux, which is efficiently blocked in the \mch\ model owing
to the presence of lines from once-ionized IGEs (Sc, Ti, Fe, and
Co). These lines contribute less to the opacity of the more ionized
sub-\mch\ model. Thus, despite the increase in temperature in the
spectrum-formation region at early times in both models, the $B-V$
colour index only reaches $\sim1.2$\,mag in the \mch\ model, but is
$\sim0.7$\,mag bluer in the sub-\mch\ model shortly before maximum
(Fig.~\ref{fig:temp_ionfrac_bmv}, bottom panel). This minimum in the
$B-V$ colour coincides with the maxima in the temperature and Co
ionization fraction at $\varv_{\rm 1/2, opt}$.

%%% FIGURE: Comparison optical spectra phasebol
%%% NOTE: HERE FOR PLACEMENT WRT TEXT
\begin{figure*}
\centering
\begin{minipage}[]{0.475\linewidth}
\includegraphics{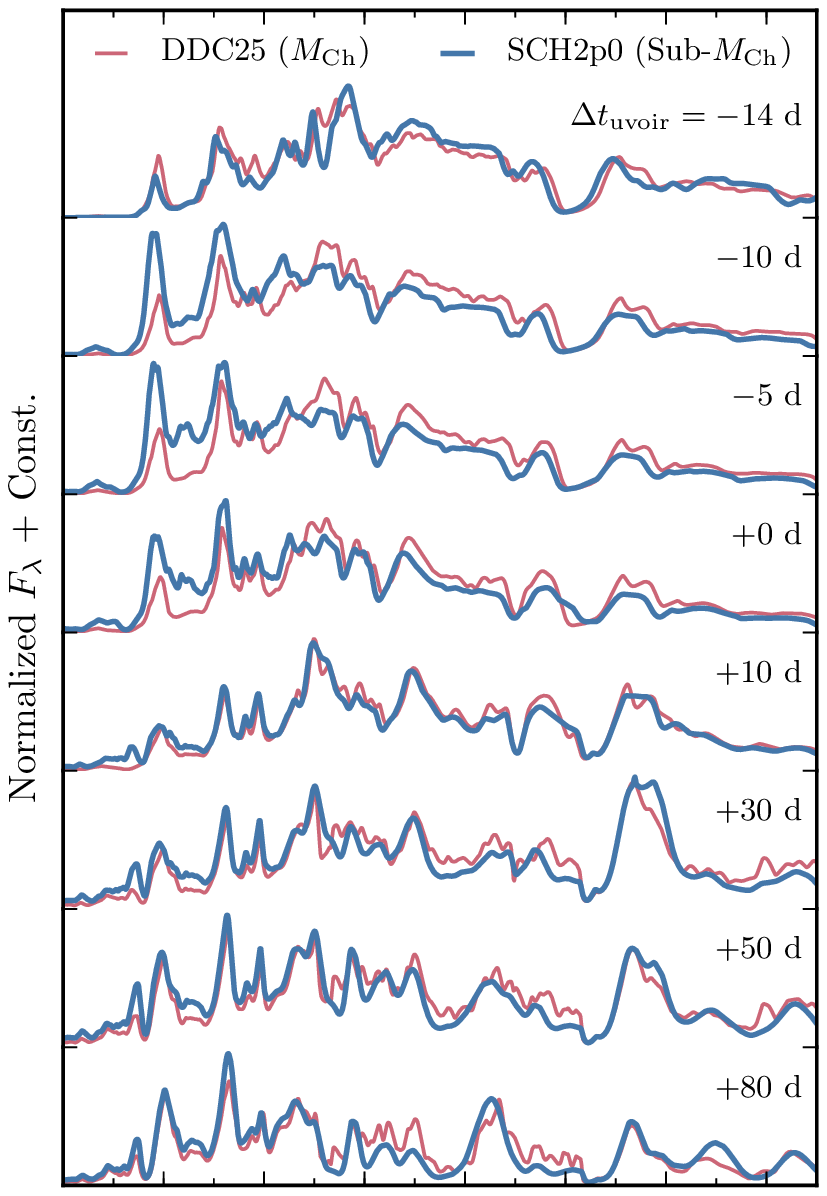}\vspace{-.5cm}
\includegraphics{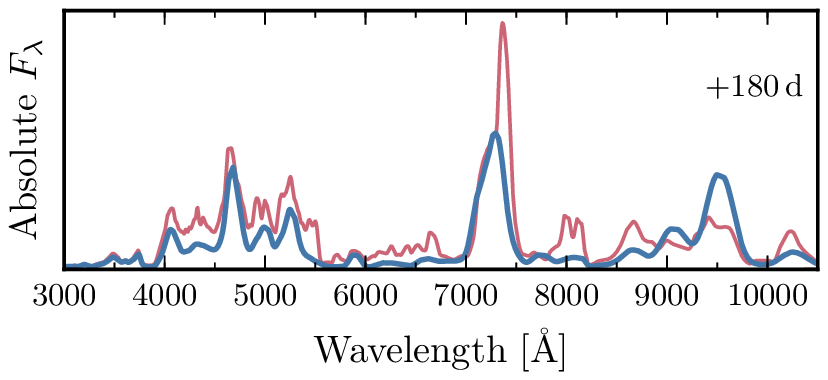}
\end{minipage}
\hspace*{.5cm}
\begin{minipage}[]{0.475\linewidth}
\includegraphics{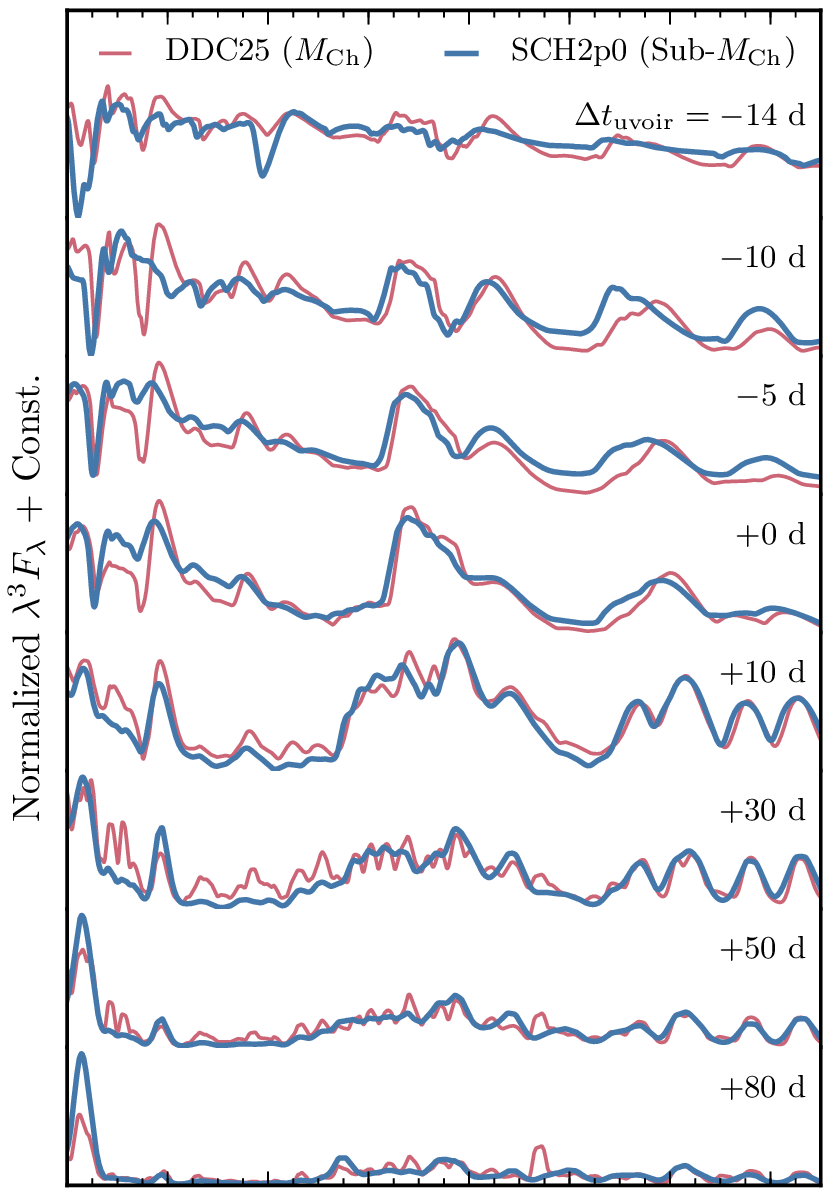}\vspace{-.5cm}
\includegraphics{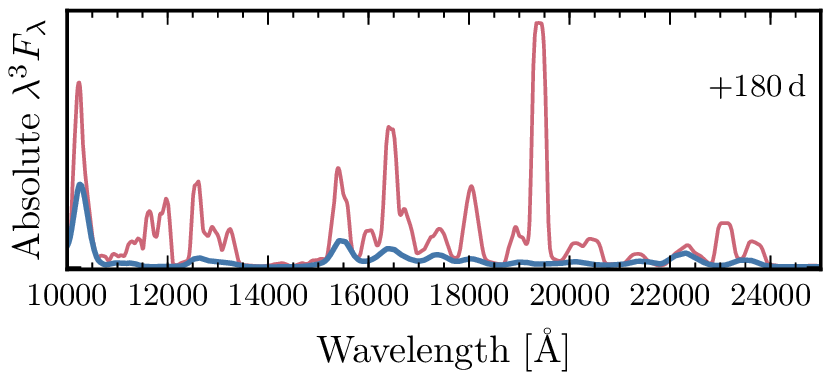}
\end{minipage}
\caption{\label{fig:comp_spec}
Optical (left) and NIR (right) spectroscopic evolution of the
\mch\ delayed-detonation model DDC25 (thin red line) and the
sub-\mch\ model SCH2p0 (thick blue line), between $-15$\,d and
$+180$\,d from uvoir maximum. The NIR flux has been scaled by
$\lambda^3$ for better visibility. In the top panels ($-15 \le \Delta
t_{\rm uvoir} \le +80$\,d), the optical spectra have been normalized
to the same mean flux in the range 3000--10000\,\AA\ (the tickmarks on
the ordinate give the zero-flux level), while the NIR spectra have
been scaled to the same $H$-band magnitude at a given time. Those in
the bottom panel ($\Delta t_{\rm uvoir} = +180$\,d) are on an absolute
flux scale.
}
\end{figure*}

After its pre-maximum peak, the temperature in the spectrum-formation
region then decreases by $\sim1500$\,K over a $\sim20$\,d timescale in
both models, but the impact on the ionization state is more important
in the sub-\mch\ model and results in a large {\sc
  iii}$\rightarrow${\sc ii} ionization shift of IGEs. The opacity in
the $B$ band is increased accordingly, and the $B-V$ colour rapidly
rises to a value comparable to the \mch\ model around $10$\,d past
maximum. At later times, the temperature evolution is more gradual and
starts to decline, yet the Co ionization steadily increases in both
models due to non-thermal processes associated with \cofs\ decay
\citep[see][]{D14_Tech}.

A lower ejecta mass thus not only affects the rate at which radiation
energy escapes the ejecta (which impacts the uvoir evolution), but
also the thermodynamic state of the gas (temperature and ionization
state). The higher temperature and higher ionization state of the
sub-\mch\ model in the spectrum-formation region also applies to the
ejecta as a whole, except in the inner ejecta layers around $\sim
3000$\,\kms, where the \nifs\ mass fraction is higher in the
\mch\ model (see Fig.~\ref{fig:elem_distrib}). At lower velocities
($\lesssim 2500$\,\kms), the larger \nifs\ mass fraction contributes
to maintain a higher temperature and ionization in the sub-\mch\ model
at late times, through (local) energy deposition by positrons emitted
in \cofs\ $\beta +$ decays (see discussion in
\citealt{Wilk/etal:2017}).

%%%%%%%%%%%%%%%%%%%%%%%%%%%%%%%%%%%%%%%%%%%%%%%%%%%%%%%%%%%%%%%%%%%%%%
\subsection{Impact on the spectroscopic evolution}\label{sect:spec}

The spectroscopic evolution of both models at selected times from
uvoir maximum is shown in Fig.~\ref{fig:comp_spec}. The overall
similarity in spectral morphology is striking, the most notable
differences occurring at the latest time shown ($180$\,d past uvoir
maximum). Subtle differences in the SED are nonetheless visible
(especially at the earliest times), as well as in the presence and
shapes of certain lines. These differences are discussed in
chronological order in what follows.

At the earliest time ($\Delta t_{\rm uvoir} = -14$\,d), the
temperature in the spectrum-formation region of the sub-\mch\ model is
lower than in the \mch\ model at the same location (see
Fig.~\ref{fig:temp_ionfrac_bmv}, second panel), since this time
corresponds to only $\sim 2$ days past explosion (cf. $\sim 7$ days
for the \mch\ model), when energy input from \nifs\ decay (with a
$\sim 6$\,d half-life) has not yet heated the layers from which the
flux emerges. As a result, the SED of the sub-\mch\ model is redder
(as is the $B-V$ colour; Fig.~\ref{fig:temp_ionfrac_bmv}, bottom
panel), with stronger absorption features resulting from neutral
species (e.g. Na\one~D, O\one~7773\,\AA, C\one~1.07\,$\mu$m doublet).

As time progresses ($\Delta t_{\rm uvoir} \gtrsim -10$\,d), the hotter
ejecta of the sub-\mch\ model produces an SED with a greater flux in
the blue (primarily the $B$ band) compared to the \mch\ model. The
spectrum is increasingly influenced by lines: the continuum flux
essentially vanishes by maximum light in both models.  The higher {\sc
  iii}/{\sc ii} ionization ratio in the sub-\mch\ model leads to less
line-blanketing from once-ionized IGEs shortward of
$\sim5000$\,\AA\ and a lower emissivity at redder wavelengths and in
the NIR (see also \citealt{Wilk/etal:2017}).  Conversely, the lower
ejecta temperatures of the \mch\ model result in a strong P-Cygni
profile due to the Ca\two\ 1.19\,$\mu$m doublet that is largely absent
from the sub-\mch\ model, and to a more pronounced emission complex
around 1.7\,$\mu$m due to Si\two.

Sub-dominant species (Sc, Ti, Cr) play a crucial role in shaping the
SED in the blue part of the optical spectrum. In their once-ionized
state, they cause strong line blanketing, even at low abundance.
Multiple weak lines of Sc\two\ and Ti\two\ (and to a lesser extent
Cr\two/Fe\two/Co\two) cause the prominent absorption trough around
4000--4500\,\AA\ characteristic of 91bg-like \sneia, despite their low
mass fractions in the spectrum-formation region ($<10^{-4}$ and
$<10^{-7}$ at maximum light for Ti and Sc, respectively).

The larger outward extent of the \nifs\ distribution in the
sub-\mch\ model results in a spectrum-formation region located at
higher velocities. Thus, even at post-maximum epochs where both models
display very similar SEDs, the spectral features of the
sub-\mch\ model are broader and more strongly influenced by line
overlap compared to their counterparts in the \mch\ model (see
e.g. the 6000--8000\,\AA\ region in the spectra at 10--80\,d past
maximum). Individual spectral lines are also broader and more
blueshifted at a given time, as is the case for the
Si\two~6355\,\AA\ line whose absorption velocity is 20 per cent larger
(in absolute value) at uvoir maximum (see
Table~\ref{tab:sch2p0_ddc25_comp}). However, the higher Ca ionization
of the sub-\mch\ model limits the formation of high-velocity
absorption features in the Ca\two~8500\,\AA\ triplet around maximum
light, in better agreement with observations of low-luminosity
\sneia\ (see Section~\ref{sect:99by}).

Both models predict weak features in the NIR associated with
Co\two\ before maximum light, although they are initially weaker in
the sub-\mch\ model due to the higher {\sc iii}/{\sc ii} ionization
ratio in the spectrum-formation region. Shortly after maximum light,
these lines become stronger in the sub-\mch\ model through both a
decreasing Co ionization and an increasing Co abundance (almost
exclusively \cofs\ from \nifs\ decay) in the NIR spectrum-formation
region. As noted by \cite{Hoeflich/etal:2002}, the Co\two/{\sc iii}
emission complex in the range 1.5--1.8\,$\mu$m characteristic of
post-maximum \snia\ spectra is weaker compared to more luminous
events, due to the more centrally-concentrated Co distribution and
hence its lower relative abundance in the spectrum-formation region.

Combined with the higher ionization state, the lower ejecta density of
the sub-\mch\ model also favours the emergence of strong forbidden
line transitions, in particular [Co\three]~5888\,\AA, as early as
$\sim10$\,d past maximum. These forbidden lines are essential in
cooling the ejecta at post-maximum epochs, and contribute to the
temperature decline in the spectrum-formation region of the
sub-\mch\ model at $\gtrsim30$\,d past maximum
(Fig.~\ref{fig:temp_ionfrac_bmv}; see also \citealt{D14_CoIII}). At
such times, the 5888\,\AA\ transition {\it alone} contributes $\sim30$
per cent of the total Co\three\ cooling rate at depths
2000--3000\,\kms, which is $\sim3$ times larger in the sub-\mch\ model
than in the \mch\ model. This line, sometimes mistaken for Na\one~D,
is present in observed \snia\ spectra at post-maximum epochs,
regardless of their luminosity.\footnote{We predict a Na\one~D feature
  in the \mch\ model DDC25 starting around $10$\,d past maximum,
  although its contribution to the emission at $\sim 5900$\,\AA\ is
  minor relative to [Co\three] 5888\,\AA.}

At later times ($\gtrsim 80$\,d past maximum), the spectrum-formation
region probes the innermost ejecta layers ($\lesssim 5000$\,\kms).
Because of the lower density of the sub-\mch\ ejecta in these layers
(Fig.~\ref{fig:elem_distrib}, lower panel), the {\sc
  iii}$\rightarrow${\sc ii} recombination is only partial.  Moreover,
the presence of \nifs\ in the inner ejecta of the sub-\mch\ model
results in local energy deposition from positrons (from \cofs\ decay),
which further contributes to the higher Fe\three/Fe\two\ ionization
ratio \citep{Wilk/etal:2017}. As a result, the NIR emissivity is
significantly reduced compared to the \mch\ model \citep[see
  also][]{Mazzali/Hachinger:2012}.

The broad emission feature spanning 7000--7500\,\AA\ in the synthetic
spectra at 180\,d past maximum has a distinct morphology in each
model, with an additional emission component in the \mch\ model due to
[Ni\two] 7378, 7412\,\AA. This feature is absent from the
sub-\mch\ model due to the lack of stable Ni isotopes synthesized in
the explosion, and could in principle serve as a diagnostic for the
progenitor scenario (see Section~\ref{sect:nihole} below). However,
overlap with [Fe\two] 7155\,\AA, [Ca\two] 7291, 7324\,\AA, and
[Ar\three] 7136, 7751\,\AA\ complicates the secure identification of
this line in practice.

The three-peaked emission complex in the range 8500-9500\,\AA\ is due
to the [Ca\two] 8500\,\AA\ triplet and the [S\three] 9068,
9530\,\AA\ doublet. The [S\three] lines are stronger in the
sub-\mch\ model, where \cite{Wilk/etal:2017} found an opposite
correlation with ejecta mass for their more luminous \snia\ models.

%%%%%%%%%%%%%%%%%%%%%%%%%%%%%%%%%%%%%%%%%%%%%%%%%%%%%%%%%%%%%%%%%%%%%%
\subsection{Nebular [Ni\two] lines as a diagnostic of the progenitor mass}\label{sect:nihole}

%%% FIGURE: [NiII] 1.939 micron line at 200d
%%% NOTE: HERE FOR PLACEMENT WRT TEXT
\begin{figure}
\centering
\includegraphics{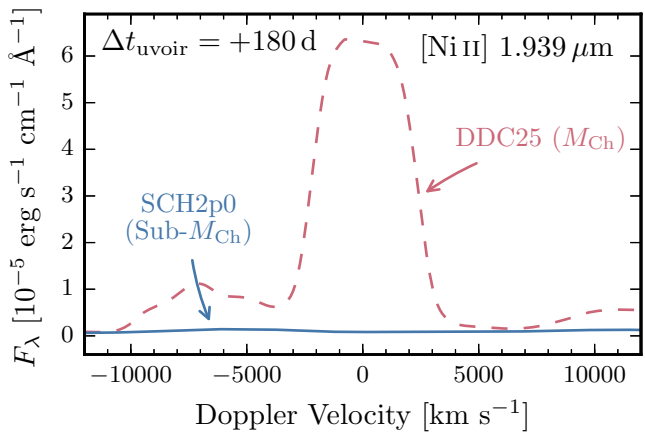}
\caption{\label{fig:ni2}
Predicted [Ni\two] 1.939\,$\mu$m line at 180\,d past uvoir maximum,
for the \mch\ delayed-detonation model DDC25 (dashed line) and the
sub-\mch\ model SCH2p0 (solid line).
}
\end{figure}

The isotopic composition of the innermost ejecta layers is sensitive
to the density of the progenitor WD star, which is set by its mass at
the time of explosion. In a sub-\mch\ WD, fusion proceeds up the
$\alpha$ chain to the radioactive isotope \nifs. In a \mch\ WD,
however, the central density is $> 2 \times 10^9$\,\gcc\ at the time
of explosion, which leads to electron captures during the explosive
phase and the subsequent production of neutron-rich {\it stable}
isotopes of IGEs (e.g. $^{54}$Fe, $^{58}$Ni; see
Fig.~\ref{fig:elem_distrib}). An observational consequence is the
presence of forbidden lines of Ni in late-time spectra ($> 150$ days
or so past explosion) when the spectrum-formation region reaches down
to $v \lesssim 2000$\,\kms.  A sub-\mch\ model, on the other hand,
should exhibit no such Ni lines as the innermost ejecta is dominated
by radioactive \nifs, which will have decayed to \fefs\ by that time.
This is also the case for the violent merger of two sub-\mch\ WDs,
whose total mass can exceed \mch\ but whose remnant has a central
density characteristic of single sub-\mch\ WDs ($\lesssim 10^7$\,\gcc;
\citealt{Pakmor/etal:2010}).

The detection of forbidden Ni lines in late-time \snia\ spectra thus
offers a powerful diagnostic of the progenitor WD mass. Tentative
identifications of [Ni\,{\sc ii}--{\sc iv}] lines have been previously
reported in mid-IR spectra of the low-luminosity SN~2005df
\citep{Gerardy/etal:2007}, and a more systematic study of the presence
of such lines would contribute greatly to our understanding of \sneia.

As noted above, the optical [Ni\two] 7378, 7412\,\AA\ lines predicted
in the \mch\ model (and absent from the sub-\mch\ model) suffers from
overlap with the neighbouring [Ca\two] 7291, 7324\,\AA\ doublet. In
the NIR, however, we predict a strong [Ni\two] 1.939\,$\mu$m line in
the \mch\ model, in a region largely uncontaminated by other lines,
and as early as $\sim 50$\,d past maximum (see
Fig.~\ref{fig:comp_spec}).  The line is absent from the
sub-\mch\ model, whose absolute flux level is a factor $\sim80$ lower
at the line's central wavelength at +180\,d (Fig.~\ref{fig:ni2}). This
line extends to $\pm 3000$\,\kms\ in velocity space, and hence probes
the inner ejecta layers containing stable Ni isotopes (predominantly
$^{58}$Ni, but also $^{60}$Ni and $^{62}$Ni). The flat-topped profile
extending over $\pm 1000$\,\kms\ is an artefact of the spatial grid
used for the radiative-transfer simulations, which is truncated at
$\varv \lesssim 1000$\,\kms\ (and hence has a hollow core).

We also identify two weaker lines in the \mch\ model due to [Ni\two]
2.308, 2.369\,$\mu$m (see Fig.~\ref{fig:comp_spec}, lower-right
panel), and another further to the red due to [Ni\two] 2.911\,$\mu$m
(not shown).

In theory, the lack of [Ni\two] lines in the sub-\mch\ model could be
due to a higher Ni ionization fraction (see previous
section). However, we predict weak lines of [Ni\three] 3.393,
3.801\,$\mu$m in both models, which are an order of magnitude
{\it stronger} in the \mch\ model despite the lower
Ni\three/Ni\two\ ratio compared to the sub-\mch\ model.

%%% FIGURE: Comparison Luvoir - twocolumn + late-time point
%%% NOTE: HERE FOR PLACEMENT WRT TEXT
\begin{figure*}
\centering
\includegraphics{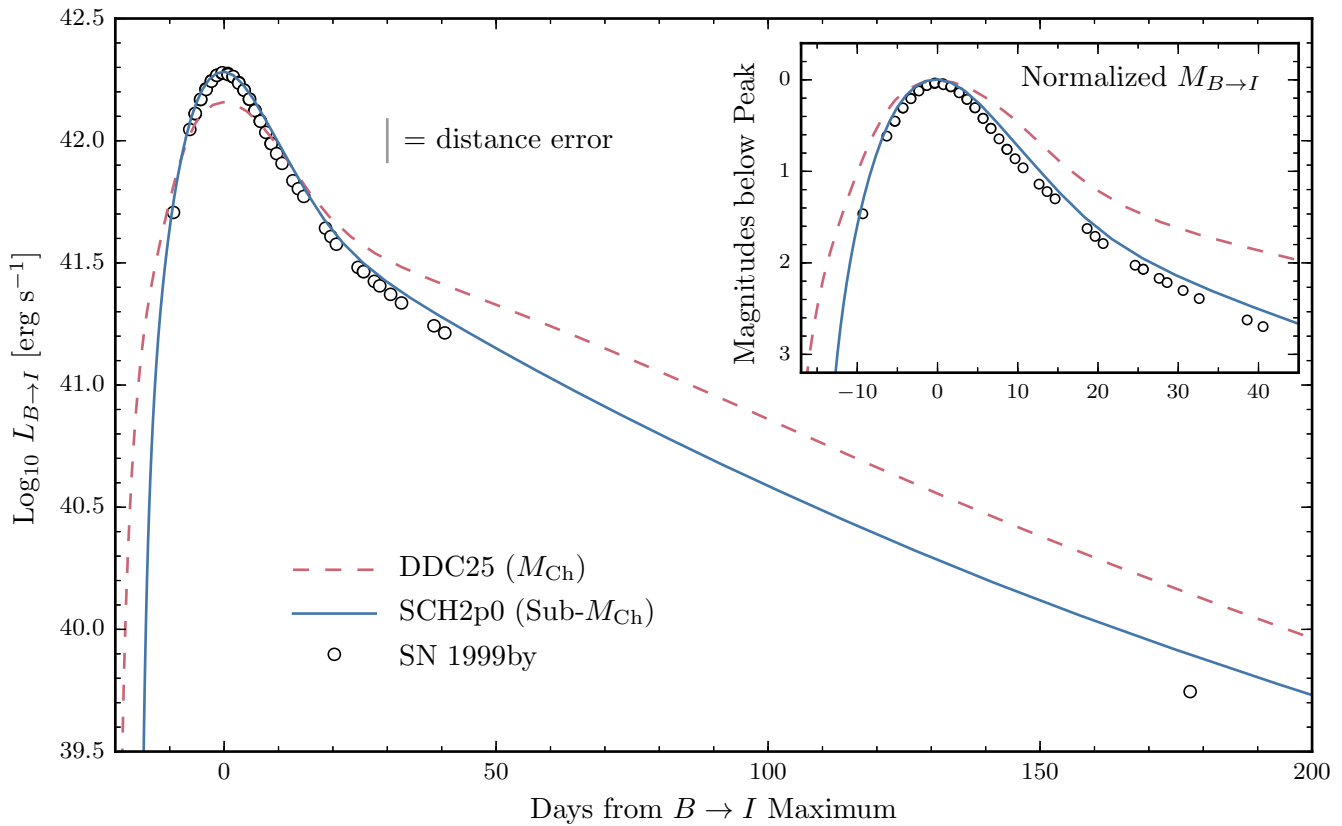}
\caption{\label{fig:comp_luvoir_inclate}
Evolution of the integrated $B\rightarrow I$ luminosity for the
\mch\ delayed-detonation model DDC25 (dashed line) and the
sub-\mch\ model SCH2p0 (solid line), compared to the low-luminosity
SN~1999by (open circles). The vertical grey bar indicates the error
resulting from the uncertainty in the assumed distance modulus
($30.97\pm0.23$\,mag).  The inset shows a close-up view around maximum
light of the $B\rightarrow I$ magnitude evolution normalized to its
peak value.
}
\end{figure*}

The prediction of a strong [Ni\two] 1.939\,$\mu$m line is thus mainly
an abundance effect: at 180\,d past maximum (corresponding to $\sim
33$ half-lives of \nifs), the Ni abundance is set by the amount of
stable isotopes (mostly $^{58}$Ni), which is a factor $\sim 17$ higher
in the \mch\ model (see Table~\ref{tab:sch2p0_ddc25_comp}). The ratio
of [Ni\two] 1.939\,$\mu$m line fluxes is significantly larger than
this (see above), so the higher density and lower ionization of the
\mch\ inner ejecta seem to enhance the observed signature, and
complicates the determination of the (stable) Ni abundance based on
this feature \citep[see also][]{Wilk/etal:2017}.

The detection of this [Ni\two] 1.939\,$\mu$m line in a \snia\ would
nonetheless strongly favour a \mch\ progenitor for that event. The
required late-time NIR spectroscopy could be obtained with the
upcoming JWST, since the location of this line at the blue edge of the
$K$ band renders ground-based observations particularly
challenging. Tentative identifications of this line have been reported
in spectra of SN~2003du, SN~2011fe, and SN~2014J around 75--100\,d
past explosion by \cite{Friesen/etal:2014}, although the emission is
offset by +6000\,\kms\ (at 1.98\,$\mu$m) from its expected
rest-wavelength location, and is associated with multiple lines of
[Co\three] in our models of comparable luminosity
\citep[see][]{Blondin/etal:2015}.

Multi-dimensional simulations of \mch\ delayed-detonation explosions
do not predict an inner core dominated by stable IGEs: these are
rapidly transported outward through buoyancy during the initial
deflagration phase, and the inward mixed C/O fuel is mainly burnt to
\nifs\ in the subsequent detonation \citep[see 
  e.g.][]{Seitenzahl/etal:2013}. However, the reported identification
of Ni lines in late-time mid-IR spectra of SN~2005df
\citep{Gerardy/etal:2007} and SN~2014J \citep{Telesco/etal:2015}
suggests a lower level of mixing between the stable IGE core and the
outer \nifs-rich layers than currently predicted during the initial
deflagration phase in 3D hydrodynamical simulations.  The velocity
extent of Ni lines in nebular spectra could be used to constrain the
level of outward mixing of stable IGEs during the explosion, and a
velocity shift would indicate an explosion offset from the WD center
\citep[see  e.g.][]{Maeda/etal:2010a}.

%%% FIGURE: Comparison UBVRIJHK LCs 
%%% NOTE: HERE FOR PLACEMENT WRT TEXT
\begin{figure*}
\centering
\includegraphics{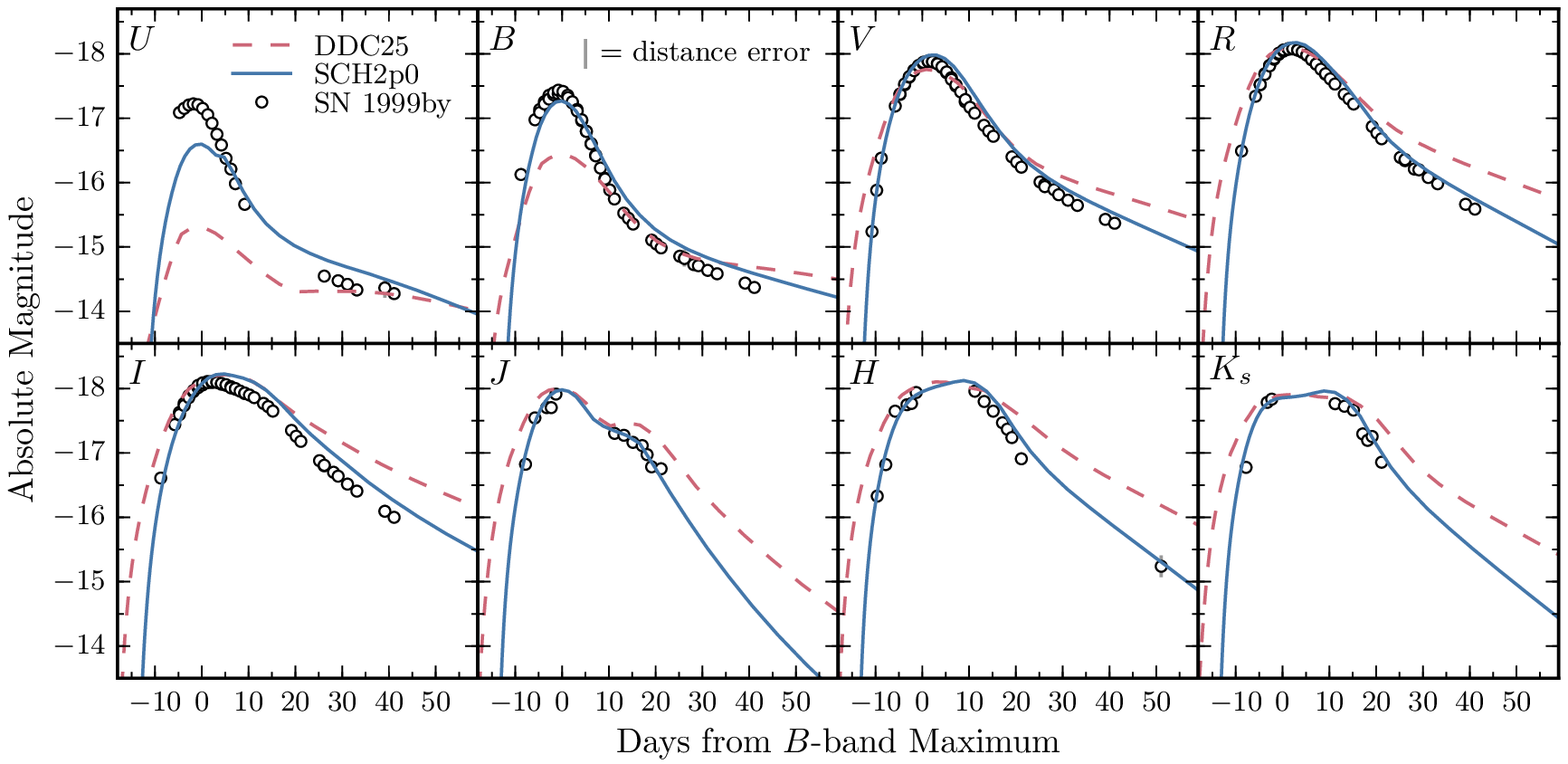}\vspace{.5cm}
\includegraphics{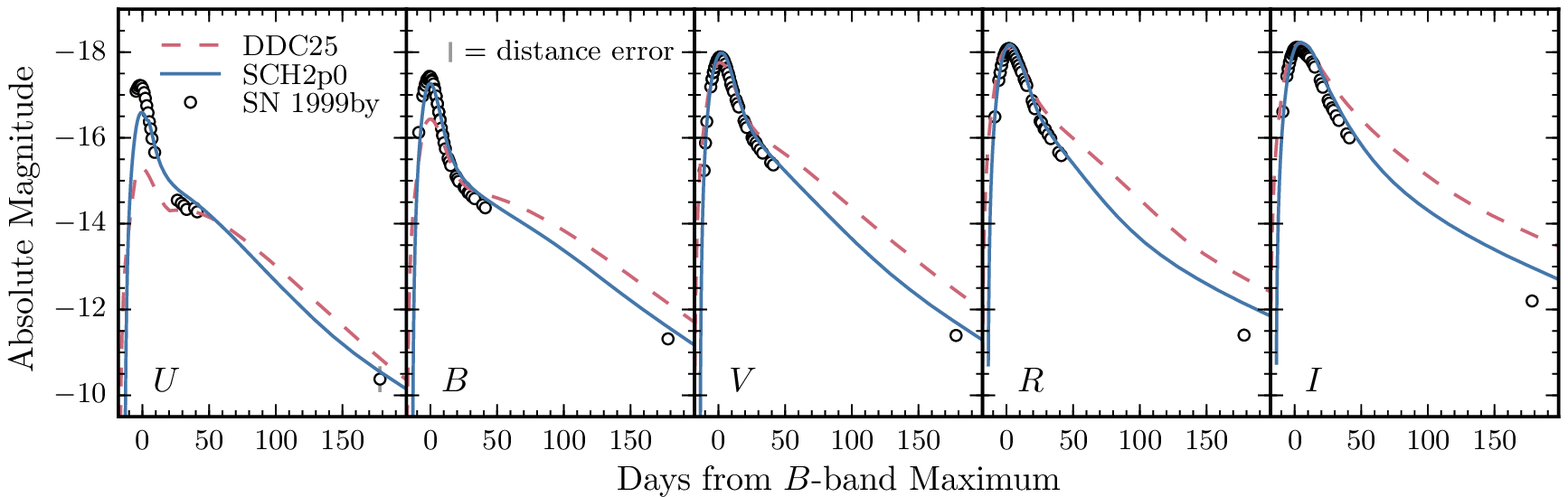}
\caption{\label{fig:comp_lc}
Top: Optical ($UBVRI$) and NIR ($JHK_s$) light curves for the
\mch\ delayed-detonation model DDC25 (dashed line) and the
sub-\mch\ model SCH2p0 (solid line), compared to the low-luminosity
SN~1999by (open circles).
The error bar on individual data points (when visible) correspond to
the measurement error only. 
The vertical grey bar in the panel showing the $B$-band light curve
indicates the error resulting from the uncertainty in the assumed
distance modulus ($30.97\pm0.23$\,mag).
Bottom: same as above
including the late-time $UBVRI$ measurements around +180\,d past
$B$-band maximum.
}
\end{figure*}

%%% FIGURE: Comparison colors 
%%% NOTE: HERE FOR PLACEMENT WRT TEXT
\begin{figure}
\centering
\includegraphics{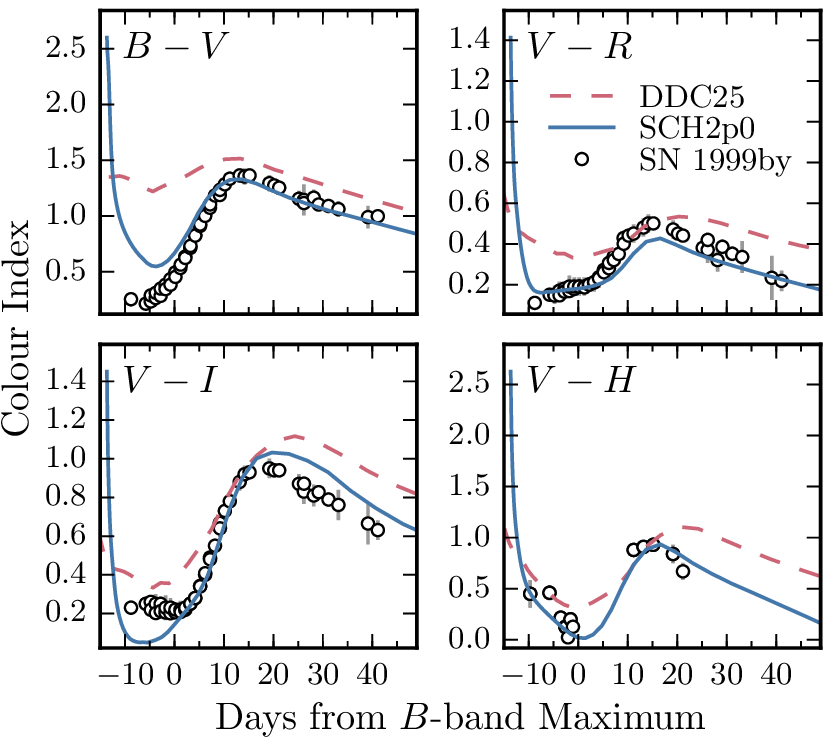}
\caption{\label{fig:comp_colors}
Comparison of $B-V$, $V-R$, $V-I$, and $V-H$ colour curves for the
\mch\ delayed-detonation model DDC25 (dashed line) and the
sub-\mch\ model SCH2p0 (solid line) with those of the low-luminosity
SN~1999by (open circles).  The error bars correspond to measurement
errors only.
}
\end{figure}

%%%%%%%%%%%%%%%%%%%%%%%%%%%%%%%%%%%%%%%%%%%%%%%%%%%%%%%%%%%%%%%%%%%%%%

\section{Comparison to the low-luminosity SN~1999\lowercase{by}}\label{sect:99by}

%%% FIGURE: Comparison optical spectra
%%% NOTE: HERE FOR PLACEMENT WRT TEXT
\begin{figure*}
\centering
\begin{minipage}[]{0.475\linewidth}  
\includegraphics{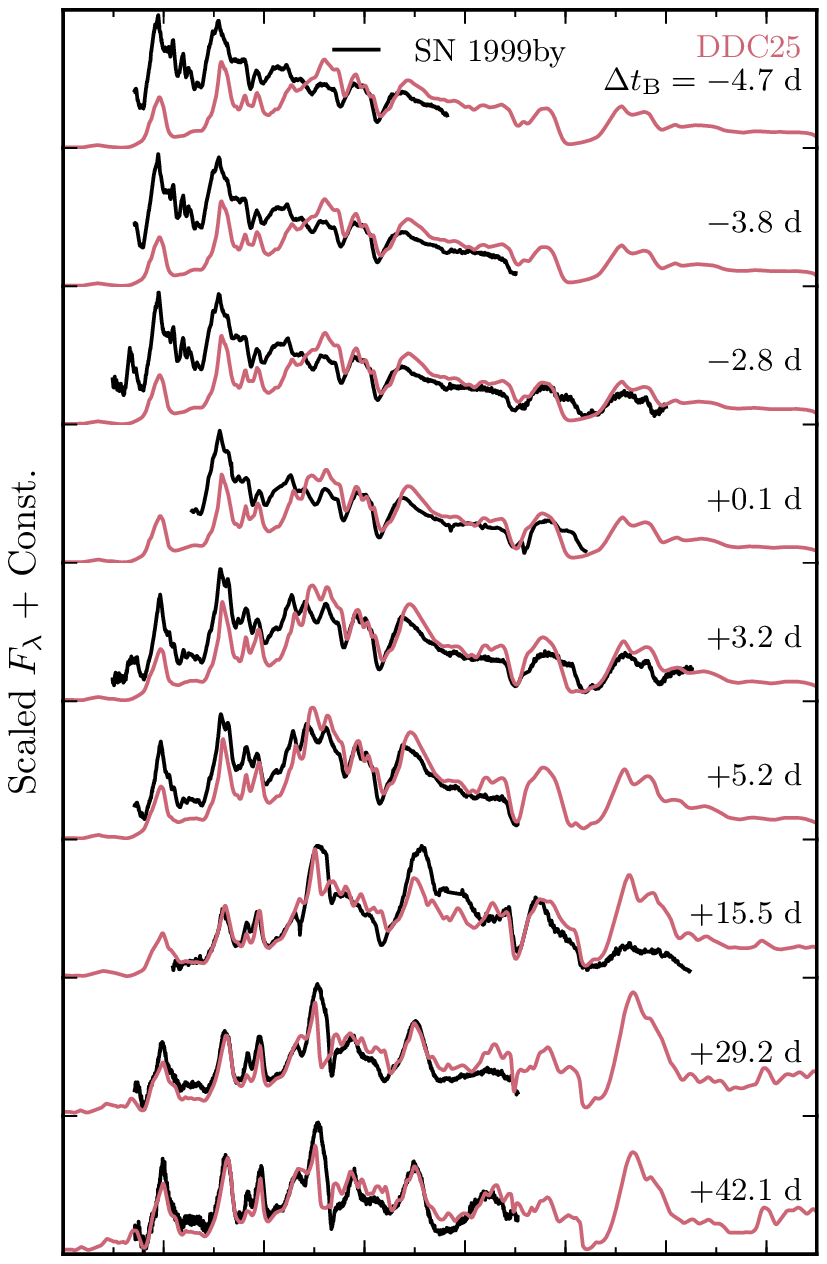}\vspace{-1cm}
\includegraphics{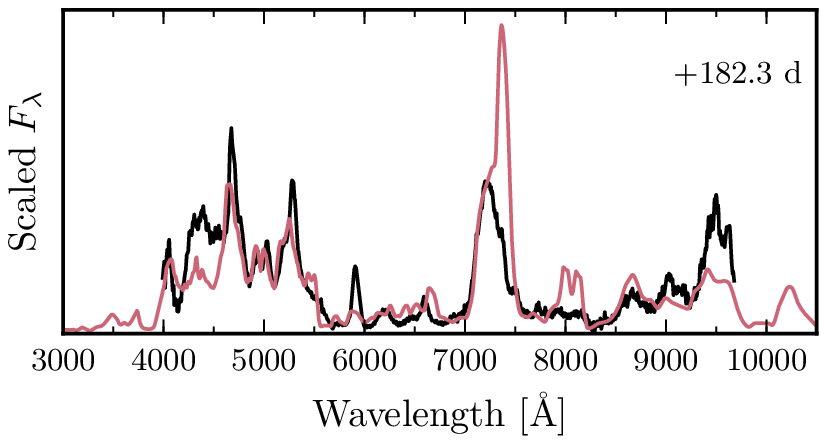}
\end{minipage}
\hspace*{.5cm}
\begin{minipage}[]{0.475\linewidth}  
\includegraphics{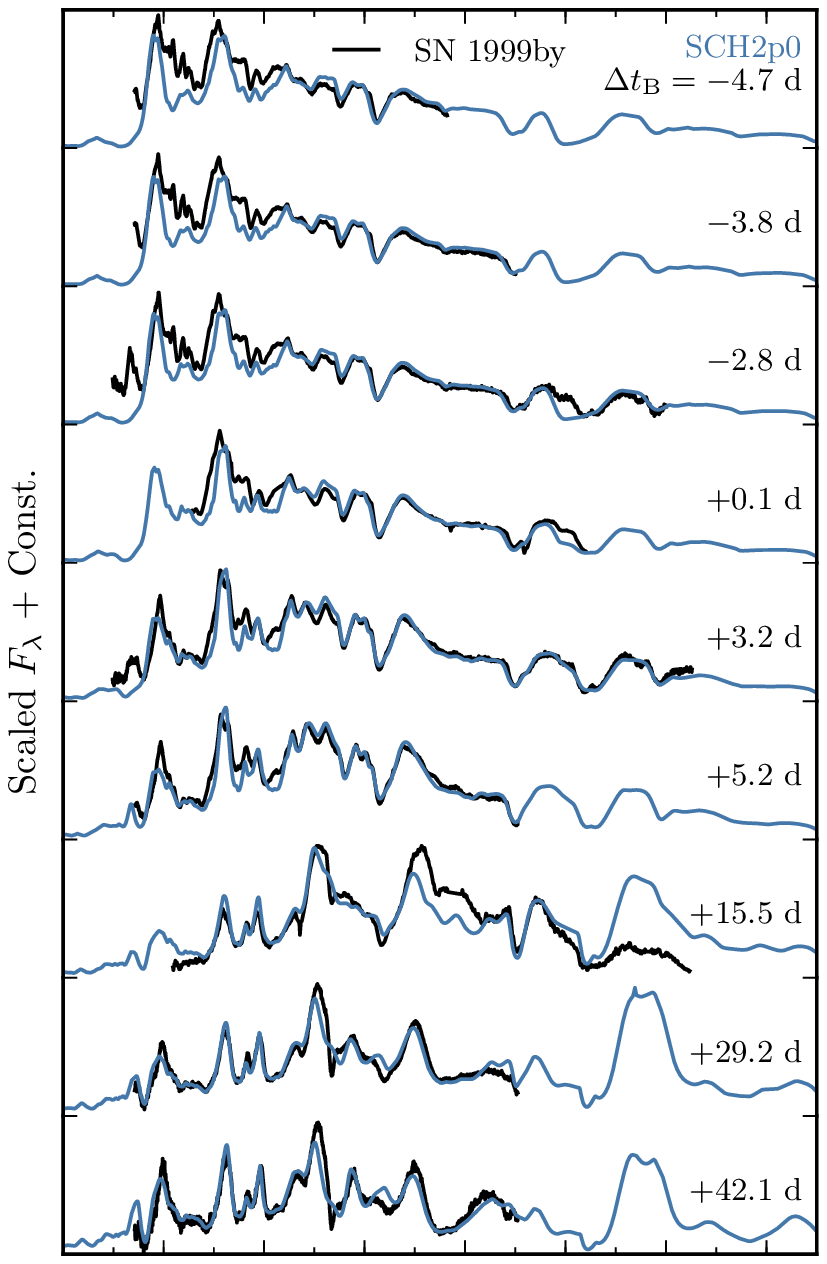}\vspace{-1cm}
\includegraphics{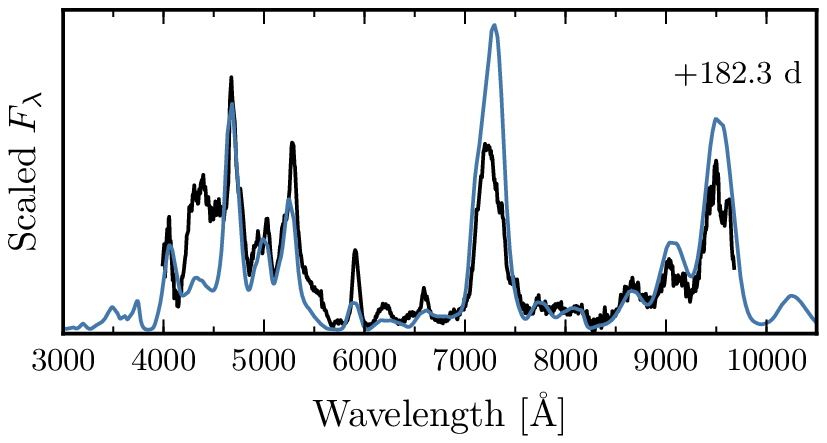}
\end{minipage}
\caption{\label{fig:comp_spec_opt}
Optical spectroscopic evolution of the \mch\ delayed-detonation model
DDC25 (left; red line) and the sub-\mch\ model SCH2p0 (right; blue
line) compared to SN~1999by (black line), between $-4.7$\,d and
$+182.3$\,d from $B$-band maximum. The tickmarks on the ordinate give
the zero-flux level. The observed spectra have been de-redshifted,
de-reddened, and scaled to match the absolute $V$-band magnitude
inferred from the corresponding photometry. An additional scaling has
been applied to the synthetic spectra to reproduce the mean observed
flux in the range 5000--6500~\AA\ (top panels) or in the range
3000--10000~\AA\ (bottom panels showing the spectra at $+182.3$\,d).
}
\end{figure*}

We confront both models to the low-luminosity SN~1999by, as was
previously done for the \mch\ model at maximum light in
\cite{Blondin/etal:2013}. SN~1999by was discovered independently on
April 30$^\mathrm{th}$, 1999 by R.~Arbour, South Wonston, Hampshire,
England, and by the Lick Observatory Supernova Search
\citep{IAUC_1999by_disc}. It exploded in the Sb galaxy NGC 2841,
well-known for having hosted three other supernovae (SN~1912A,
SN~1957A, and SN~1972R). An optical spectrum taken three days after
discovery secured its classification as a Type Ia supernova
\citep{IAUC_1999by_specid}, while a report based on a second optical
spectrum taken on May 6$^\mathrm{th}$ by \cite{IAUC_1999by_specid2}
not only confirmed the \snia\ classification, but also noted the large
strength of the Si\two\,5972\,\AA\ absorption feature relative to the
neighbouring Si\two\,6355\,\AA\ absorption, characteristic of
low-luminosity, 91bg-like \sneia\ \citep{Nugent/etal:1995}.

SN~1999by is one of the best-observed low-luminosity \sneia, with a
wealth of publicly-available data covering pre-maximum to nebular
phases, at optical and NIR wavelengths.  We use optical/NIR photometry
and optical spectra published by \cite{Garnavich/etal:2004}, with
additional NIR photometry and spectroscopy from
\cite{Hoeflich/etal:2002}.  We assume a distance modulus of
30.97\,mag, which is 1$\sigma$ above the Cepheid distance to the host
galaxy NGC 2841 inferred by \cite{Macri/etal:2001}
($30.74\pm0.23$\,mag), to ensure a good match to the peak luminosity
of the sub-\mch\ model (see below). We neglect extinction in the
host-galaxy, as inferred by \citealt{Garnavich/etal:2004} based on the
late-time $B-V$ colour evolution. The Galactic reddening estimate of
$E(B-V)_{\rm Gal}=0.016$\,mag (i.e. $A_{V, {\rm Gal}}\approx
0.05$\,mag for $R_V=3.1$) is derived from the IR dust maps of
\cite{SFD98}.  In what follows we correct the magnitudes and spectra
of SN~1999by for reddening based on this value, assuming the standard
extinction law of \cite{CCM89}.

\subsection{UV-optical-IR evolution}

Due to the lack of UV observations of SN~1999by and the absence of NIR
measurements beyond $\sim20$\,d past $B$-band maximum (except for a
single $H$-band measurement around $+50$\,d), we approximate the uvoir
luminosity by integrating $BVRI$ fluxes using the same procedure as in
\cite{Blondin/etal:2013}. The resulting light curve for SN~1999by is
shown in Fig.~\ref{fig:comp_luvoir_inclate}, along with the integrated
$B \rightarrow I$ light curves of the \mch\ (DDC25; dashed line) and
sub-\mch\ (SCH2p0; solid line) models.  As expected from our earlier
analysis of the uvoir luminosity (Section~\ref{sect:luvoir}), the
sub-\mch\ model reaches a $\gtrsim30$ per cent higher peak
$B\rightarrow I$ luminosity compared to the \mch\ model despite having
the same \nifs\ mass. The peak uvoir luminosity (taking into account
the full UV-optical-IR range) is only $\sim20$ per cent larger (see
Table~\ref{tab:sch2p0_ddc25_comp}), owing to the larger fraction of
flux emitted redward of the $I$ band in the cooler \mch\ model.

Clearly, the sub-\mch\ model is a better match to the overall
luminosity evolution of SN~1999by.  The peak luminosity of SN~1999by
would be better matched by the \mch\ model DDC25 with a lower assumed
distance modulus (see above and \citealt{Blondin/etal:2013}), although
this would only exacerbate the mismatch at earlier and later times
(see Fig.~\ref{fig:comp_luvoir_inclate} inset).  The $B\rightarrow I$
magnitude post-maximum decline rate is $\Delta M_{15}(B\rightarrow I)
\approx 1.3$\,mag for SN~1999by, and compares well with the
sub-\mch\ model ($\Delta M_{15}(B\rightarrow I) = 1.22$\,mag). The
\mch\ model, on the other hand, displays a $\sim 0.3$\,mag smaller
decline rate ($\Delta M_{15}(B\rightarrow I) = 0.87$\,mag).  The
low-luminosity SN~1991bg also displays a larger post-maximum
$B\rightarrow I$ decline rate compared to more luminous events
\citep{Contardo/Leibundgut/Vacca:2000,Stritzinger/etal:2006b}.

The sub-\mch\ model reproduces better the late-time luminosity decline
of SN~1999by compared to the \mch\ model.  This provides an additional
argument in favour of an ejecta mass significantly below \mch\ for
this low-luminosity \snia\ (see Section~\ref{sect:luvoir}).

\subsection{Colour evolution}\label{sect:comp_col}

The faster uvoir evolution of 91bg-like \sneia\ compared to more
luminous events results in narrower light curves in individual
photometric bands (Fig.~\ref{fig:comp_lc}). This effect is modulated
by the gradual reddening of the SED around maximum light, which causes
a decrease in the post-maximum decline rate along the sequence
$B\rightarrow V\rightarrow R\rightarrow I$ (larger than the uvoir
decline rate in all cases). The SN~1999by data illustrate the faster
colour evolution in low-luminosity \sneia\ around maximum light, and
are well matched by the sub-\mch\ model
(Fig.~\ref{fig:comp_colors}). This model also reproduces the steep
early rise in the $V$-band light curve, which is the only band with
measurements earlier than 10\,d before $B$-band maximum.

The sub-\mch\ model nonetheless overestimates the flux in the $R$ and
$I$ bands at late times (Fig.~\ref{fig:comp_lc}, bottom panel),
resulting in $V-R$ and $V-I$ colour indices that are $\sim 0.5$\,mag
too red compared to SN~1999by. This colour mismatch at late times
explains the excess $B \rightarrow I$ luminosity at $\sim 180$\,d past
maximum (see Fig.~\ref{fig:comp_luvoir_inclate}).

In the NIR, the earlier {\sc iii}$\rightarrow${\sc ii} recombination
of IGEs in low-luminosity \sneia\ leads to an earlier secondary
maximum in $JHK_s$ compared to more luminous events \citep[see
  e.g.][]{Dhawan/etal:2017}. For SN~1999by, the secondary maximum is
barely noticeable in the $J$ band, resulting in a ``shoulder'' around
10--15\,d past $B$-band maximum, which is present in the
sub-\mch\ model where the \mch\ model displays a genuine second peak.
In the $H$ and $K_s$ bands, the secondary maximum appears to merge
with the primary maximum. It is also more prominent, and better
reproduced by the sub-\mch\ model, owing to the larger
Co\two\ emissivity in the NIR at these epochs (see
Section~\ref{sect:col}; the same applies to the more luminous models
of \citealt{Wilk/etal:2017}). The apparent broadness of these light
curves around maximum thus results from a colour shift that is not
apparent in the evolution of the integrated $B\rightarrow I$
luminosity.

The success of the sub-\mch\ model in reproducing the optical and NIR
light curves of SN~1999by is nonetheless undermined by the obvious
flux deficit in the $U$ band (and to a lesser extent in the $B$ band)
at pre-maximum epochs, although the \mch\ model displays an even
larger flux deficit. In Section~\ref{sect:col} we highlighted the
impact of a low ejecta mass on the gas properties (higher temperature
and ionization state) in the spectrum-formation region, which results
in a bluer $B-V$ colour at maximum light. In \cite{Blondin/etal:2017a}
we found this to be the key ingredient to reproduce the faint end of
the observed $B$-band WLR.  Despite the higher ejecta temperature, the
$B-V$ colour of the sub-\mch\ model is still too red compared to
SN~1999by at pre-maximum epochs, by $\sim0.2$\,mag at $-5$\,d and
$\lesssim0.5$\,mag at $-10$\,d (Fig.~\ref{fig:comp_colors}).  Such
offsets are not surprising given the artificial setup of our explosion
models (see Section~\ref{sect:model}). More relevant to the present
study are the {\it relative} differences between both models, and the
closer match of the sub-\mch\ model to SN~1999by throughout its
evolution. In Section~\ref{sect:ccl} we discuss ways in which our
sub-\mch\ model could be tuned to produce an even better match to
SN~1999by.

\subsection{Optical spectroscopic evolution}

We compare the optical spectroscopic evolution of both models to 
SN~1999by in Fig.~\ref{fig:comp_spec_opt}.  The prominent absorption
trough around 4000--4500\,\AA\ is clearly visible, becoming stronger
on its way to maximum light. This spectroscopic hallmark of 91bg-like
\sneia\ is present in both models, although it remains significantly
too strong in the \mch\ model until a few days past maximum.  The
sub-\mch\ model fares better in this respect, but fails to reproduce
the narrow absorption features in this region present in SN~1999by up
until maximum light. The sub-\mch\ model also underestimates
the flux blueward of $\sim5000$\,\AA\ owing to the low predicted
ionization level and the resulting absorption by several lines of
Sc\two\ and Ti\two\ (and to a lesser extent Cr\two). Redward of the
$B$ band, the match to the SN~1999by spectra is far more satisfactory,
in part due to the dominance of strong lines of IMEs, less sensitive
to the ionization balance compared to weaker lines of heavier
ions. This explains the good correspondence of the sub-\mch\ model
with the $V-\{R,I,H\}$ colour evolution of SN~1999by at early times,
while the $B-V$ colour remains too red (Fig.~\ref{fig:comp_colors}).

%%% FIGURE: Comparison NIR spectra
\begin{figure*}
\centering
\includegraphics{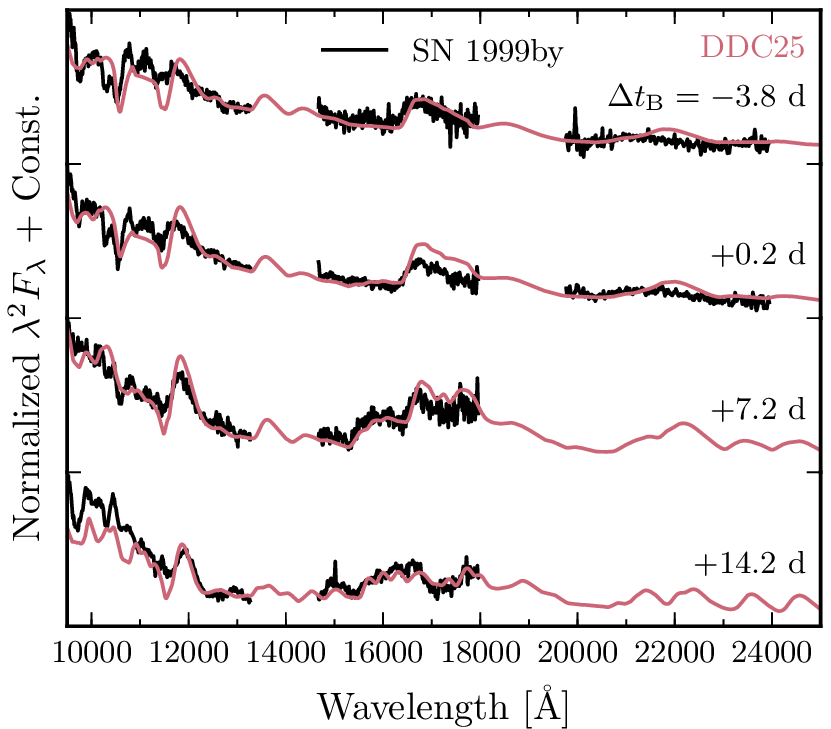}\hspace{.5cm}
\includegraphics{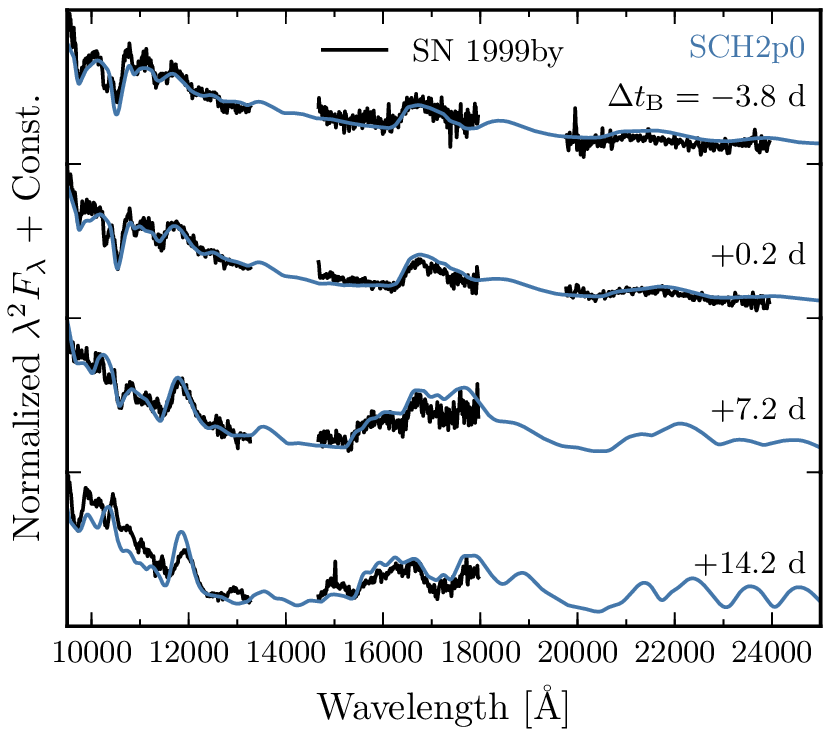}
\caption{\label{fig:comp_spec_nir}
NIR spectroscopic evolution of the \mch\ delayed-detonation model
DDC25 (left; red line) and the sub-\mch\ model SCH2p0 (right; blue
line) compared to SN~1999by (black line), between $-3.8$\,d and
$+14.2$\,d from $B$-band maximum, corresponding to the earliest and
latest NIR spectra for this event. The flux has been scaled by
$\lambda^2$ for better visibility, and the tickmarks on the ordinate
give the zero-flux level. The observed spectra have been
de-redshifted, de-reddened, and scaled to the same absolute $J$-band
magnitude as the corresponding model.
}
\end{figure*}

\subsubsection{Absence of C\two\ lines}

Due to the low carbon abundance in the spectrum-formation region,
combined with the low ionization level and low temperature at early
times, we do not predict optical lines of C\two\ (e.g. the
6580\,\AA\ doublet) whose presence was inferred in spectra of the
low-luminosity SN~2005bl \citep{Taubenberger/etal:2008}. The small dip
in the Si\two~6355\,\AA\ emission profile in the earliest optical
spectrum of SN~1999by (Fig.~\ref{fig:comp_spec_opt}), associated with
C\two~6580\,\AA\ by \cite{Taubenberger/etal:2008} in the case of
SN~2005bl, is due to absorption by the Mg\two~6347\,\AA\ doublet in
our sub-\mch\ model (see also Fig.~\ref{fig:ladder_1}).

\subsubsection{High-velocity absorption in the Ca\two~8500\,\AA\ triplet}

High-velocity absorption features (HVFs) in the
Ca\two~8500\,\AA\ triplet are common in early-time spectra of
\sneia\ \citep{Mazzali/etal:2005b}, sometimes causing a detached
absorption feature blueward of the main component.  A weak absorption
in the blue wing of the main component is visible at $\sim
8000$\,\AA\ in the earliest SN~1999by spectrum covering this
wavelength range (at $-2.8$\,d from $B$-band maximum; see also
\citealt{Blondin/etal:2012}, their Fig.~20), and could correspond to a
weak Ca\two\ HVF. Both models predict the presence of HVFs extending
blueward of 8000\,\AA\ (corresponding to Doppler velocities $\lesssim
-20000$\,\kms) up until maximum light, but they are suppressed at an
earlier stage in the sub-\mch\ model due to the increase in the Ca
ionization level at high velocities caused by non-local energy
deposition \citep[see also][]{Blondin/etal:2013}. The weak absorption
at $\sim 8000$\,\AA\ in the sub-\mch\ model synthetic spectrum at
+0.1\,d is indeed due to Ca\two\ (see Appendix~\ref{sect:ladder}), and
supports the association of this feature with an HVF in the SN~1999by
spectra at this and earlier times.

The double-absorption morphology of the main
Ca\two~8500\,\AA\ component results from overlap of the individual
transitions composing the triplet (8498, 8542, and 8662\,\AA). By
maximum light, the width of the Ca\two~8500\,\AA\ absorption in
SN~1999by is well matched by the sub-\mch\ model. Despite a possible
mis-calibration of the +15.5\,d spectrum redward of 6000\,\AA, the
emission component of the Ca\two~8500\,\AA\ feature is obviously
over-estimated in our models, possibly due to scattering of Co\two\ photons
emitted in underlying layers \cite[see][]{Blondin/etal:2015}.

\subsubsection{Si\two~6355\,\AA\ absorption velocity}

At maximum light, the Si\two~6355\,\AA\ absorption velocity for
SN~1999by is approximately $-10000$\,\kms, as in the
sub-\mch\ model. The \mch\ model, however, displays a
$\sim1500$\,\kms\ lower blueshift. As noted in
Section~\ref{sect:spec}, the higher velocity location of the
spectrum-formation region in the sub-\mch\ model leads to broader and
more blueshifted absorption profiles, in better agreement with the
line-profile morphology of SN~1999by.

\subsubsection{Post-maximum evolution out to the nebular phase}

At later times, both models reproduce the overall SED of SN~1999by, as
expected from the comparison of their broad-band colours in
Fig.~\ref{fig:comp_colors}, but the \mch\ model displays small-scale
variations not present in the data. The sub-\mch\ model matches better
the typical widths of spectral features, although SN~1999by displays
even narrower features, e.g., on either side of the Cr\two\ emission
peak at $\sim 5500$\,\AA\ between +15.5\,d and +42.1\,d. The strength
of the [Co\three]~5888\,\AA\ emission feature in the SN~1999by
spectrum at +42.1\,d pleads in favour of the higher ionization state
and lower ejecta densities of the sub-\mch\ model.

The nebular-phase spectrum of SN~1999by at +182\,d past maximum
(Fig.~\ref{fig:comp_spec_opt}, lower panel) does not appear to display
the predicted narrow emission feature due to [Ni\two] 7378,
7412\,\AA\ in the \mch\ model (see Section~\ref{sect:spec}), although
the flux scaling exacerbates the mismatch. The sub-\mch\ model does
not display [Ni\two] lines (see Sect.~\ref{sect:nihole}), but the
[Ca\two] 7291, 7324\,\AA\ doublet contributes too much flux around
7300\,\AA\ relative to other lines, and is largely responsible for the
excess flux in the $R$- and $I$-band light curves at this time
(Fig.~\ref{fig:comp_lc}, bottom panel).

The double-peak emission profile due to [Cr\two] 8000, 8125\,\AA\ in
the \mch\ model is absent from the SN~1999by spectrum, which is then
better matched by the sub-\mch\ model in this range.  The high
ionization of the sub-\mch\ model also results in the strong [S\three]
9068, 9530\,\AA\ lines (with a $\sim 1:2.5$ ratio) that are clearly
visible in SN~1999by and nebular-phase spectra of other low-luminosity
\sneia\ (e.g. SN~1991bg, \citealt{Ruiz-Lapuente/etal:1993}; SN~2003gs,
\citealt{Silverman/etal:2012a}). These [S\three] lines are also
predicted in the higher-luminosity models of \cite{Wilk/etal:2017},
but seem to be absent from nebular spectra of more luminous \sneia.
  
Both models underestimate the [Fe\two]-dominated emission feature
around 4350\,\AA. As noted by \cite{Wilk/etal:2017}, this feature is
particularly sensitive to the Fe$^{+}$/Fe$^{2+}$ ionization ratio,
which we compute consistently via a solution to the non-LTE
time-dependent rate equations.
Furthermore, neither model fully reproduces the narrow cores (${\rm
  FWHM} < 4000$\,\kms) of the [Fe\three]~4700\,\AA\ emission feature
and of the [Co\three]~5888\,\AA\ line. To confine the line-emitting
region to $\varv \lesssim 4000$\,\kms, one would need to artificially
reduce the density in the overlying layers (and hence increase the
density in the underlying layers to conserve mass), or reduce the
velocity extent of the \nifs\ distribution (and hence reduce the
\nifs\ yield), since the Fe and Co abundances at this time are largely
dominated by the \nifs\ decay products \fefs\ and \cofs,
respectively. However, such arbitrary adjustments would violate the
physical consistency of our input hydrodynamical model.

\subsection{NIR spectral evolution}

The NIR spectroscopic evolution of SN~1999by between $-3.8$\,d and
$+14.2$\,d from $B$-band maximum is shown in
Fig.~\ref{fig:comp_spec_nir}. Up until maximum light, the spectrum is
mostly shaped by lines of IMEs (in particular, S\two, Si\two, and
Mg\two), while lines of IGEs leave their imprint shortly thereafter
(see Appendix~\ref{sect:ladder}).  At these times, the differences in
the overall SED between the two models are less apparent in the NIR
range than in the optical, and it is difficult to gauge the quality of
their respective match to the SN~1999by data.

\subsubsection{C\one\ lines past maximum light?}

Both models fail to reproduce the narrow absorption feature at
$\sim1.03$\,$\mu$m present in the SN~1999by spectra up until $\lesssim
7$\,d past maximum. This feature is associated with the
C\one\ 1.07$\mu$m doublet in both models at early times, but is
largely gone by $-10$\,d from maximum, once the spectrum-formation
region no longer probes the carbon-rich ejecta layers at $\varv
\gtrsim 15000$\,\kms\ (see Fig.~\ref{fig:elem_distrib}).
\cite{Hoeflich/etal:2002}, however, were able to reproduce this line
in their maximum-light synthetic spectra. Their 5p0z22.8 model is
similar to our \mch\ DDC25 model (in particular, both have the same
deflagration-to-detonation transition density $\rho_{\rm tr}=8\times
10^6$\,\gcc), but differences in treatment of the burning front during
the deflagration phase lead to less efficient burning in their model
and to the presence of unburnt carbon at lower velocities. The carbon
mass fraction reaches a maximum value of $\sim 0.5$ at
$\sim15000$\,\kms\ in their model, favouring the emergence of
C\one\ lines, while it is an order of magnitude less in our DDC25
model at the same velocity.

%%%%%%%%%%%%%%%%%%%%%%%%%%%%%%%%%%%%%%%%%%%%%%%%%%%%%%%%%%%%%%%%%%%%%%

\section{Discussion and Conclusions}\label{sect:ccl}

We have studied the impact of the progenitor WD mass on the radiative
display of low-luminosity \sneia, illustrated here with the
\mch\ delayed-detonation model DDC25 of \cite{Blondin/etal:2013} and a
sub-\mch\ model resulting from the pure central detonation of a
0.90\,\msun\ WD progenitor (SCH2p0;
\citealt{Blondin/etal:2017a}). Both models have the same \nifs\ yield
of 0.12\,\msun, and hence differ in their \mratio\ ratio.
Although the setup for the progenitor and explosion scenario is
somewhat artificial (1D, numerically-triggered explosion etc.; see
Section~\ref{sect:model}), it allows us to assess the impact of the
ejecta mass on the resulting light curves and spectra.

The lower ejecta mass of the sub-\mch\ model results in a larger
outward extent of the \nifs\ distribution for a given \nifs\ mass, and
hence to a faster rise to peak luminosity and a more rapid
post-maximum decline. The larger \nifs-to-total mass ratio leads to
bluer colours at all times.  Moreover, the higher \nifs\ mass fraction
at larger velocities enhances the ionization state of the gas in the
outer ejecta (ions primarily twice ionized rather than once ionized)
where the flux emerges at early times, and limits line blanketing from
once-ionized IGEs compared to the \mch\ model. The spectrum-formation
region is also located at higher velocities, giving rise to broader
and more blueshifted absorption features. Taken together, the match to
the UV-optical-IR luminosity, broad-band colours, and spectral
evolution of the low-luminosity SN~1999by argues in favour of a
sub-\mch\ progenitor for this event and, by analogy, for other
91bg-like \sneia.

\citealt{Hoeflich/etal:2002} were able to reproduce optical and
near-infrared observations of SN~1999by with a standard
\mch\ delayed-detonation model similar to our DDC25 model. In
particular, their 5p0z22.8 model predicts fast-evolving light curves
in $B$ and $V$, with a $B$-band decline rate $\dmft = 1.73$\,mag,
similar to our sub-\mch\ model SCH2p0 ($\dmft = 1.64$\,mag). Moreover,
their synthetic maximum-light optical spectrum matches reasonably well
the SN~1999by data (their Fig. 9). However, their light-curve
calculations rely on averaged LTE opacities, and their spectral
calculations do not include time-dependent terms in the
radiative-transfer and rate equations, where only a subset of atomic
levels are treated in non-LTE. Our more elaborate and consistent
treatment of the radiation transport is probably more suited in this
respect, although we cannot exclude differences in the input
hydrodynamical models as contributing to our distinct model
predictions. Such issues can only be resolved with detailed
code-comparison studies based on benchmarked models, which are
currently lacking for \sneia.

Our sub-\mch\ model from a 0.90\,\msun\ WD progenitor SCH2p0 is
similar in many respects to the 0.88\,\msun\ model of
\cite{Sim/etal:2010}. However, the \nifs\ yield for their model is
lower (0.07\,\msun, cf. 0.12\,\msun\ for our SCH2p0 model), resulting
in a less luminous and faster-evolving light curve, redder colours,
and less blueshifted spectroscopic absorption features. Moreover,
their maximum-light spectrum does not display the prominent absorption
trough around 4000--4500\,\AA\ characteristic of 91bg-like
\sneia\ (see their Fig.~3), which could be due to the quasi-absence of
Ti/Sc above $\sim 10000$\,\kms\ in their model (Sim, priv. comm.).

By construction, our spherically-symmetric ejecta cannot account for
potential ejecta asymmetries resulting from the explosion. Given the
supersonic propagation of the detonation front in our models, the 1D
approximation is not really a limitation in this context
\citep[see][]{Livne/Arnett:1995}. However, spectro-polarimetric
observations of SN~1999by near maximum light have revealed an
intrinsic polarization at the $\lesssim1$ per cent level, suggestive
of a well-defined axis of symmetry according to
\cite{Howell/etal:2001}. Given the overall agreement of the
sub-\mch\ model SCH2p0 with optical and NIR observations of SN~1999by,
the impact of such asymmetries is not expected to dramatically affect
the radiative display.

The recent detailed nucleosynthesic calculations of
\cite{Shen/etal:2017} predict a significantly higher \nifs\ yield for
a given WD mass, particularly for their low-mass models.  The larger
specific heating rate from the higher \nifs-to-total mass ratio should
lead to bluer colours at early times, in better agreement with
observations.

We reiterate the potential of the [Ni\two] 1.939\,$\mu$m line to
constrain the presence and distribution of stable Ni isotopes
predicted in the innermost ejecta of 1D \mch\ models. While stable
IGEs might be mixed outwards during the initial deflagration phase
(see Section~\ref{sect:nihole}), detection of this line in nebular
\snia\ spectra would provide an unambiguous signature of burning at
high densities during the explosion, which can only be realized in a
denser, near-\mch\ WD progenitor.

The question remains whether plausible progenitor scenarios leading to
detonations of sub-\mch\ WDs exist in Nature. A generic problem of
double-detonation models resides in the IGE-rich composition of the
detonated accreted He shell, needed to trigger a secondary detonation
in the C-O core \citep[see  e.g.][]{Kromer/etal:2010}. However,
\cite{Shen/Moore:2014} succeeded in detonating a 0.005\,\msun\ He
shell on a 1.0\,\msun\ WD, producing only $^{28}$Si and $^{4}$He. The
subsequent detonation of the C-O core would in this case result in an
ejecta with similar properties as our sub-\mch\ model. This is also
true of the head-on collision of two 0.5\,\msun\ WDs proposed by
\cite{Kushnir/etal:2013}, with a similar \nifs\ yield of 0.11\,\msun,
although such collisions are predicted to account for at most a few
per cent of the observed \snia\ rate \citep[see
  e.g.][]{Papish/Perets:2016}.

Regardless of the precise ignition mechanism, the results presented in
this paper strongly suggest that low-luminosity \sneia\ similar to
SN~1999by result from the explosion of sub-\mch\ WDs.
In an upcoming paper we will study the feasibility of such
sub-\mch\ models in reproducing the observed properties of more
luminous events, to address the question of multiple progenitor
channels for Type Ia supernovae.

%%%%%%%%%%%%%%%%%%%%%%%%%%%%%%%%%%%%%%%%%%%%%%%%%%%%%%%%%%%%%%%%%%%%%%

\section*{acknowledgments}

SB acknowledges helpful discussions with Roland Diehl, Marten van
Kerkwijk, Ken Shen, Stuart Sim, and Jason Spyromilio. Part of this
work was realized during a one-month visit of SB to ESO as part of the
ESO Scientific Visitor Programme. LD and SB acknowledge financial
support from the Programme National de Physique Stellaire (PNPS) of
CNRS/INSU, France. DJH acknowledges support from STScI theory grant
HST-AR-12640.01, and NASA theory grant NNX14AB41G. This work was
granted access to the HPC resources of CINES under the allocation
c2014046608 made by GENCI (Grand Equipement National de Calcul
Intensif). This work also used computing resources of the mesocentre
SIGAMM, hosted by the Observatoire de la C\^ote d'Azur, Nice, France.
This research was supported by the DFG cluster of excellence ``Origin
and Structure of the Universe''. We thank the anonymous referee for
a thorough review and useful suggestions that improved the quality of
this manuscript.

%%%%%%%%%%%%%%%%%%%%%%%%%%%%%%%%%%%%%%%%%%%%%%%%%%%%%%%%%%%%%%%%%%%%%%

\bibliographystyle{mnras}
\bibliography{ms_99by} % if your bibtex file is called example.bib

%%%%%%%%%%%%%%%%%%%%%%%%%%%%%%%%%%%%%%%%%%%%%%%%%%

%%%%%%%%%%%%%%%%% APPENDICES %%%%%%%%%%%%%%%%%%%%%

\appendix

\section{Synthetic light curves}\label{sect:lctabs}

Tables~\ref{tab:ddc25_mags} and \ref{tab:sch2p0_mags} give the
integrated UV-optical-IR luminosity ($L_\mathrm{uvoir}$), the
$\gamma$-ray luminosity ($L_\gamma$), and the (true) bolometric
luminosity ($L_{\rm bol} = L_{\rm uvoir} + L_{\gamma}$), as well as
the absolute optical ($UBVRI$) and NIR ($JHK_s$) magnitudes of the
\mch\ model DDC25 (Table~\ref{tab:ddc25_mags}) and the sub-\mch\ model
SCH2p0 (Table~\ref{tab:sch2p0_mags}), as a function of time since
explosion and from uvoir maximum.

Also included is the conversion constant between the (absolute)
$V$-band magnitude and the uvoir magnitude, such that:

\begin{eqnarray}
M_\mathrm{uvoir} &=& M_{\mathrm{uvoir},\sun} - 2.5 \log_{10} \left(
\frac{L_\mathrm{uvoir}}{L_{\mathrm{uvoir},\sun}} \right) \label{eqn:muvoir} \\
                &=& M_V + (V \rightarrow \mathrm{uvoir}), \label{eqn:mvconv}
\end{eqnarray}

\noindent
where we have assumed that $M_{\mathrm{uvoir},\sun} =
M_{\mathrm{bol},\sun} = -4.75$\,mag and $L_{\mathrm{uvoir},\sun} =
L_{\mathrm{bol},\sun} = 3.826 \times 10^{33}$\,\ergs, i.e. neglecting
the Sun's high-energy radiative output (X-rays and $\gamma$-rays).
Knowing the peak $V$-band magnitude, one can compute the peak absolute
uvoir magnitude (equation~\ref{eqn:mvconv}), and hence the peak uvoir
luminosity (using equation~\ref{eqn:muvoir}), which can then be used
to infer the \nifs\ mass.

\begin{table*}
\footnotesize
\caption{Light curves for model DDC25.}\label{tab:ddc25_mags}
\begin{tabular}{r@{\hspace{3.3mm}}r@{\hspace{3.3mm}}r@{\hspace{0.0mm}}r@{\hspace{3.3mm}}r@{\hspace{3.3mm}}r@{\hspace{3.3mm}}r@{\hspace{3.3mm}}r@{\hspace{3.3mm}}r@{\hspace{3.3mm}}r@{\hspace{3.3mm}}r@{\hspace{3.3mm}}r@{\hspace{3.3mm}}r@{\hspace{3.3mm}}r}
\hline
\multicolumn{1}{c}{$t_\mathrm{exp}$} & \multicolumn{1}{c}{$\Delta t_\mathrm{uvoir}$} & \multicolumn{1}{c}{$L_\mathrm{uvoir}$} & \multicolumn{1}{c}{$V \rightarrow \mathrm{uvoir}$} & \multicolumn{1}{c}{$L_{\gamma}$} & \multicolumn{1}{c}{$L_\mathrm{bol}$} & \multicolumn{1}{c}{$U$} & \multicolumn{1}{c}{$B$} & \multicolumn{1}{c}{$V$} & \multicolumn{1}{c}{$R$} & \multicolumn{1}{c}{$I$} & \multicolumn{1}{c}{$J$} & \multicolumn{1}{c}{$H$} & \multicolumn{1}{c}{$K_s$} \\
\multicolumn{1}{c}{(days)} & \multicolumn{1}{c}{(days)} & \multicolumn{1}{c}{(erg s$^{-1}$)} & \multicolumn{1}{c}{(mag)} & \multicolumn{1}{c}{(erg s$^{-1}$)} & \multicolumn{1}{c}{(erg s$^{-1}$)} & \multicolumn{1}{c}{(mag)} & \multicolumn{1}{c}{(mag)} & \multicolumn{1}{c}{(mag)} & \multicolumn{1}{c}{(mag)} & \multicolumn{1}{c}{(mag)} & \multicolumn{1}{c}{(mag)} & \multicolumn{1}{c}{(mag)} & \multicolumn{1}{c}{(mag)} \\
\hline
1.30 & $-$19.74 & 2.79\,(39) & $-$1.97 & \multicolumn{1}{c}{$\cdots$} & 2.79\,(39) & $-$3.00 & $-$5.15 & $-$7.94 & $-$9.91 & $-$10.29 & $-$11.74 & $-$11.88 & $-$12.31 \\
1.43 & $-$19.61 & 3.56\,(39) & $-$1.82 & \multicolumn{1}{c}{$\cdots$} & 3.56\,(39) & $-$3.09 & $-$5.60 & $-$8.35 & $-$10.25 & $-$10.60 & $-$12.00 & $-$12.10 & $-$12.51 \\
1.57 & $-$19.47 & 4.69\,(39) & $-$1.66 & \multicolumn{1}{c}{$\cdots$} & 4.69\,(39) & $-$3.33 & $-$6.16 & $-$8.81 & $-$10.65 & $-$10.95 & $-$12.29 & $-$12.35 & $-$12.74 \\
1.73 & $-$19.31 & 6.37\,(39) & $-$1.47 & \multicolumn{1}{c}{$\cdots$} & 6.37\,(39) & $-$3.70 & $-$6.82 & $-$9.33 & $-$11.09 & $-$11.33 & $-$12.60 & $-$12.63 & $-$12.99 \\
1.90 & $-$19.14 & 8.86\,(39) & $-$1.29 & \multicolumn{1}{c}{$\cdots$} & 8.86\,(39) & $-$4.27 & $-$7.57 & $-$9.87 & $-$11.57 & $-$11.73 & $-$12.92 & $-$12.94 & $-$13.26 \\
2.09 & $-$18.95 & 1.26\,(40) & $-$1.06 & \multicolumn{1}{c}{$\cdots$} & 1.26\,(40) & $-$5.06 & $-$8.36 & $-$10.48 & $-$12.09 & $-$12.16 & $-$13.24 & $-$13.26 & $-$13.51 \\
2.30 & $-$18.74 & 1.82\,(40) & $-$0.78 & \multicolumn{1}{c}{$\cdots$} & 1.82\,(40) & $-$5.83 & $-$9.10 & $-$11.16 & $-$12.61 & $-$12.62 & $-$13.55 & $-$13.58 & $-$13.77 \\
2.53 & $-$18.51 & 2.60\,(40) & $-$0.54 & \multicolumn{1}{c}{$\cdots$} & 2.60\,(40) & $-$6.57 & $-$9.74 & $-$11.79 & $-$13.06 & $-$13.04 & $-$13.84 & $-$13.90 & $-$14.04 \\
2.78 & $-$18.26 & 3.64\,(40) & $-$0.37 & \multicolumn{1}{c}{$\cdots$} & 3.64\,(40) & $-$7.40 & $-$10.34 & $-$12.32 & $-$13.46 & $-$13.43 & $-$14.12 & $-$14.21 & $-$14.31 \\
3.06 & $-$17.98 & 5.02\,(40) & $-$0.24 & \multicolumn{1}{c}{$\cdots$} & 5.02\,(40) & $-$8.34 & $-$10.93 & $-$12.80 & $-$13.82 & $-$13.79 & $-$14.40 & $-$14.50 & $-$14.58 \\
3.37 & $-$17.67 & 6.87\,(40) & $-$0.13 & \multicolumn{1}{c}{$\cdots$} & 6.87\,(40) & $-$9.27 & $-$11.51 & $-$13.25 & $-$14.15 & $-$14.12 & $-$14.67 & $-$14.78 & $-$14.86 \\
3.70 & $-$17.34 & 9.27\,(40) & $-$0.05 & \multicolumn{1}{c}{$\cdots$} & 9.27\,(40) & $-$10.08 & $-$12.03 & $-$13.66 & $-$14.47 & $-$14.44 & $-$14.94 & $-$15.05 & $-$15.12 \\
4.07 & $-$16.97 & 1.23\,(41) & 0.02 & \multicolumn{1}{c}{$\cdots$} & 1.23\,(41) & $-$10.75 & $-$12.50 & $-$14.03 & $-$14.77 & $-$14.73 & $-$15.20 & $-$15.31 & $-$15.37 \\
4.48 & $-$16.56 & 1.61\,(41) & 0.07 & \multicolumn{1}{c}{$\cdots$} & 1.61\,(41) & $-$11.32 & $-$12.92 & $-$14.38 & $-$15.06 & $-$15.01 & $-$15.45 & $-$15.55 & $-$15.61 \\
4.93 & $-$16.11 & 2.07\,(41) & 0.11 & \multicolumn{1}{c}{$\cdots$} & 2.07\,(41) & $-$11.80 & $-$13.28 & $-$14.69 & $-$15.32 & $-$15.27 & $-$15.69 & $-$15.78 & $-$15.84 \\
5.42 & $-$15.62 & 2.63\,(41) & 0.15 & \multicolumn{1}{c}{$\cdots$} & 2.63\,(41) & $-$12.22 & $-$13.61 & $-$14.99 & $-$15.58 & $-$15.52 & $-$15.92 & $-$16.00 & $-$16.05 \\
5.96 & $-$15.08 & 3.29\,(41) & 0.18 & \multicolumn{1}{c}{$\cdots$} & 3.29\,(41) & $-$12.58 & $-$13.91 & $-$15.27 & $-$15.81 & $-$15.75 & $-$16.13 & $-$16.21 & $-$16.25 \\
6.56 & $-$14.48 & 4.06\,(41) & 0.21 & \multicolumn{1}{c}{$\cdots$} & 4.06\,(41) & $-$12.89 & $-$14.17 & $-$15.52 & $-$16.03 & $-$15.98 & $-$16.34 & $-$16.41 & $-$16.45 \\
7.22 & $-$13.82 & 4.95\,(41) & 0.23 & \multicolumn{1}{c}{$\cdots$} & 4.95\,(41) & $-$13.15 & $-$14.41 & $-$15.76 & $-$16.24 & $-$16.20 & $-$16.54 & $-$16.60 & $-$16.63 \\
7.94 & $-$13.10 & 6.01\,(41) & 0.24 & \multicolumn{1}{c}{$\cdots$} & 6.01\,(41) & $-$13.39 & $-$14.63 & $-$15.98 & $-$16.44 & $-$16.41 & $-$16.73 & $-$16.78 & $-$16.82 \\
8.73 & $-$12.31 & 7.29\,(41) & 0.26 & \multicolumn{1}{c}{$\cdots$} & 7.29\,(41) & $-$13.62 & $-$14.85 & $-$16.21 & $-$16.65 & $-$16.63 & $-$16.93 & $-$16.96 & $-$17.00 \\
9.60 & $-$11.44 & 8.75\,(41) & 0.28 & \multicolumn{1}{c}{$\cdots$} & 8.75\,(41) & $-$13.86 & $-$15.08 & $-$16.43 & $-$16.86 & $-$16.84 & $-$17.09 & $-$17.11 & $-$17.14 \\
10.56 & $-$10.48 & 1.07\,(42) & 0.30 & \multicolumn{1}{c}{$\cdots$} & 1.07\,(42) & $-$14.12 & $-$15.33 & $-$16.66 & $-$17.08 & $-$17.07 & $-$17.29 & $-$17.29 & $-$17.30 \\
11.62 & $-$9.42 & 1.31\,(42) & 0.31 & \multicolumn{1}{c}{$\cdots$} & 1.31\,(42) & $-$14.40 & $-$15.59 & $-$16.90 & $-$17.30 & $-$17.29 & $-$17.47 & $-$17.47 & $-$17.47 \\
12.78 & $-$8.26 & 1.58\,(42) & 0.34 & \multicolumn{1}{c}{$\cdots$} & 1.58\,(42) & $-$14.68 & $-$15.85 & $-$17.12 & $-$17.51 & $-$17.49 & $-$17.63 & $-$17.62 & $-$17.59 \\
14.06 & $-$6.98 & 1.89\,(42) & 0.35 & 1.25\,(40) & 1.90\,(42) & $-$14.96 & $-$16.09 & $-$17.34 & $-$17.71 & $-$17.69 & $-$17.80 & $-$17.79 & $-$17.73 \\
15.47 & $-$5.57 & 2.18\,(42) & 0.38 & 1.48\,(40) & 2.19\,(42) & $-$15.20 & $-$16.29 & $-$17.51 & $-$17.87 & $-$17.85 & $-$17.91 & $-$17.90 & $-$17.80 \\
17.02 & $-$4.02 & 2.40\,(42) & 0.39 & 1.88\,(40) & 2.41\,(42) & $-$15.27 & $-$16.38 & $-$17.63 & $-$17.99 & $-$17.99 & $-$17.98 & $-$17.97 & $-$17.82 \\
18.72 & $-$2.32 & 2.54\,(42) & 0.42 & 2.01\,(40) & 2.56\,(42) & $-$15.30 & $-$16.43 & $-$17.72 & $-$18.06 & $-$18.08 & $-$18.00 & $-$18.04 & $-$17.89 \\
20.59 & $-$0.45 & 2.61\,(42) & 0.42 & 3.06\,(40) & 2.64\,(42) & $-$15.29 & $-$16.44 & $-$17.76 & $-$18.09 & $-$18.15 & $-$17.98 & $-$18.08 & $-$17.90 \\
22.65 & +1.61 & 2.59\,(42) & 0.42 & 4.35\,(40) & 2.63\,(42) & $-$15.22 & $-$16.37 & $-$17.74 & $-$18.09 & $-$18.20 & $-$17.92 & $-$18.10 & $-$17.91 \\
24.91 & +3.87 & 2.46\,(42) & 0.40 & 6.42\,(40) & 2.53\,(42) & $-$15.10 & $-$16.25 & $-$17.67 & $-$18.04 & $-$18.22 & $-$17.78 & $-$18.10 & $-$17.91 \\
27.40 & +6.36 & 2.24\,(42) & 0.38 & 8.41\,(40) & 2.33\,(42) & $-$14.94 & $-$16.07 & $-$17.55 & $-$17.93 & $-$18.19 & $-$17.55 & $-$18.05 & $-$17.87 \\
30.14 & +9.10 & 2.00\,(42) & 0.30 & 1.16\,(41) & 2.11\,(42) & $-$14.77 & $-$15.84 & $-$17.34 & $-$17.77 & $-$18.13 & $-$17.42 & $-$18.01 & $-$17.86 \\
33.15 & +12.11 & 1.74\,(42) & 0.18 & 1.51\,(41) & 1.90\,(42) & $-$14.59 & $-$15.56 & $-$17.07 & $-$17.56 & $-$18.00 & $-$17.48 & $-$17.97 & $-$17.88 \\
36.46 & +15.42 & 1.44\,(42) & 0.09 & 1.92\,(41) & 1.63\,(42) & $-$14.42 & $-$15.30 & $-$16.78 & $-$17.30 & $-$17.79 & $-$17.43 & $-$17.80 & $-$17.73 \\
40.11 & +19.07 & 1.18\,(42) & 0.03 & 2.37\,(41) & 1.42\,(42) & $-$14.30 & $-$15.09 & $-$16.50 & $-$17.04 & $-$17.59 & $-$17.25 & $-$17.61 & $-$17.52 \\
44.12 & +23.08 & 9.75\,(41) & 0.03 & 2.83\,(41) & 1.26\,(42) & $-$14.31 & $-$14.93 & $-$16.29 & $-$16.82 & $-$17.41 & $-$16.89 & $-$17.38 & $-$17.17 \\
48.53 & +27.49 & 8.14\,(41) & 0.05 & 3.28\,(41) & 1.14\,(42) & $-$14.31 & $-$14.82 & $-$16.12 & $-$16.63 & $-$17.21 & $-$16.47 & $-$17.12 & $-$16.80 \\
53.38 & +32.34 & 6.93\,(41) & 0.09 & 3.73\,(41) & 1.07\,(42) & $-$14.31 & $-$14.76 & $-$15.99 & $-$16.45 & $-$17.01 & $-$16.10 & $-$16.88 & $-$16.51 \\
58.72 & +37.68 & 5.92\,(41) & 0.14 & 4.14\,(41) & 1.01\,(42) & $-$14.28 & $-$14.71 & $-$15.87 & $-$16.29 & $-$16.80 & $-$15.72 & $-$16.65 & $-$16.25 \\
64.59 & +43.55 & 5.03\,(41) & 0.19 & 4.47\,(41) & 9.51\,(41) & $-$14.22 & $-$14.65 & $-$15.74 & $-$16.13 & $-$16.60 & $-$15.35 & $-$16.42 & $-$16.00 \\
71.05 & +50.01 & 4.24\,(41) & 0.23 & 4.75\,(41) & 8.99\,(41) & $-$14.13 & $-$14.59 & $-$15.59 & $-$15.95 & $-$16.39 & $-$14.96 & $-$16.18 & $-$15.73 \\
78.16 & +57.12 & 3.53\,(41) & 0.27 & 4.91\,(41) & 8.44\,(41) & $-$14.02 & $-$14.51 & $-$15.43 & $-$15.76 & $-$16.17 & $-$14.57 & $-$15.90 & $-$15.45 \\
86.00 & +64.96 & 2.89\,(41) & 0.29 & 5.01\,(41) & 7.90\,(41) & $-$13.86 & $-$14.40 & $-$15.24 & $-$15.54 & $-$15.95 & $-$14.20 & $-$15.59 & $-$15.13 \\
94.60 & +73.56 & 2.34\,(41) & 0.30 & 4.98\,(41) & 7.32\,(41) & $-$13.66 & $-$14.28 & $-$15.02 & $-$15.31 & $-$15.72 & $-$13.87 & $-$15.24 & $-$14.77 \\
104.10 & +83.06 & 1.87\,(41) & 0.31 & 4.86\,(41) & 6.72\,(41) & $-$13.43 & $-$14.12 & $-$14.78 & $-$15.02 & $-$15.48 & $-$13.62 & $-$14.86 & $-$14.37 \\
114.50 & +93.46 & 1.47\,(41) & 0.29 & 4.68\,(41) & 6.15\,(41) & $-$13.16 & $-$13.94 & $-$14.51 & $-$14.72 & $-$15.23 & $-$13.44 & $-$14.47 & $-$13.96 \\
126.00 & +104.96 & 1.14\,(41) & 0.27 & 4.42\,(41) & 5.56\,(41) & $-$12.85 & $-$13.72 & $-$14.21 & $-$14.38 & $-$14.97 & $-$13.32 & $-$14.10 & $-$13.55 \\
138.60 & +117.56 & 8.76\,(40) & 0.23 & 4.09\,(41) & 4.97\,(41) & $-$12.50 & $-$13.46 & $-$13.88 & $-$14.02 & $-$14.71 & $-$13.22 & $-$13.79 & $-$13.15 \\
152.50 & +131.46 & 6.65\,(40) & 0.18 & 3.74\,(41) & 4.40\,(41) & $-$12.11 & $-$13.16 & $-$13.53 & $-$13.65 & $-$14.45 & $-$13.09 & $-$13.55 & $-$12.79 \\
167.80 & +146.76 & 5.00\,(40) & 0.11 & 3.36\,(41) & 3.86\,(41) & $-$11.69 & $-$12.83 & $-$13.15 & $-$13.30 & $-$14.19 & $-$12.93 & $-$13.35 & $-$12.46 \\
184.60 & +163.56 & 3.71\,(40) & 0.03 & 2.96\,(41) & 3.33\,(41) & $-$11.22 & $-$12.44 & $-$12.75 & $-$12.97 & $-$13.94 & $-$12.73 & $-$13.18 & $-$12.15 \\
203.10 & +182.06 & 2.72\,(40) & $-$0.06 & 2.57\,(41) & 2.84\,(41) & $-$10.73 & $-$12.03 & $-$12.32 & $-$12.65 & $-$13.68 & $-$12.50 & $-$13.02 & $-$11.83 \\
223.40 & +202.36 & 1.97\,(40) & $-$0.16 & 2.18\,(41) & 2.38\,(41) & $-$10.22 & $-$11.58 & $-$11.87 & $-$12.32 & $-$13.40 & $-$12.26 & $-$12.87 & $-$11.50 \\
\hline
\end{tabular}
\flushleft
{\bf Notes:}
Numbers in parenthesis correspond to powers of ten.
$\Delta t_\mathrm{uvoir}$ corresponds to the age in days from maximum UV-optical-IR luminosity ($L_\mathrm{uvoir}$).
%$L_{B\rightarrow I}$ is the integrated $BVRI$ luminosity.
$V \rightarrow \mathrm{uvoir}$ is the conversion constant between the
(absolute) $V$-band magnitude and the uvoir magnitude, such that:
$M_\mathrm{uvoir} = M_V + (V \rightarrow \mathrm{uvoir})$.
$L_\gamma$ is the $\gamma$-ray luminosity; a ``$\cdots$'' entry means
that the energy from radioactive decays was assumed to be deposited
locally at this time. Small fluctuations in $L_\gamma$ are
possible at early times and result from low photon statistics in the
Monte Carlo $\gamma$-ray transport calculation.
The true bolometric luminosity is: $L_\mathrm{bol} = L_\mathrm{uvoir} + L_{\gamma}$.
$UBVRI$ magnitudes are based on the passbands of \cite{Bessell:1990}.
$JHK_s$ magnitudes are in the 2MASS system \citep{Cohen/etal:2003}.

\end{table*}

\begin{table*}
\footnotesize
\caption{Light curves for model SCH2p0. See Table~\ref{tab:ddc25_mags} for notes.}\label{tab:sch2p0_mags}
\begin{tabular}{r@{\hspace{3.3mm}}r@{\hspace{3.3mm}}r@{\hspace{0.0mm}}r@{\hspace{3.3mm}}r@{\hspace{3.3mm}}r@{\hspace{3.3mm}}r@{\hspace{3.3mm}}r@{\hspace{3.3mm}}r@{\hspace{3.3mm}}r@{\hspace{3.3mm}}r@{\hspace{3.3mm}}r@{\hspace{3.3mm}}r@{\hspace{3.3mm}}r}
\hline
\multicolumn{1}{c}{$t_\mathrm{exp}$} & \multicolumn{1}{c}{$\Delta t_\mathrm{uvoir}$} & \multicolumn{1}{c}{$L_\mathrm{uvoir}$} & \multicolumn{1}{c}{$V \rightarrow \mathrm{uvoir}$} & \multicolumn{1}{c}{$L_{\gamma}$} & \multicolumn{1}{c}{$L_\mathrm{bol}$} & \multicolumn{1}{c}{$U$} & \multicolumn{1}{c}{$B$} & \multicolumn{1}{c}{$V$} & \multicolumn{1}{c}{$R$} & \multicolumn{1}{c}{$I$} & \multicolumn{1}{c}{$J$} & \multicolumn{1}{c}{$H$} & \multicolumn{1}{c}{$K_s$} \\
\multicolumn{1}{c}{(days)} & \multicolumn{1}{c}{(days)} & \multicolumn{1}{c}{(erg s$^{-1}$)} & \multicolumn{1}{c}{(mag)} & \multicolumn{1}{c}{(erg s$^{-1}$)} & \multicolumn{1}{c}{(erg s$^{-1}$)} & \multicolumn{1}{c}{(mag)} & \multicolumn{1}{c}{(mag)} & \multicolumn{1}{c}{(mag)} & \multicolumn{1}{c}{(mag)} & \multicolumn{1}{c}{(mag)} & \multicolumn{1}{c}{(mag)} & \multicolumn{1}{c}{(mag)} & \multicolumn{1}{c}{(mag)} \\
\hline
0.91 & $-$15.03 & 3.67\,(39) & $-$0.90 & \multicolumn{1}{c}{$\cdots$} & 3.67\,(39) & $-$2.95 & $-$6.70 & $-$9.30 & $-$10.71 & $-$10.75 & $-$11.90 & $-$11.93 & $-$12.21 \\
1.00 & $-$14.94 & 5.24\,(39) & $-$0.63 & \multicolumn{1}{c}{$\cdots$} & 5.24\,(39) & $-$3.69 & $-$7.48 & $-$9.96 & $-$11.22 & $-$11.19 & $-$12.21 & $-$12.23 & $-$12.47 \\
1.10 & $-$14.84 & 7.44\,(39) & $-$0.41 & \multicolumn{1}{c}{$\cdots$} & 7.44\,(39) & $-$4.43 & $-$8.15 & $-$10.56 & $-$11.68 & $-$11.62 & $-$12.49 & $-$12.55 & $-$12.72 \\
1.21 & $-$14.73 & 1.04\,(40) & $-$0.25 & \multicolumn{1}{c}{$\cdots$} & 1.04\,(40) & $-$5.24 & $-$8.75 & $-$11.08 & $-$12.09 & $-$12.00 & $-$12.77 & $-$12.85 & $-$12.98 \\
1.33 & $-$14.61 & 1.43\,(40) & $-$0.14 & \multicolumn{1}{c}{$\cdots$} & 1.43\,(40) & $-$6.24 & $-$9.34 & $-$11.54 & $-$12.45 & $-$12.35 & $-$13.05 & $-$13.14 & $-$13.24 \\
1.46 & $-$14.48 & 1.94\,(40) & $-$0.04 & \multicolumn{1}{c}{$\cdots$} & 1.94\,(40) & $-$7.41 & $-$9.96 & $-$11.97 & $-$12.79 & $-$12.68 & $-$13.31 & $-$13.42 & $-$13.50 \\
1.61 & $-$14.33 & 2.65\,(40) & 0.04 & \multicolumn{1}{c}{$\cdots$} & 2.65\,(40) & $-$8.58 & $-$10.61 & $-$12.39 & $-$13.12 & $-$13.01 & $-$13.59 & $-$13.70 & $-$13.78 \\
1.77 & $-$14.17 & 3.57\,(40) & 0.09 & \multicolumn{1}{c}{$\cdots$} & 3.57\,(40) & $-$9.47 & $-$11.17 & $-$12.77 & $-$13.43 & $-$13.32 & $-$13.85 & $-$13.97 & $-$14.04 \\
1.95 & $-$13.99 & 4.78\,(40) & 0.13 & \multicolumn{1}{c}{$\cdots$} & 4.78\,(40) & $-$10.20 & $-$11.68 & $-$13.12 & $-$13.74 & $-$13.61 & $-$14.12 & $-$14.23 & $-$14.30 \\
2.15 & $-$13.79 & 6.35\,(40) & 0.17 & \multicolumn{1}{c}{$\cdots$} & 6.35\,(40) & $-$10.82 & $-$12.13 & $-$13.47 & $-$14.03 & $-$13.89 & $-$14.38 & $-$14.48 & $-$14.56 \\
2.37 & $-$13.57 & 8.36\,(40) & 0.20 & \multicolumn{1}{c}{$\cdots$} & 8.36\,(40) & $-$11.37 & $-$12.56 & $-$13.80 & $-$14.31 & $-$14.17 & $-$14.64 & $-$14.73 & $-$14.81 \\
2.61 & $-$13.33 & 1.09\,(41) & 0.24 & \multicolumn{1}{c}{$\cdots$} & 1.09\,(41) & $-$11.86 & $-$12.95 & $-$14.12 & $-$14.58 & $-$14.43 & $-$14.88 & $-$14.97 & $-$15.05 \\
2.87 & $-$13.07 & 1.41\,(41) & 0.27 & \multicolumn{1}{c}{$\cdots$} & 1.41\,(41) & $-$12.32 & $-$13.32 & $-$14.43 & $-$14.83 & $-$14.68 & $-$15.11 & $-$15.20 & $-$15.28 \\
3.16 & $-$12.78 & 1.81\,(41) & 0.29 & \multicolumn{1}{c}{$\cdots$} & 1.81\,(41) & $-$12.76 & $-$13.68 & $-$14.72 & $-$15.08 & $-$14.94 & $-$15.34 & $-$15.42 & $-$15.50 \\
3.48 & $-$12.46 & 2.30\,(41) & 0.31 & \multicolumn{1}{c}{$\cdots$} & 2.30\,(41) & $-$13.19 & $-$14.02 & $-$15.01 & $-$15.32 & $-$15.19 & $-$15.56 & $-$15.64 & $-$15.72 \\
3.83 & $-$12.11 & 2.90\,(41) & 0.33 & \multicolumn{1}{c}{$\cdots$} & 2.90\,(41) & $-$13.57 & $-$14.34 & $-$15.28 & $-$15.55 & $-$15.43 & $-$15.77 & $-$15.85 & $-$15.93 \\
4.21 & $-$11.73 & 3.63\,(41) & 0.34 & \multicolumn{1}{c}{$\cdots$} & 3.63\,(41) & $-$13.93 & $-$14.64 & $-$15.53 & $-$15.77 & $-$15.66 & $-$15.98 & $-$16.06 & $-$16.13 \\
4.63 & $-$11.30 & 4.51\,(41) & 0.35 & \multicolumn{1}{c}{$\cdots$} & 4.51\,(41) & $-$14.26 & $-$14.93 & $-$15.78 & $-$15.99 & $-$15.88 & $-$16.19 & $-$16.26 & $-$16.33 \\
5.09 & $-$10.85 & 5.57\,(41) & 0.36 & \multicolumn{1}{c}{$\cdots$} & 5.57\,(41) & $-$14.56 & $-$15.21 & $-$16.01 & $-$16.20 & $-$16.10 & $-$16.39 & $-$16.45 & $-$16.52 \\
5.60 & $-$10.34 & 6.87\,(41) & 0.36 & \multicolumn{1}{c}{$\cdots$} & 6.87\,(41) & $-$14.85 & $-$15.48 & $-$16.25 & $-$16.42 & $-$16.31 & $-$16.60 & $-$16.64 & $-$16.72 \\
6.16 & $-$9.78 & 8.43\,(41) & 0.36 & \multicolumn{1}{c}{$\cdots$} & 8.43\,(41) & $-$15.13 & $-$15.74 & $-$16.47 & $-$16.63 & $-$16.53 & $-$16.80 & $-$16.83 & $-$16.90 \\
6.78 & $-$9.16 & 1.03\,(42) & 0.36 & \multicolumn{1}{c}{$\cdots$} & 1.03\,(42) & $-$15.38 & $-$16.00 & $-$16.68 & $-$16.85 & $-$16.74 & $-$16.99 & $-$17.01 & $-$17.08 \\
7.46 & $-$8.48 & 1.25\,(42) & 0.36 & \multicolumn{1}{c}{$\cdots$} & 1.25\,(42) & $-$15.63 & $-$16.24 & $-$16.89 & $-$17.05 & $-$16.94 & $-$17.18 & $-$17.19 & $-$17.25 \\
8.21 & $-$7.72 & 1.50\,(42) & 0.35 & \multicolumn{1}{c}{$\cdots$} & 1.50\,(42) & $-$15.85 & $-$16.47 & $-$17.09 & $-$17.25 & $-$17.14 & $-$17.36 & $-$17.35 & $-$17.40 \\
9.03 & $-$6.91 & 1.79\,(42) & 0.35 & 7.88\,(39) & 1.80\,(42) & $-$16.07 & $-$16.70 & $-$17.28 & $-$17.44 & $-$17.33 & $-$17.52 & $-$17.50 & $-$17.54 \\
9.93 & $-$6.01 & 2.10\,(42) & 0.35 & 2.87\,(39) & 2.11\,(42) & $-$16.26 & $-$16.90 & $-$17.45 & $-$17.62 & $-$17.51 & $-$17.67 & $-$17.63 & $-$17.66 \\
10.92 & $-$5.02 & 2.41\,(42) & 0.35 & 1.44\,(40) & 2.43\,(42) & $-$16.41 & $-$17.06 & $-$17.60 & $-$17.78 & $-$17.67 & $-$17.80 & $-$17.74 & $-$17.74 \\
12.01 & $-$3.93 & 2.70\,(42) & 0.37 & 5.49\,(39) & 2.71\,(42) & $-$16.52 & $-$17.18 & $-$17.74 & $-$17.91 & $-$17.82 & $-$17.89 & $-$17.83 & $-$17.80 \\
13.21 & $-$2.72 & 2.95\,(42) & 0.38 & 4.79\,(40) & 3.00\,(42) & $-$16.59 & $-$17.25 & $-$17.85 & $-$18.03 & $-$17.95 & $-$17.96 & $-$17.90 & $-$17.84 \\
14.53 & $-$1.41 & 3.11\,(42) & 0.41 & 6.04\,(40) & 3.17\,(42) & $-$16.60 & $-$17.27 & $-$17.93 & $-$18.11 & $-$18.07 & $-$17.98 & $-$17.95 & $-$17.86 \\
15.98 & +0.04 & 3.17\,(42) & 0.43 & 8.61\,(40) & 3.25\,(42) & $-$16.54 & $-$17.22 & $-$17.98 & $-$18.16 & $-$18.16 & $-$17.96 & $-$17.99 & $-$17.87 \\
17.58 & +1.64 & 3.10\,(42) & 0.46 & 1.18\,(41) & 3.22\,(42) & $-$16.43 & $-$17.11 & $-$17.98 & $-$18.18 & $-$18.22 & $-$17.87 & $-$18.03 & $-$17.88 \\
19.34 & +3.40 & 2.92\,(42) & 0.47 & 1.57\,(41) & 3.08\,(42) & $-$16.39 & $-$16.90 & $-$17.93 & $-$18.14 & $-$18.23 & $-$17.71 & $-$18.06 & $-$17.90 \\
21.27 & +5.33 & 2.63\,(42) & 0.46 & 2.02\,(41) & 2.83\,(42) & $-$16.16 & $-$16.64 & $-$17.80 & $-$18.04 & $-$18.20 & $-$17.52 & $-$18.10 & $-$17.93 \\
23.40 & +7.46 & 2.32\,(42) & 0.40 & 2.55\,(41) & 2.58\,(42) & $-$15.86 & $-$16.34 & $-$17.61 & $-$17.89 & $-$18.16 & $-$17.42 & $-$18.12 & $-$17.96 \\
25.74 & +9.80 & 2.02\,(42) & 0.29 & 3.15\,(41) & 2.34\,(42) & $-$15.59 & $-$16.03 & $-$17.35 & $-$17.71 & $-$18.10 & $-$17.34 & $-$18.09 & $-$17.94 \\
28.31 & +12.37 & 1.73\,(42) & 0.19 & 3.78\,(41) & 2.10\,(42) & $-$15.36 & $-$15.74 & $-$17.07 & $-$17.48 & $-$17.97 & $-$17.27 & $-$17.96 & $-$17.82 \\
31.14 & +15.21 & 1.42\,(42) & 0.11 & 4.43\,(41) & 1.86\,(42) & $-$15.18 & $-$15.50 & $-$16.79 & $-$17.22 & $-$17.79 & $-$17.15 & $-$17.72 & $-$17.57 \\
34.26 & +18.32 & 1.11\,(42) & 0.11 & 5.01\,(41) & 1.61\,(42) & $-$15.02 & $-$15.28 & $-$16.51 & $-$16.91 & $-$17.55 & $-$16.79 & $-$17.37 & $-$17.17 \\
37.68 & +21.74 & 8.74\,(41) & 0.13 & 5.59\,(41) & 1.43\,(42) & $-$14.90 & $-$15.11 & $-$16.27 & $-$16.63 & $-$17.30 & $-$16.38 & $-$17.02 & $-$16.78 \\
41.45 & +25.52 & 7.05\,(41) & 0.16 & 6.10\,(41) & 1.32\,(42) & $-$14.79 & $-$14.96 & $-$16.07 & $-$16.40 & $-$17.06 & $-$15.96 & $-$16.72 & $-$16.44 \\
45.60 & +29.66 & 5.77\,(41) & 0.19 & 6.51\,(41) & 1.23\,(42) & $-$14.69 & $-$14.83 & $-$15.88 & $-$16.18 & $-$16.81 & $-$15.52 & $-$16.43 & $-$16.13 \\
50.16 & +34.22 & 4.73\,(41) & 0.23 & 6.83\,(41) & 1.16\,(42) & $-$14.59 & $-$14.71 & $-$15.71 & $-$15.97 & $-$16.54 & $-$15.08 & $-$16.16 & $-$15.83 \\
55.18 & +39.24 & 3.89\,(41) & 0.26 & 7.06\,(41) & 1.10\,(42) & $-$14.46 & $-$14.60 & $-$15.53 & $-$15.76 & $-$16.28 & $-$14.63 & $-$15.88 & $-$15.51 \\
60.70 & +44.77 & 3.20\,(41) & 0.29 & 7.16\,(41) & 1.04\,(42) & $-$14.32 & $-$14.48 & $-$15.35 & $-$15.54 & $-$16.01 & $-$14.17 & $-$15.57 & $-$15.18 \\
66.77 & +50.83 & 2.61\,(41) & 0.32 & 7.17\,(41) & 9.78\,(41) & $-$14.15 & $-$14.35 & $-$15.15 & $-$15.31 & $-$15.74 & $-$13.71 & $-$15.24 & $-$14.83 \\
73.45 & +57.52 & 2.12\,(41) & 0.33 & 7.08\,(41) & 9.20\,(41) & $-$13.96 & $-$14.22 & $-$14.93 & $-$15.04 & $-$15.48 & $-$13.25 & $-$14.87 & $-$14.44 \\
80.80 & +64.86 & 1.71\,(41) & 0.33 & 6.89\,(41) & 8.59\,(41) & $-$13.74 & $-$14.07 & $-$14.70 & $-$14.76 & $-$15.21 & $-$12.79 & $-$14.47 & $-$14.04 \\
88.88 & +72.94 & 1.37\,(41) & 0.31 & 6.63\,(41) & 8.00\,(41) & $-$13.48 & $-$13.91 & $-$14.45 & $-$14.47 & $-$14.96 & $-$12.33 & $-$14.05 & $-$13.64 \\
97.77 & +81.83 & 1.09\,(41) & 0.28 & 6.31\,(41) & 7.40\,(41) & $-$13.20 & $-$13.74 & $-$14.17 & $-$14.16 & $-$14.71 & $-$11.88 & $-$13.64 & $-$13.24 \\
107.55 & +91.61 & 8.68\,(40) & 0.22 & 5.93\,(41) & 6.80\,(41) & $-$12.88 & $-$13.53 & $-$13.86 & $-$13.86 & $-$14.47 & $-$11.47 & $-$13.27 & $-$12.86 \\
118.31 & +102.38 & 6.86\,(40) & 0.15 & 5.51\,(41) & 6.20\,(41) & $-$12.54 & $-$13.29 & $-$13.53 & $-$13.56 & $-$14.23 & $-$11.13 & $-$12.94 & $-$12.50 \\
130.14 & +114.20 & 5.40\,(40) & 0.07 & 5.05\,(41) & 5.59\,(41) & $-$12.17 & $-$13.02 & $-$13.19 & $-$13.27 & $-$13.99 & $-$10.86 & $-$12.65 & $-$12.19 \\
143.15 & +127.22 & 4.21\,(40) & $-$0.02 & 4.57\,(41) & 4.99\,(41) & $-$11.78 & $-$12.71 & $-$12.84 & $-$12.99 & $-$13.75 & $-$10.65 & $-$12.40 & $-$11.88 \\
157.47 & +141.53 & 3.28\,(40) & $-$0.10 & 4.08\,(41) & 4.41\,(41) & $-$11.38 & $-$12.37 & $-$12.48 & $-$12.72 & $-$13.51 & $-$10.45 & $-$12.15 & $-$11.59 \\
173.22 & +157.28 & 2.55\,(40) & $-$0.20 & 3.60\,(41) & 3.85\,(41) & $-$10.98 & $-$12.01 & $-$12.11 & $-$12.45 & $-$13.26 & $-$10.29 & $-$11.93 & $-$11.30 \\
190.54 & +174.60 & 1.97\,(40) & $-$0.29 & 3.13\,(41) & 3.32\,(41) & $-$10.59 & $-$11.64 & $-$11.74 & $-$12.17 & $-$13.01 & $-$10.12 & $-$11.69 & $-$10.99 \\
209.59 & +193.66 & 1.50\,(40) & $-$0.39 & 2.67\,(41) & 2.82\,(41) & $-$10.20 & $-$11.23 & $-$11.35 & $-$11.89 & $-$12.74 & $-$9.97 & $-$11.44 & $-$10.66 \\
230.00 & +214.06 & 1.14\,(40) & $-$0.49 & 2.25\,(41) & 2.36\,(41) & $-$9.82 & $-$10.82 & $-$10.95 & $-$11.59 & $-$12.47 & $-$9.82 & $-$11.20 & $-$10.31 \\
\hline
\end{tabular}
\end{table*}

\section{Contribution of individual ions to the total optical and near-infrared
  flux}\label{sect:ladder} 

\begin{figure*}
\centering
\includegraphics{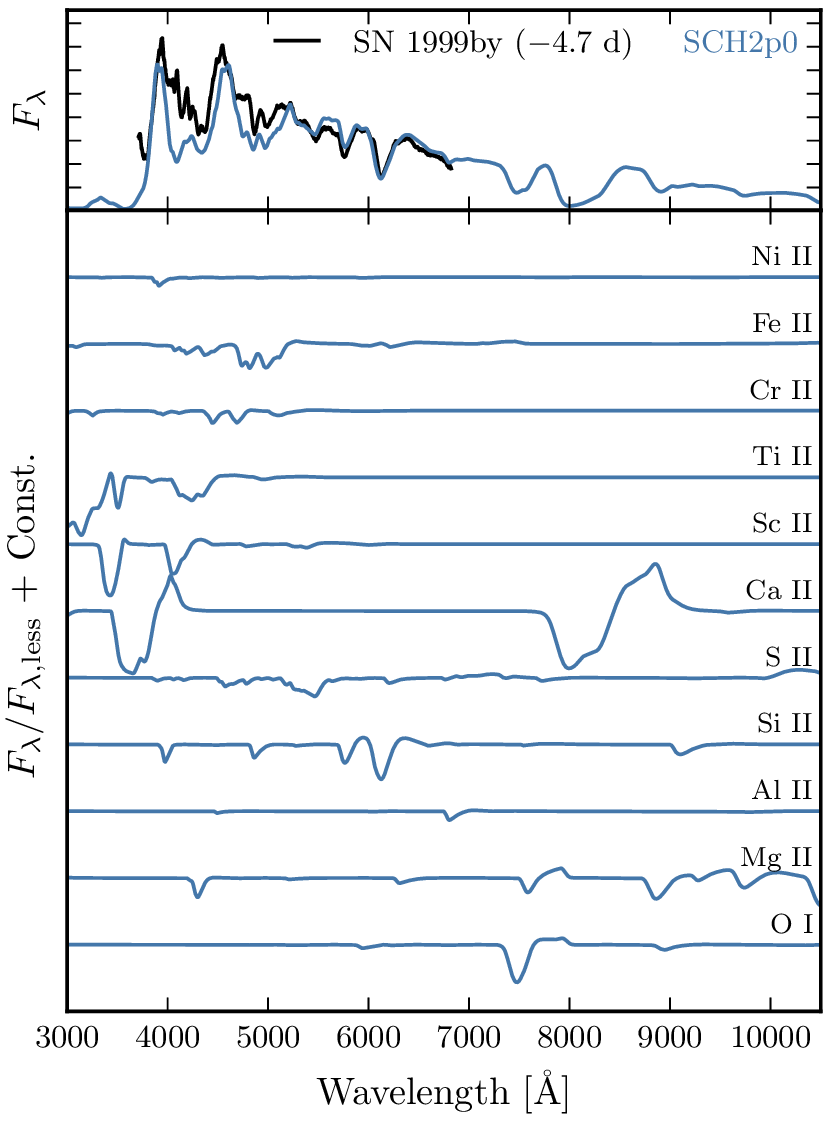}\hspace{.5cm}
\includegraphics{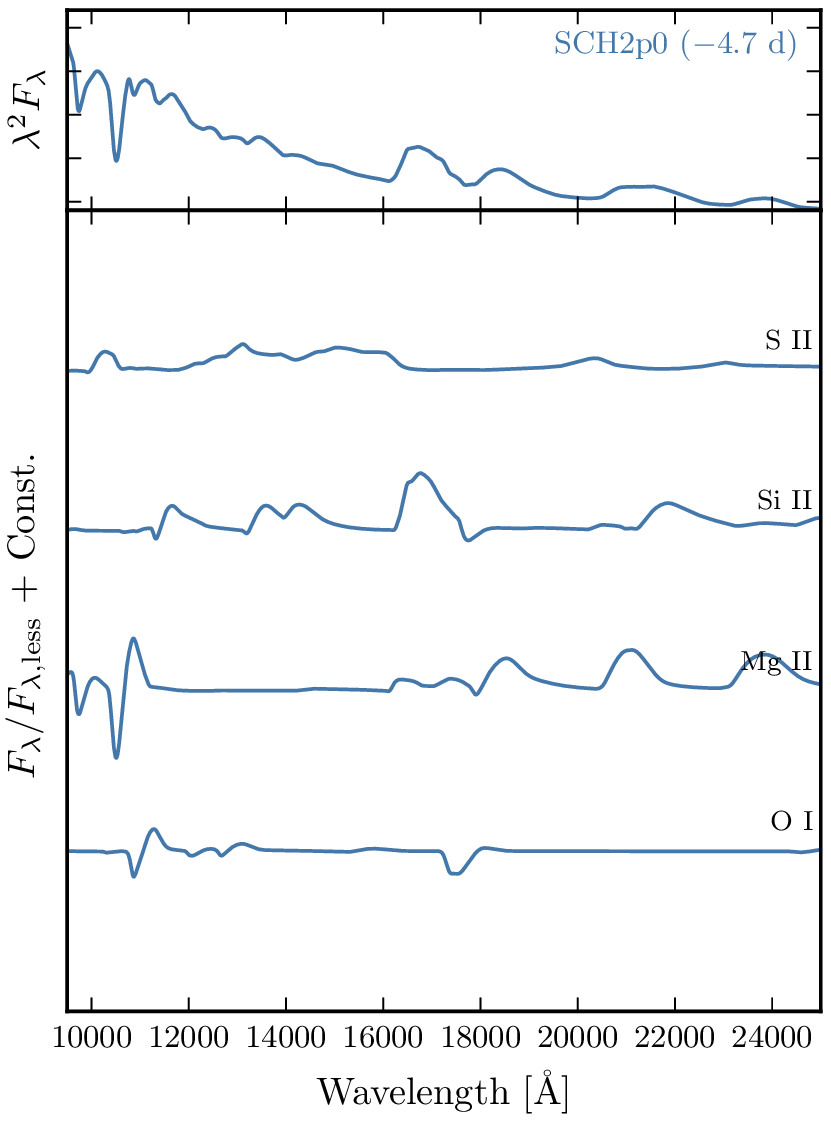}\vspace{.25cm}
\includegraphics{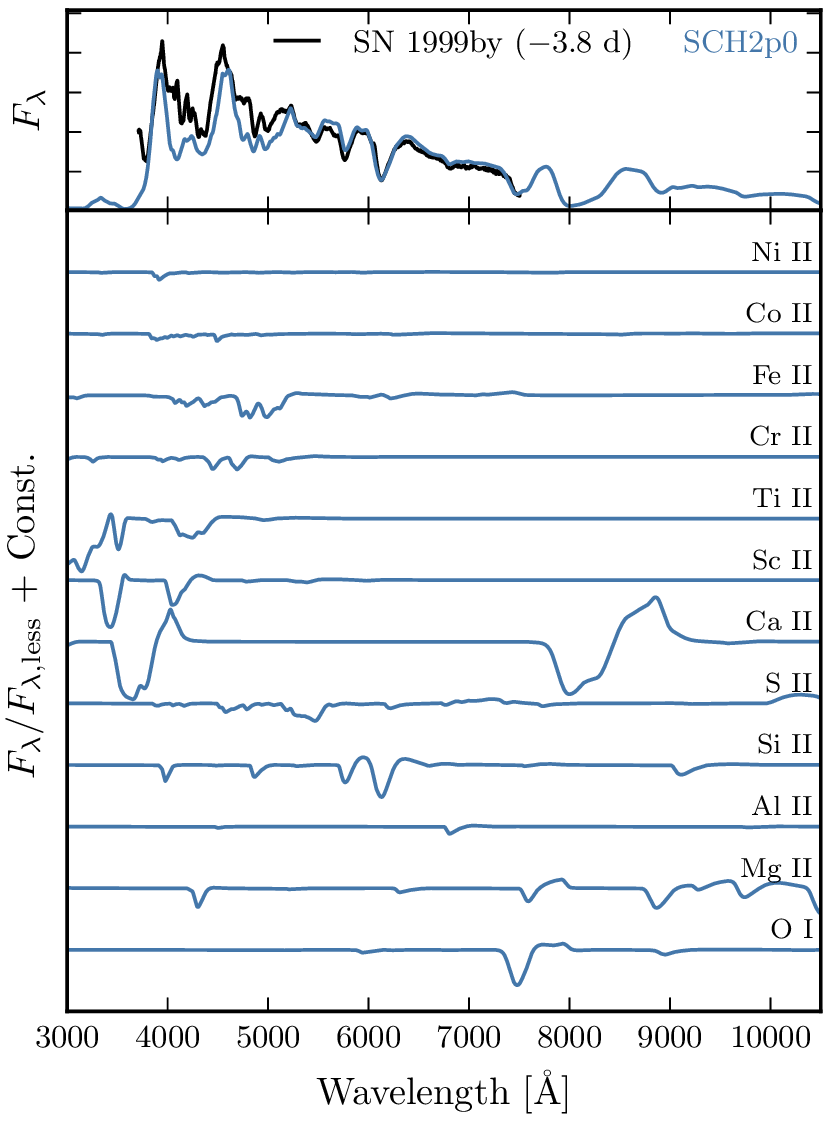}\hspace{.5cm}
\includegraphics{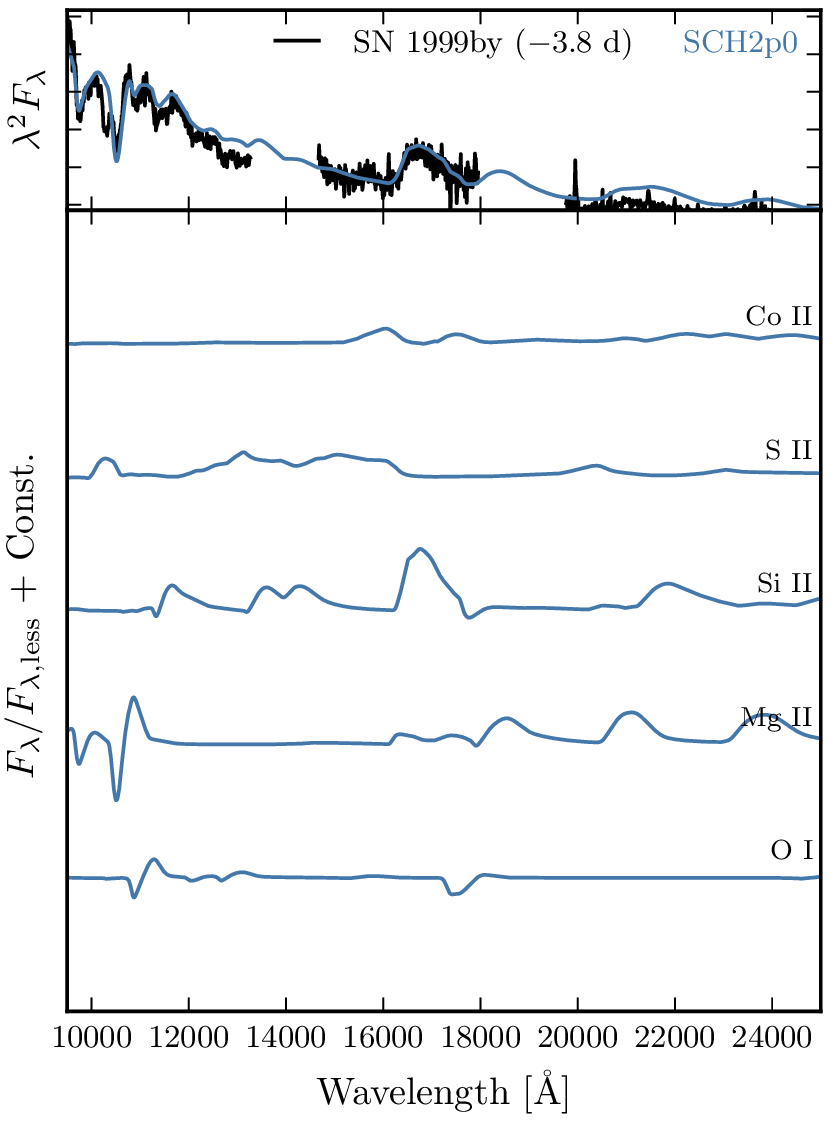}
\caption{\label{fig:ladder_1}
Contribution of individual ions (bottom panels) to the full optical
(left) and NIR (right; the flux has been scaled by $\lambda^2$ for
better visibility) synthetic spectra of the sub-\mch\ model SCH2p0
(blue line), compared to the low-luminosity SN~1999by (black line) at
$-5$\,d (top; there is no NIR spectrum of SN~1999by at this time) and
$-4$\,d (bottom) from $B$-band maximum.
}
\end{figure*}

\begin{figure*}
\centering
\includegraphics{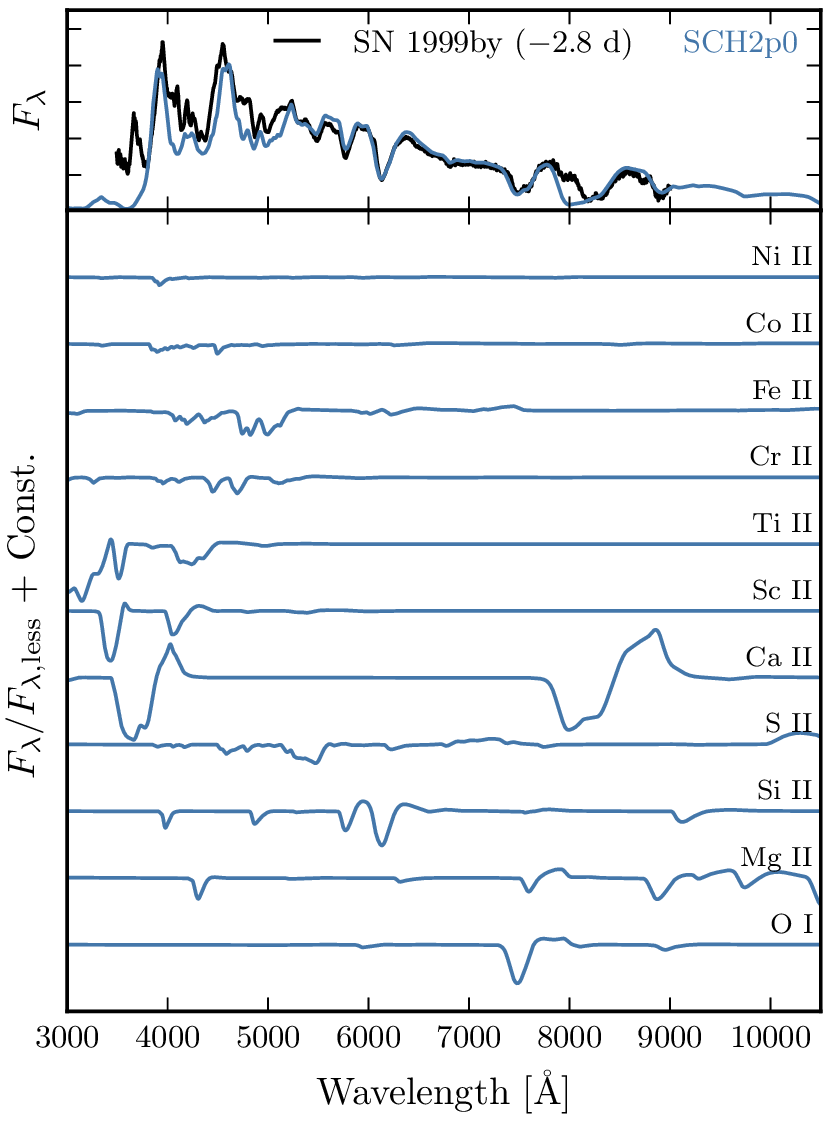}\hspace{.5cm}
\includegraphics{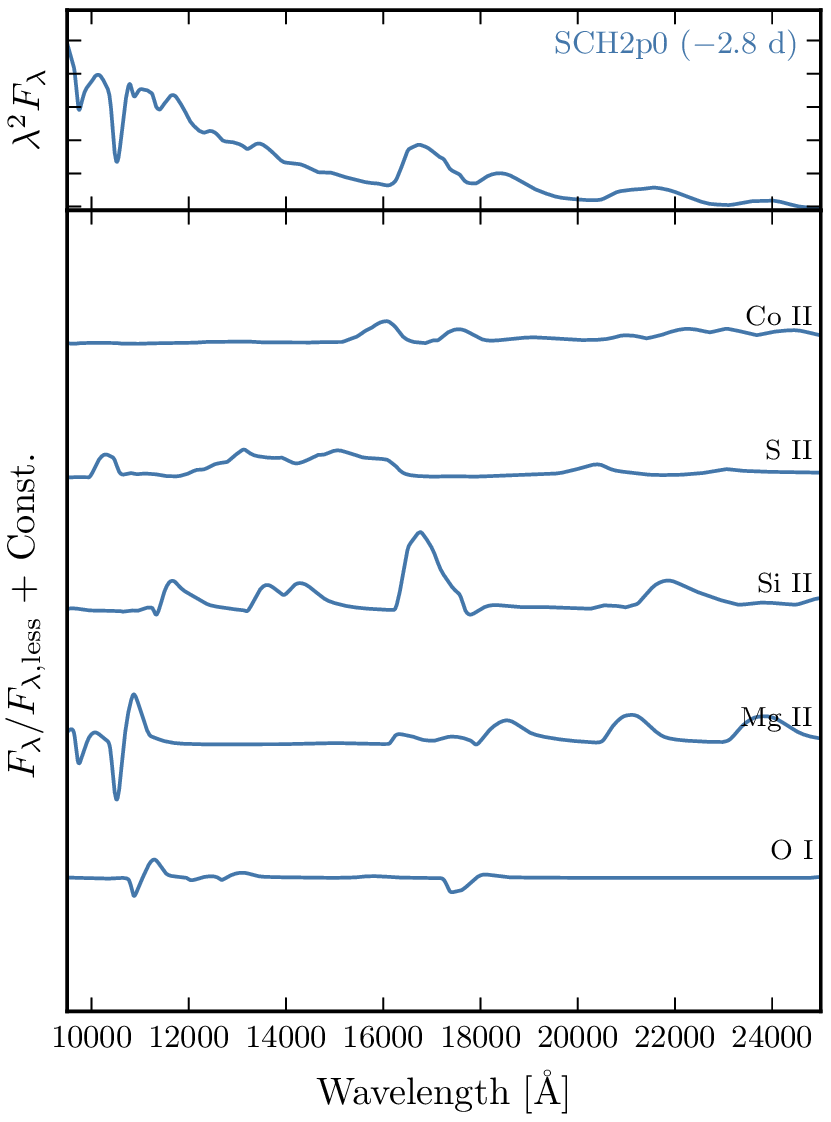}\vspace{.25cm}
\includegraphics{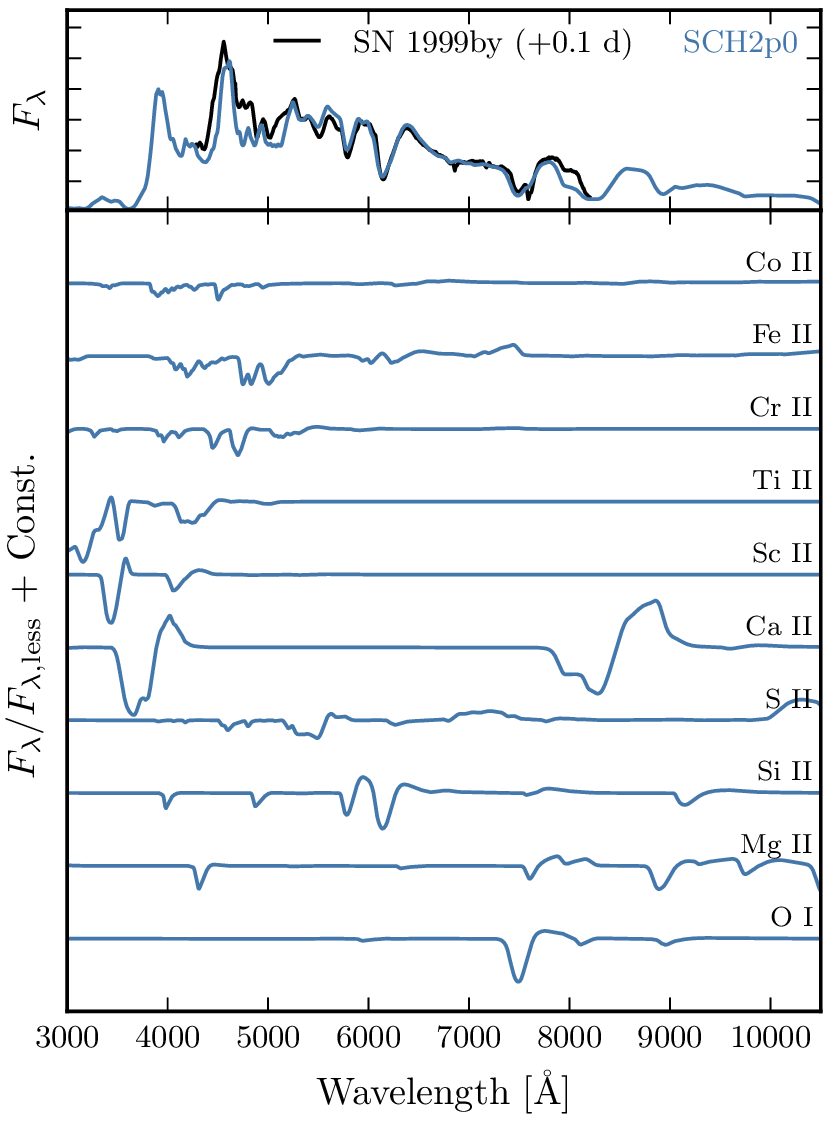}\hspace{.5cm}
\includegraphics{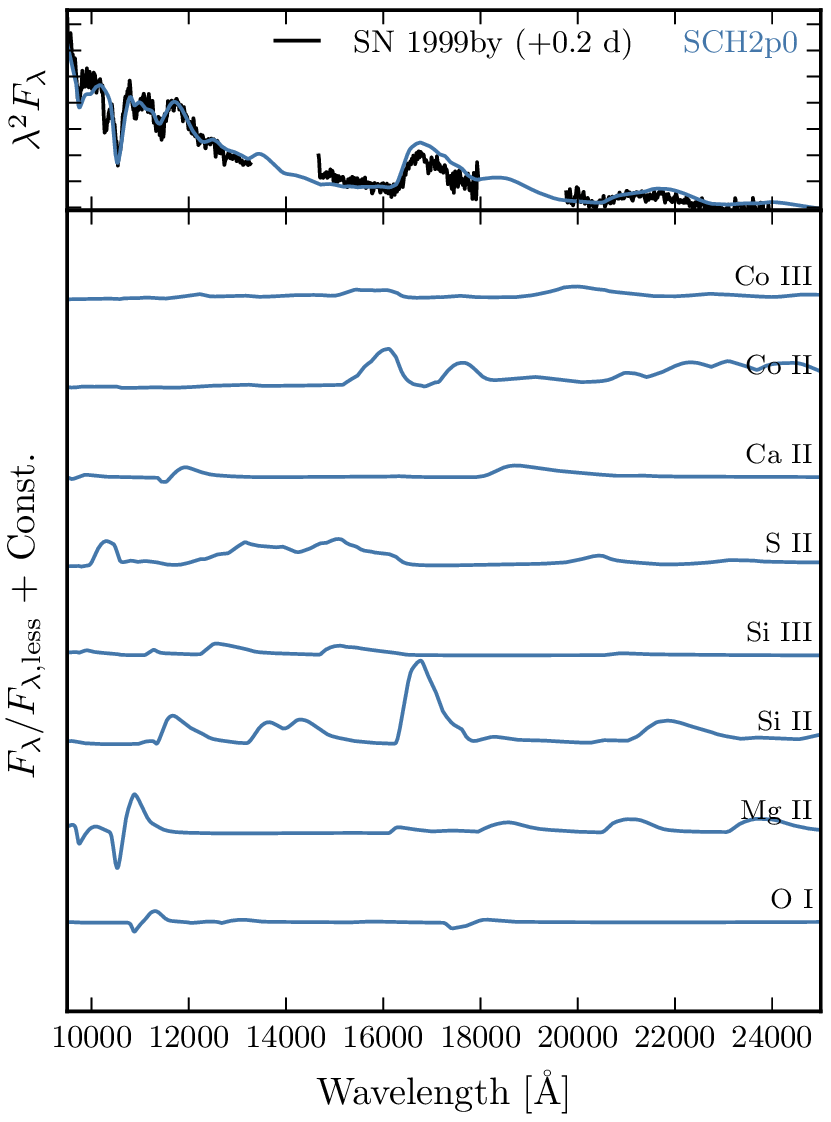}
\caption{\label{fig:ladder_2}
Same as Fig.~\ref{fig:ladder_1} for spectra at $-3$\,d from $B$-band
maximum (top; there is no NIR spectrum of SN~1999by at this time) and
at maximum light (bottom).  }
\end{figure*}

\begin{figure*}
\centering
\includegraphics{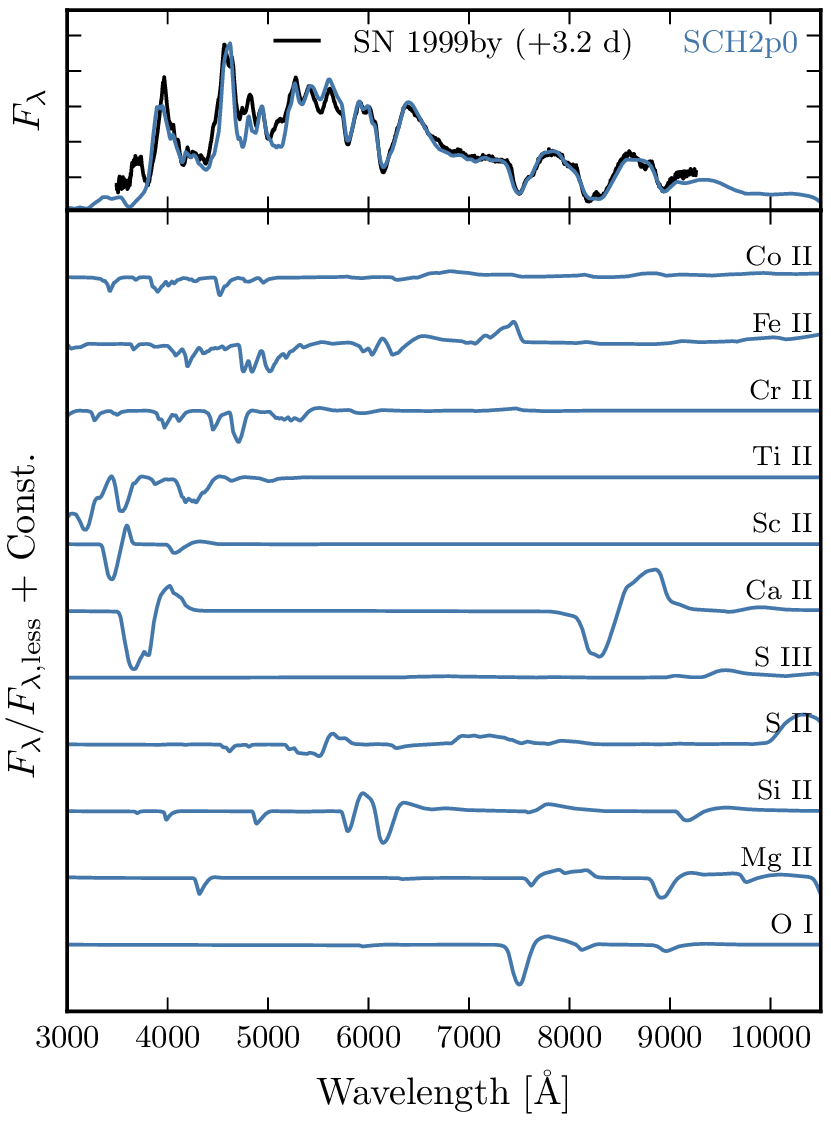}\hspace{.5cm}
\includegraphics{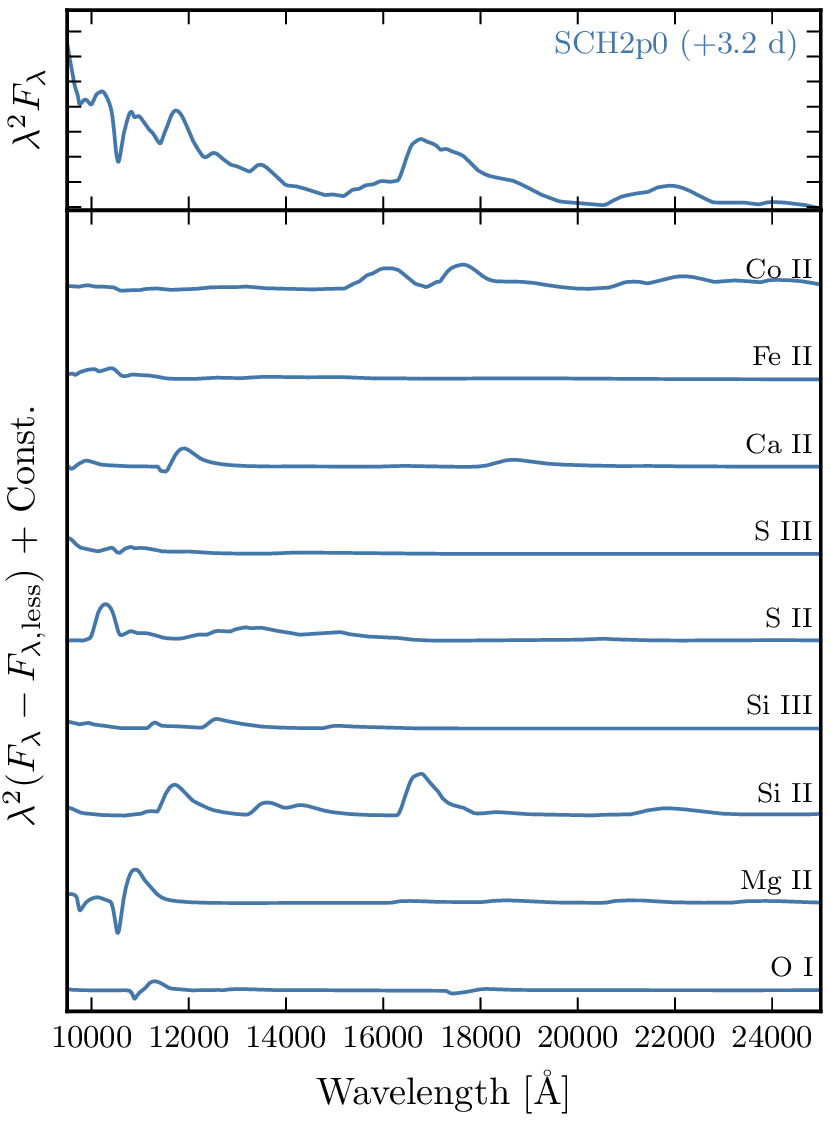}\vspace{.25cm}
\includegraphics{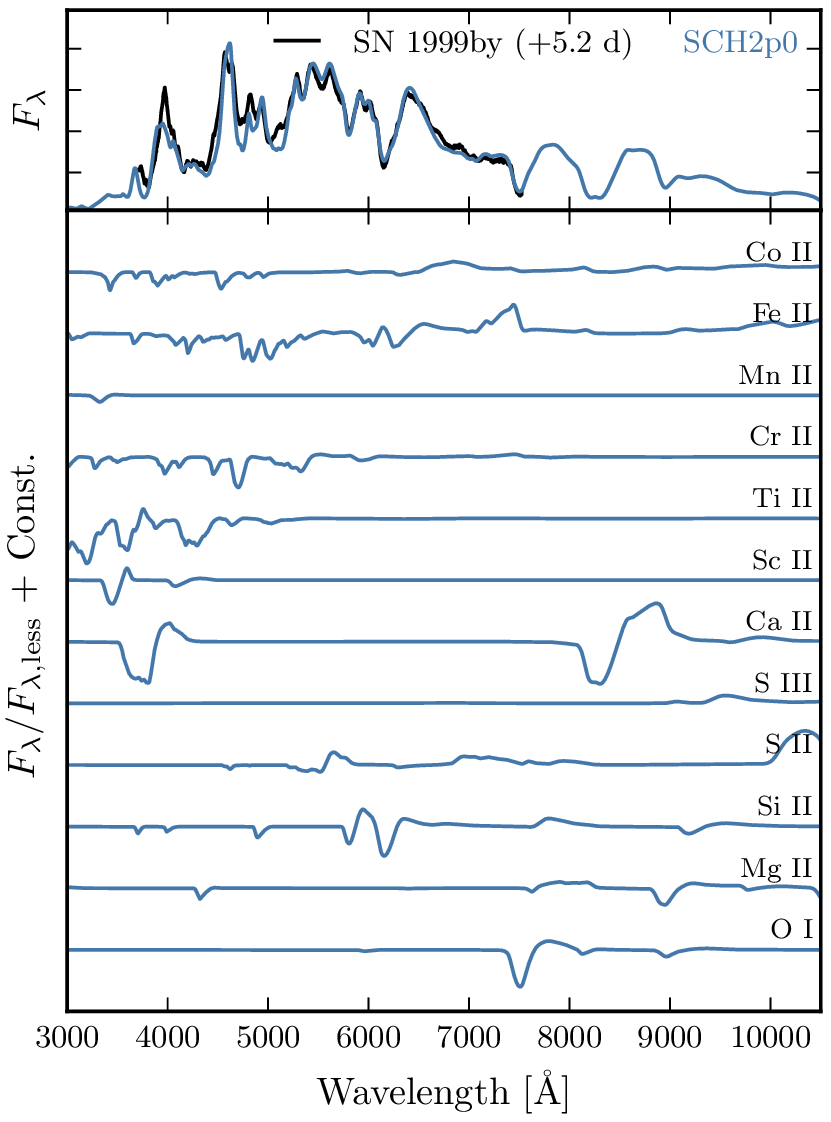}\hspace{.5cm}
\includegraphics{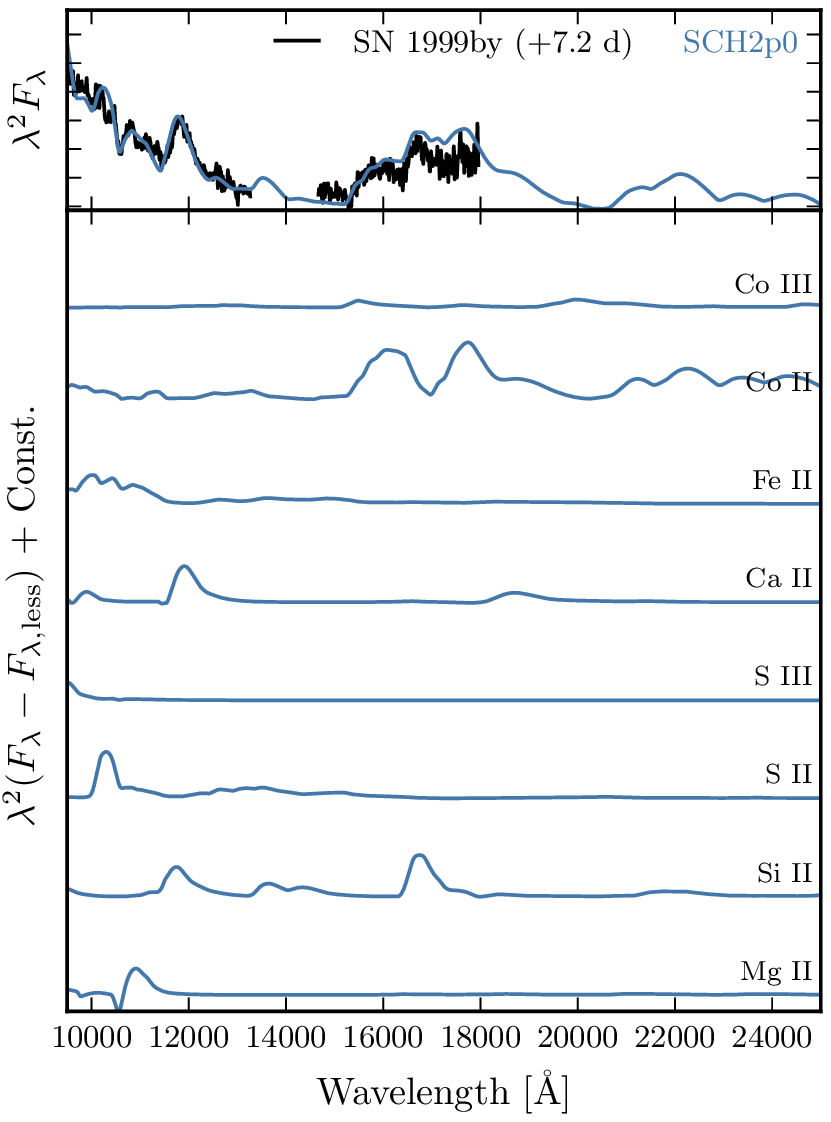}
\caption{\label{fig:ladder_3} 
Same as Fig.~\ref{fig:ladder_1} for spectra at $3$\,d (top; there is
no NIR spectrum of SN~1999by at this time) and around $5$\,d (bottom)
past $B$-band maximum.  }
\end{figure*}

\begin{figure*}
\centering
\includegraphics{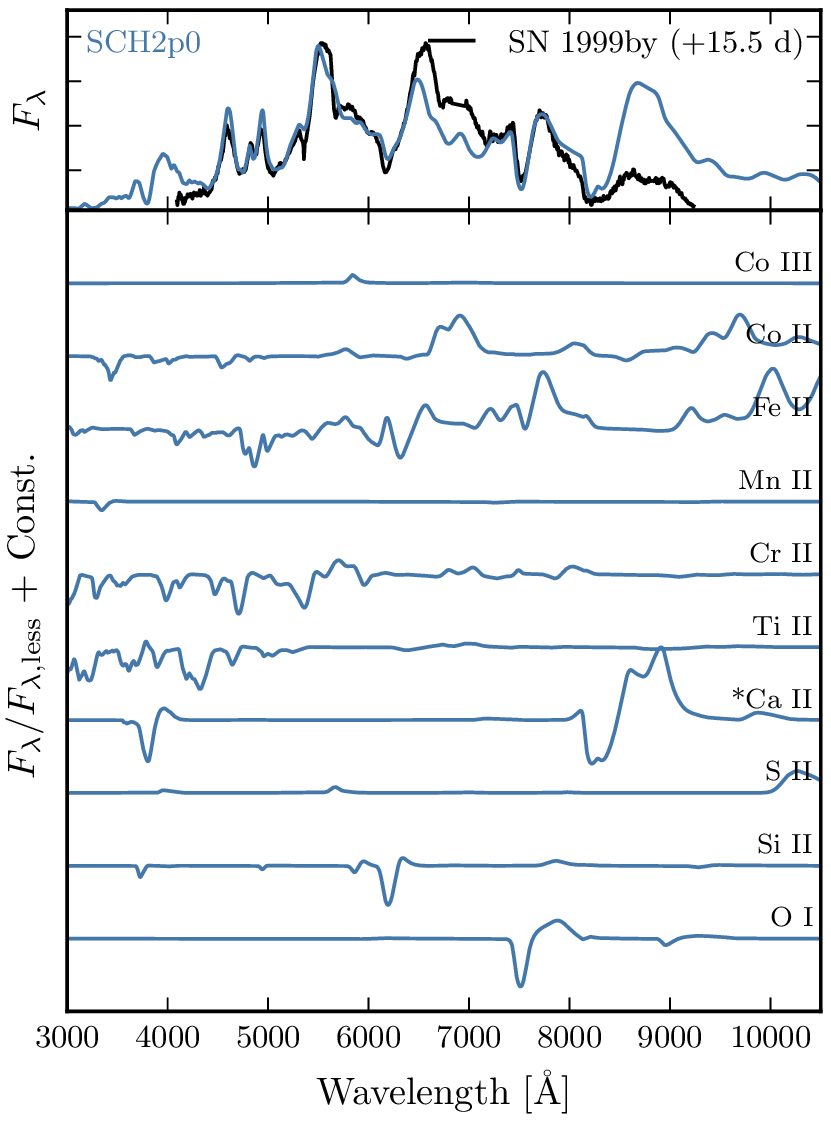}\hspace{.5cm}
\includegraphics{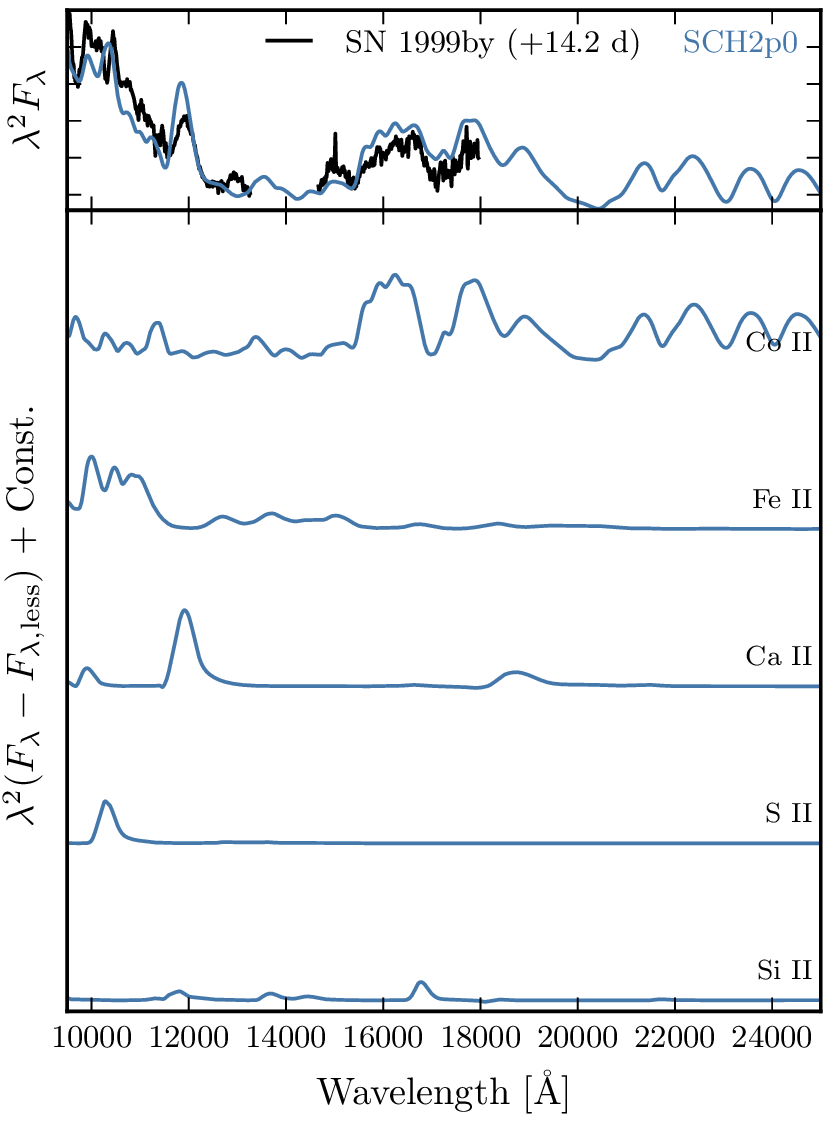}\vspace{.25cm}
\includegraphics{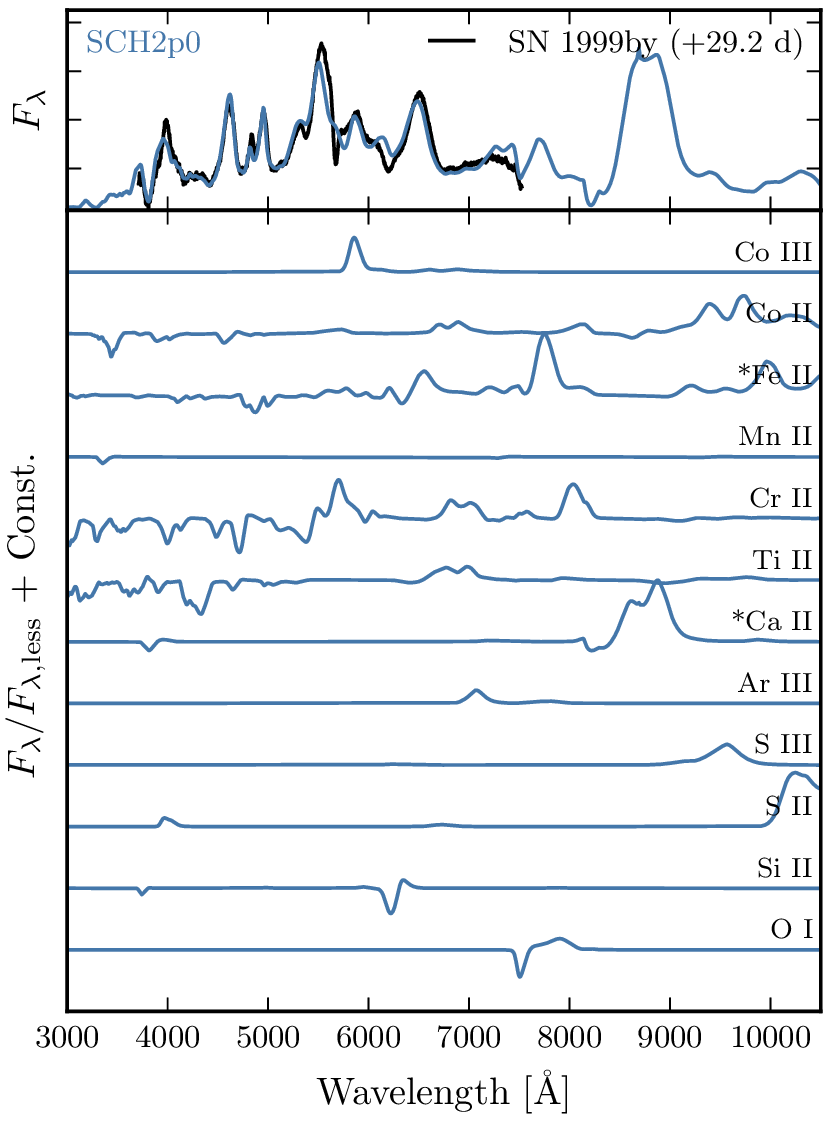}\hspace{.5cm}
\includegraphics{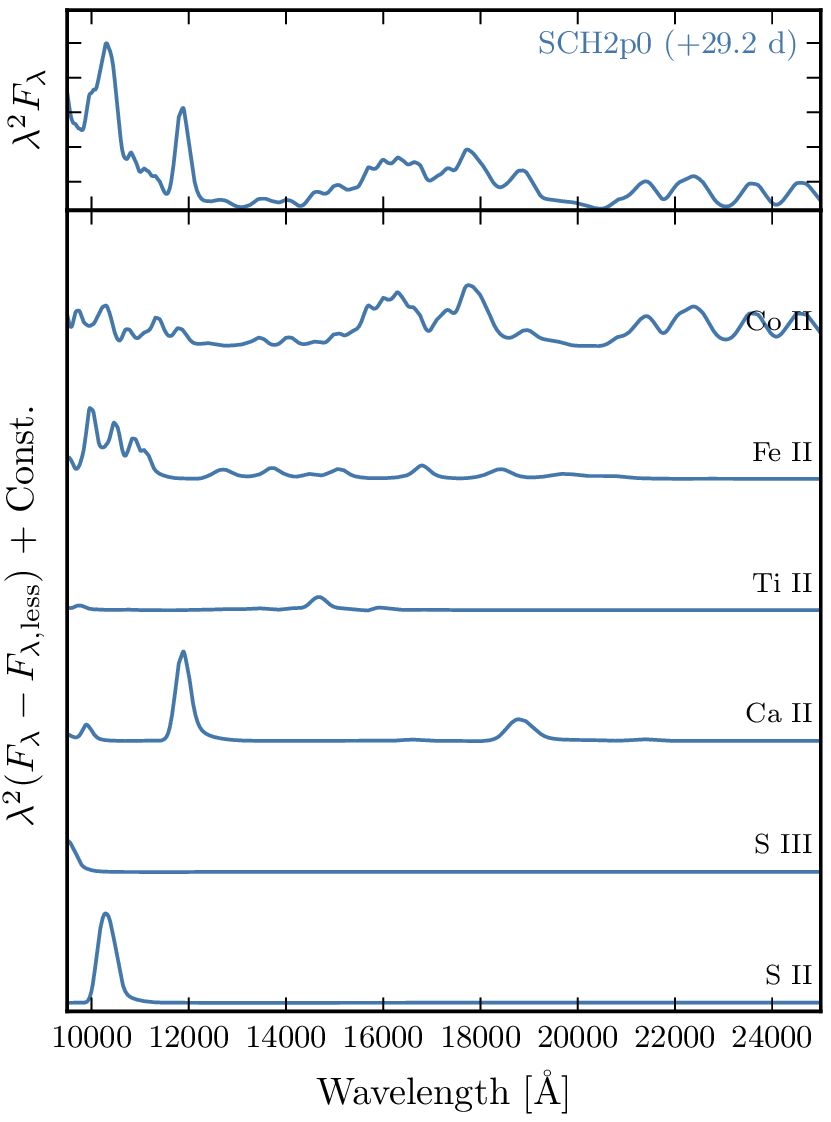}\vspace{.25cm}
\caption{\label{fig:ladder_4}
Same as Fig.~\ref{fig:ladder_1} for spectra around $15$\,d (top) and
at $29$\,d (bottom; there is no NIR spectrum of SN~1999by at this
time) past $B$-band maximum.  }
\end{figure*}

\begin{figure*}
\centering
\includegraphics{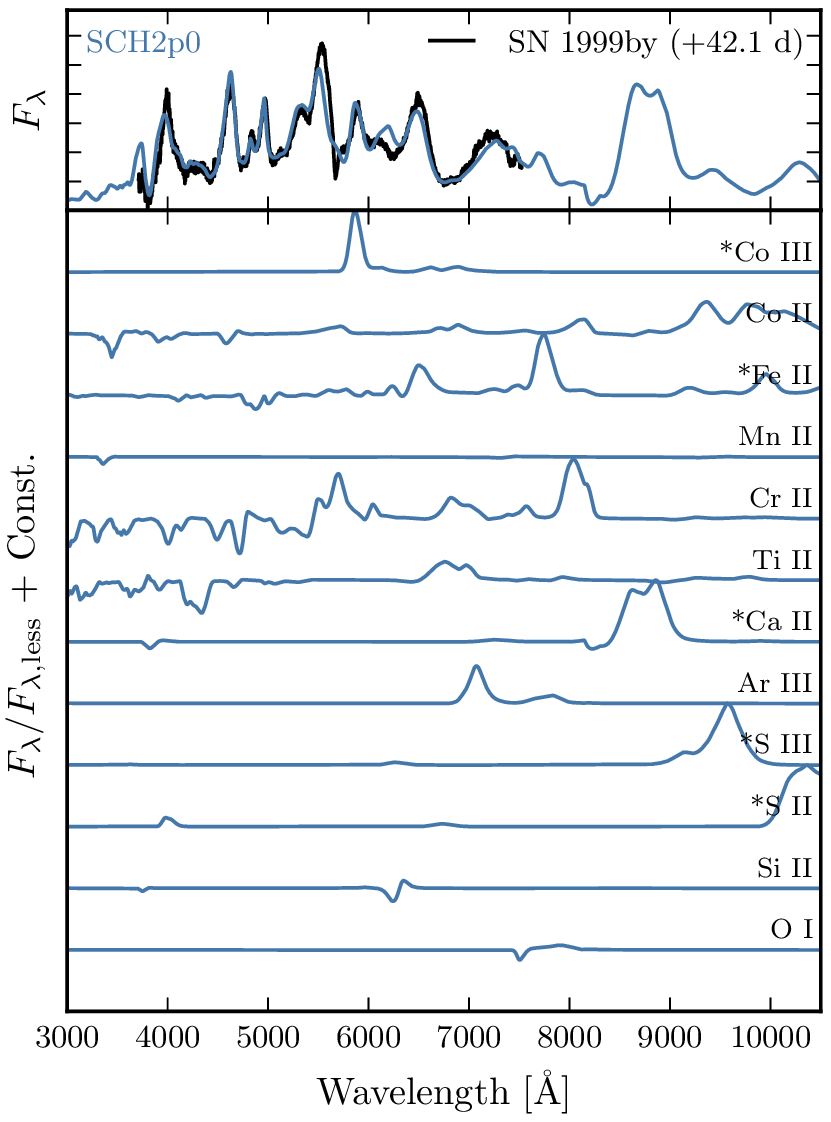}\hspace{.5cm}
\includegraphics{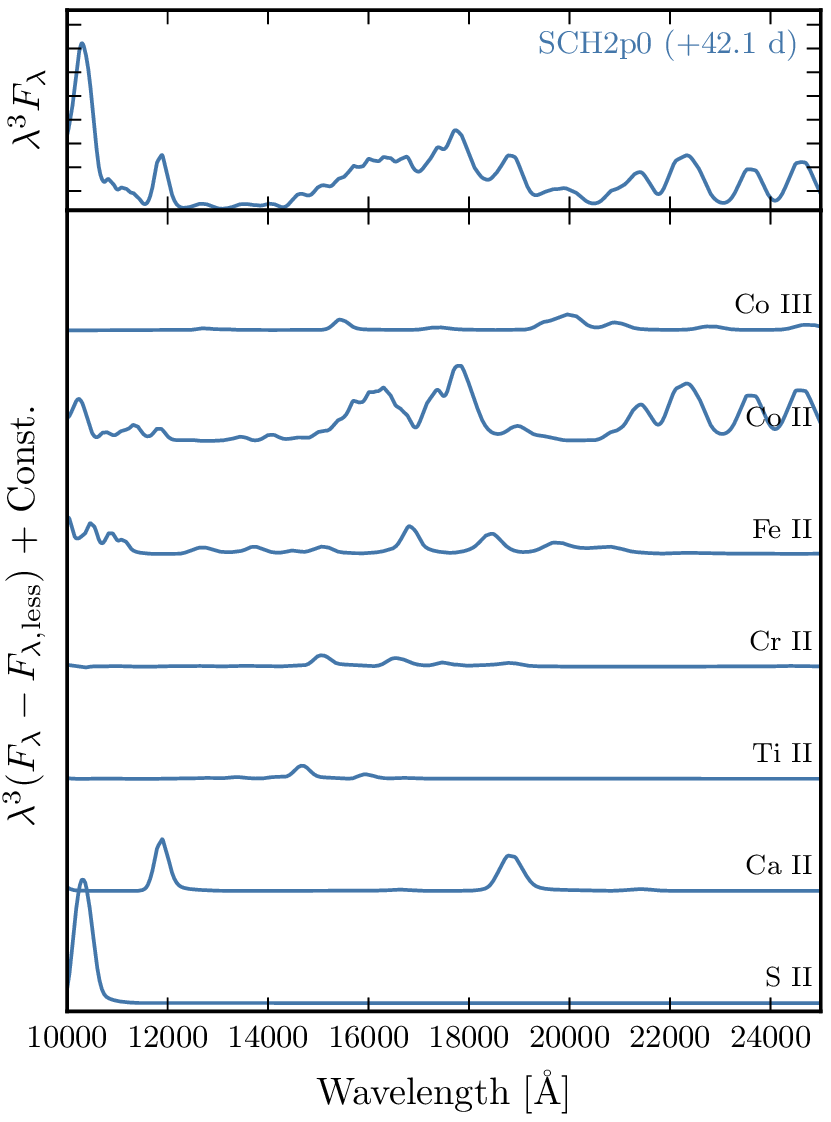}\vspace{.25cm}
\includegraphics{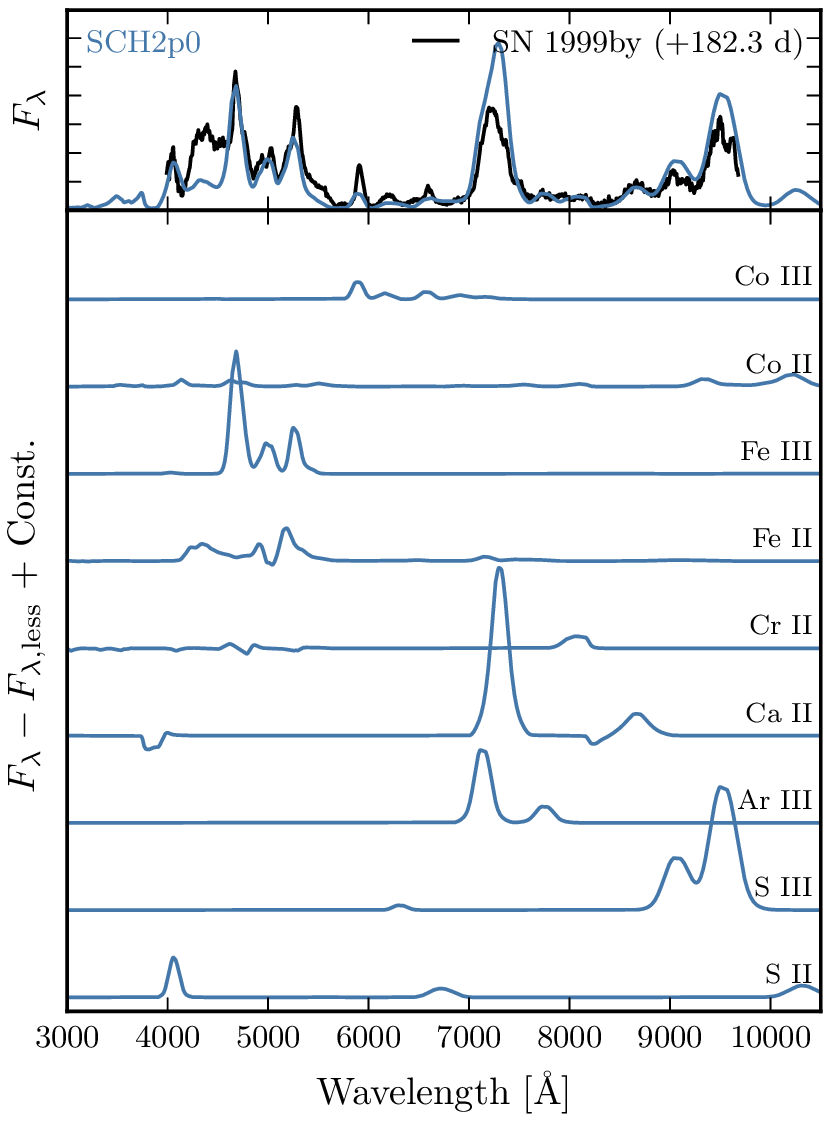}\hspace{.5cm}
\includegraphics{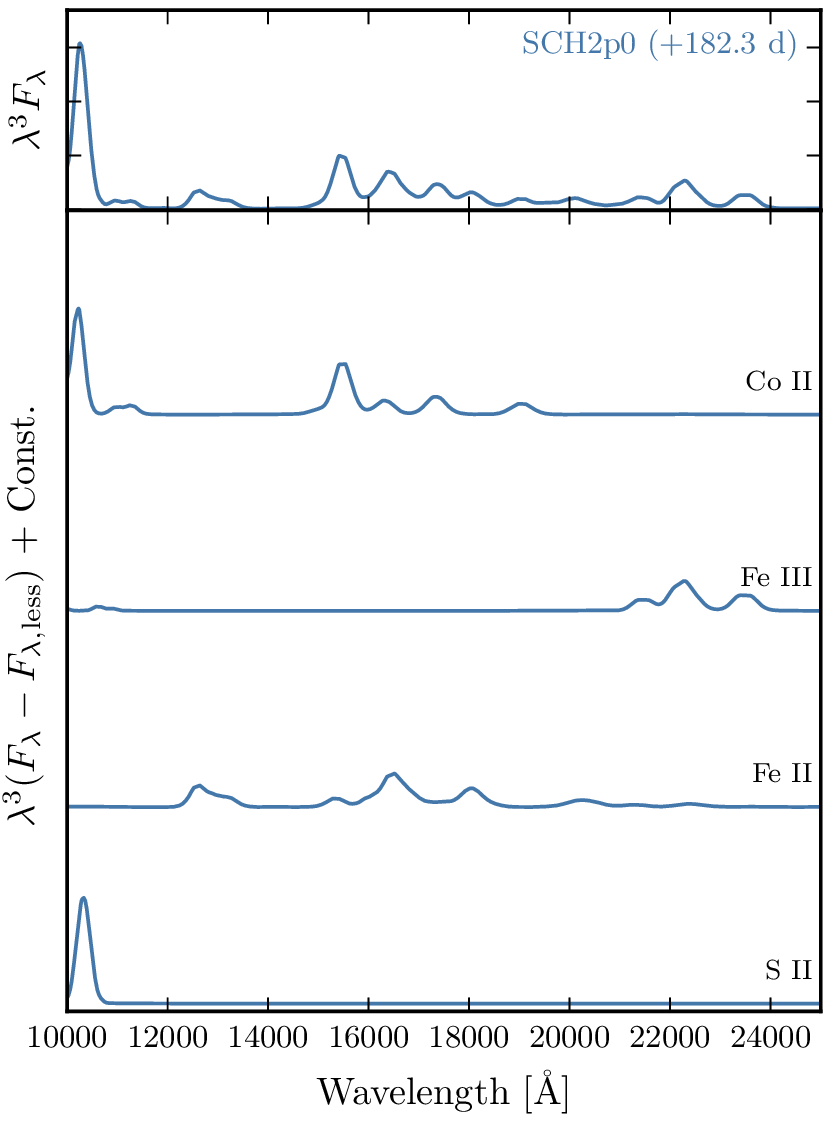}
\caption{\label{fig:ladder_5} 
Same as Fig.~\ref{fig:ladder_1} for spectra at $42$\,d (top) at
$182$\,d (bottom) past $B$-band maximum. There are no NIR spectra of
SN~1999by at these times. The NIR flux has been scaled by $\lambda^3$
for better visibility (not $\lambda^2$ as in
Figs.~\ref{fig:ladder_1}--\ref{fig:ladder_4}).  }
\end{figure*}

Figures~\ref{fig:ladder_1}--\ref{fig:ladder_5} reveal the contribution
of individual ions to the full optical and NIR synthetic spectra of
the sub-\mch\ model SCH2p0, compared to the low-luminosity SN~1999by
between $-5$\,d and $+182$\,d from $B$-band maximum.  The observed
spectra have been de-redshifted, de-reddened, and scaled to match the
absolute $V$-band (for the optical spectra) or $J$-band (for the NIR
spectra) magnitude inferred from the corresponding photometry. An
additional scaling has been applied to the optical synthetic spectra
to reproduce the mean observed flux in the range
5000--6500~\AA\ (Figs.~\ref{fig:ladder_1}--\ref{fig:ladder_4} and
upper panels of Fig.~\ref{fig:ladder_5}) or in the range
3000--10000\,\AA\ (bottom panels of Fig.~\ref{fig:ladder_5} showing
the spectra at +182.3\,d). For the NIR spectra, we use the wavelength
range 10000--11500\,\AA\ to compute the additional scale factor, and
further scale the flux by $\lambda^2$
(Figs.~\ref{fig:ladder_1}--\ref{fig:ladder_4}) or $\lambda^3$
(Figs.~\ref{fig:ladder_5}) for better visibility.

The ion spectra in the optical range up until +42.1\,d and in the NIR
range up until +0.2\,d are computed by taking the ratio of the full
spectrum ($F_\lambda$) to that excluding all bound-bound transitions
of the corresponding ion ($F_{\lambda,{\rm less}}$). Ion spectra
computed as a ratio marked with a ``*'' have been scaled down for
clarity. Only ions that impact the flux at the $>10$ per cent level at
a given phase are shown.

For the nebular optical spectrum at +182.3\,d and for the NIR spectra
from +3.2\,d onwards, the ratio $F_\lambda/F_{\lambda,{\rm less}}$
artificially enhances the strength of weak emission lines, and we
instead compute the ion spectra by taking the difference $F_\lambda -
F_{\lambda,{\rm less}}$. This approach preserves the relative
contribution of different ions to the total flux, but the global
scaling we adopt occasionally causes individual ion spectra to overlap
(see e.g. the optical nebular spectrum at +182.3\,d in
Fig.~\ref{fig:ladder_5}). Only ions that impact the flux at the $>5$
per cent level at a given phase are shown.

We caution the reader that these plots are merely illustrative. The
complex nature of the opacity in \sneia, characterized by weak
continuum opacities and numerous overlapping spectral lines, renders
problematic the quantitative evaluation of the contribution of
individual ions to the total flux at a given frequency.

%%% FIGURE: optical ladder plots

%%%%%%%%%%%%%%%%%%%%%%%%%%%%%%%%%%%%%%%%%%%%%%%%%%

% Don't change these lines
\bsp	% typesetting comment
\label{lastpage}
\end{document}